\documentclass[10.5pt]{article}
\usepackage{amsfonts,amsmath,color,graphicx,tikz,tipa,stmaryrd}

\newtheorem{theorem}{Theorem}[section]

\newtheorem{conjecture}[theorem]{Conjecture}

\newtheorem{property}[theorem]{Property}
\newtheorem{problem}[theorem]{Problem}

\newtheorem{remark}[theorem]{Remark}

\newdimen\tableauside\tableauside=1.0ex
\newdimen\tableaurule\tableaurule=0.4pt
\newdimen\tableaustep
\def\phantomhrule#1{\hbox{\vbox to0pt{\hrule height\tableaurule width#1\vss}}}
\def\phantomvrule#1{\vbox{\hbox to0pt{\vrule width\tableaurule height#1\hss}}}
\def\sqr{\vbox{%
  \phantomhrule\tableaustep
  \hbox{\phantomvrule\tableaustep\kern\tableaustep\phantomvrule\tableaustep}%
  \hbox{\vbox{\phantomhrule\tableauside}\kern-\tableaurule}}}
\def\squares#1{\hbox{\count0=#1\noindent\loop\sqr
  \advance\count0 by-1 \ifnum\count0>0\repeat}}
\def\tableau#1{\vcenter{\offinterlineskip
  \tableaustep=\tableauside\advance\tableaustep by-\tableaurule
  \kern\normallineskip\hbox
    {\kern\normallineskip\vbox
      {\gettableau#1 0 }%
     \kern\normallineskip\kern\tableaurule}%
  \kern\normallineskip\kern\tableaurule}}
\def\gettableau#1{\ifnum#1=0\let\next=\null\else
\squares{#1}\let\next=\gettableau\fi\next}

\newcommand{\ee}[1]{{{\rm e}^{#1}}}

\newcommand{\ds}{\textrm{\textsc{d}}}

\newcommand{\hc}{\textrm{\texthth}}
\newcommand{\beq}{\begin{equation}}
\newcommand{\eeq}{\end{equation}}
\newcommand{\bea}{\begin{eqnarray}}
\newcommand{\eea}{\end{eqnarray}}

\newcommand{\kt}{\underline{\kappa}(\tau)}

\renewcommand{\and}{{\qquad {\rm and} \qquad}}

\newcommand{\dd}{\mathrm{d}}

 \newcommand{\Tr}{{\,\rm Tr}\:}

\newcommand{\Res}{\mathop{\,\rm Res\,}}

\newcommand{\om}{\omega}

\newcommand{\Pint}{{\int\kern -1.em -\kern-.25em}}

\newcommand{\blue}{\color[rgb]{0.256226,0.383270,0.903061}}

\newcommand{\ab}{\mathrm{\mathbf{a}}}

\newcommand{\curve}{{\cal C}}
\newcommand{\spcurve}{{\cal S}}

\newcommand{\bcycle}{{\cal B}}

\newcommand{\Tau}{{\cal T}}

\definecolor{rouge}{rgb}{0.84,0.18,0.07}
\definecolor{bleu}{rgb}{0.22,0.41,0.74}
\definecolor{vertf}{rgb}{0.08,0.46,0.07}
\usepackage[pdftex]{hyperref}
\hypersetup{colorlinks,urlcolor=vertf,citecolor=rouge,linkcolor=bleu,filecolor=black}

\begin{document}

\begin{center}
{\large \textbf{All order asymptotics of hyperbolic knot invariants from non-perturbative topological recursion of A-polynomials}}
\end{center}

\vspace{26pt}

\begin{center}
\textsl{Ga\"etan Borot}\footnote{D\'epartement de Math\'ematiques, Universit\'e de Gen\`eve. \href{mailto:gaetan.borot@unige.ch}{\textsf{gaetan.borot@unige.ch}}},
 \textsl{Bertrand Eynard}\footnote{Institut de Physique Th\'eorique, CEA Saclay, \href{mailto:bertrand.eynard@cea.fr}{\textsf{bertrand.eynard@cea.fr}}}
\end{center}

\vspace{20pt}

\begin{abstract}
We propose a conjecture to compute the all-order asymptotic expansion of the colored Jones polynomial of the complement of a hyperbolic knot, $J_N(q = e^{\frac{2u}{N}})$ when $N \rightarrow \infty$. 
Our conjecture claims that the asymptotic expansion of the colored Jones polynomial is a formal wave function of an integrable system whose semiclassical spectral curve $\mathcal{C}$ would be the $\mathrm{SL}_2(\mathbb{C})$ character variety of the knot (the A-polynomial), and is formulated in the framework of the topological recursion. It takes as starting point the proposal made recently by Dijkgraaf, Fuji and Manabe (who kept only the perturbative part of the wave function, and found some discrepancies), but it also contains the non-perturbative parts, and solves the discrepancy problem. These non-perturbative corrections are derivatives of Theta functions associated to $\mathcal{C}$. For a large class of knots, this expansion is still in powers of $1/N$ due to the special properties of $A$-polynomials. We provide a detailed check of our proposal for the figure-eight knot and the once-punctured torus bundle $L^2R$. We also present a heuristic argument inspired from the case of torus knots, for which knot invariants can be computed by a matrix model.
\end{abstract}

\vspace{10pt}


\section{Introduction}

The asymptotic expansion of the colored Jones polynomial $J_N(\mathfrak{K},q)$ of a knot $\mathfrak{K}$ when $N \rightarrow \infty$, and more generally of invariants of $3$-manifolds, has received much attention recently. The terms of such an asymptotic expansion are also invariants of $3$-manifolds, which are interesting for themselves. They are generically called "perturbative invariants". Many intriguing properties of these expansions have been observed, first in relation with hyperbolic geometry and the volume conjecture \cite{Kashaev}, \cite{Murav}, then concerning arithmeticity \cite{GZLD}, modularity \cite{ZagierLaw} or quantum modularity \cite[Examples 4 and 5]{Zagierqmod}.

\subsection{Solutions of the A-hat recursion relation}

Garoufalidis and L\^{e} have shown that the Jones polynomial of a knot $\mathfrak{K} \subseteq \mathbb{S}_3$, denote $J_N(\mathfrak{K},q)$, is $q$-holonomic \cite{GarouLe}: if we denote by $\Delta$ the shift $N \rightarrow N + 1$, there exists an operator $\mathfrak{A}_K(q^{N/2},\Delta,q)$ which is polynomial in its three variables, so that $\widehat{\mathfrak{A}}_K\cdot J_N(\mathfrak{K},q) = 0$. More generally, one may consider the space of solutions $\mathcal{J}_{\hbar}(u)$ of the difference equation
\beq
\label{Aha}\widehat{\mathfrak{A}}_K(e^{u},e^{\hbar\partial_u},e^{2\hbar})\cdot\mathcal{J}_{\hbar}(u) = 0,
\eeq
where we replaced formally $u = N\hbar$ and $q = e^{2\hbar}$. The Jones polynomial is by construction a solution of \ref{Aha} where $u/\hbar$ is restricted to a discrete set of values. One may also look for solutions of \ref{Aha} among formal power series of the form:
\beq
\label{ququq}\mathcal{J}_{\hbar}(u) = \hbar^{\delta/2}\exp\Big(\sum_{\chi \geq -1} \hbar^{\chi}\,\jmath_{\chi}(u)\Big).
\eeq
The Wilson loops in the representation of dimension $N$ in the $\mathrm{SL}_{2}(\mathbb{C})$ Chern-Simons theory (viewed as a perturbative quantum field theory after expansion around a flat connection with meridian holonomy $u$, with coupling constant $\hbar = \mathrm{i}\pi/\mathrm{integer}$) provide such solutions, let us call them $\mathcal{J}_{\hbar}^{\mathrm{CS},(\alpha)}(u)$. For some examples of hyperbolic $3$-manifolds and a choice of triangulation, it has been observed \cite{GZLD} that the asymptotics of Hikami integral\footnote{The Hikami integral is a finite-dimensional constructed from a triangulation of a hyperbolic $3$-manifold and an elimination procedure ; however, it is not an invariant of $3$-manifolds.} $\mathcal{J}_{\hbar}^{\mathrm{H},(\alpha)}(u)$ (which depend on a choice of integration contour $\gamma_{\alpha} \subseteq \mathbb{C}$) coincides with $\mathcal{J}_{\hbar}^{\mathrm{CS},(\alpha)}(u)$. In other words, $\mathcal{J}_{\hbar}^{\mathrm{H},(\alpha)}$ are also solutions (in those examples) of \eqref{Aha}.

We would like to propose a third method which we conjecture to provide formal solutions of \ref{Aha} for any hyperbolic $3$-manifold $\mathfrak{M}$, and relies only on algebraic geometry on the $\mathrm{SL}_2(\mathbb{C})$ character variety of $\mathfrak{M}$ (Conjecture~\ref{ahconj}). The latter is a complex curve obtained as the zero locus of a polynomial $\mathfrak{A}(m,l) \in \mathbb{Z}[m,l]$, called the A-polynomial of $\mathfrak{M}$. The AJ conjecture \cite{AJconj} states that:
\beq
\label{alini}\lim_{\hbar \rightarrow 0} \widehat{\mathfrak{A}}_{\hbar}(m,l,e^{2\hbar}) \propto \mathfrak{A}(m,l),
\eeq
where $\propto$ means up to an (irrelevant) polynomial in $m$. Eqn.~\ref{alini} has been checked in numerous examples (it holds for instance for the figure-eight knot) and has been proved recently for a infinite class of knots \cite{AJLe}. In the light of the AJ conjecture, we can summarize our work by saying that we propose an algorithm to construct formal solutions of \ref{Aha} starting only from the classical limit of the operator $\widehat{\mathfrak{A}}$. For the figure-eight knot, we have checked that it gives a correct result for the first few terms.

The final goal would be to identify our series with the genuine all-order asymptotics of invariants in $3$-dimensional defined in the realm of quantum topology, like the colored Jones polynomial. This step is subtle because of wild behavior when $q$ is a root of unity, and non trivial Stokes phenomena, as one can already observe already in the case of the figure-eight knot, for which rigorous results of Murakami \cite{Murakami1} are available (see~\S~\ref{volc}). Though not completely predictive on the range of validity in $u$, the generalized volume conjecture of Gukov asserts that $J_{u/\hbar}(q = e^{\hbar})$ has an asymptotic expansion of the form \eqref{ququq} when $\hbar = {\rm i}\pi/k$ and $k$ is a integer going to infinity, and the coefficients coincide with those of $\mathcal{J}_{\hbar}^{\mathrm{CS},(\alpha)}(u)$ for some $\alpha$. In this framework, we can also reformulate our conjecture by saying that our method retrieves the coefficients in the expansion of the colored Jones polynomial (Conjecture~\ref{qiq}), and will discuss in \S~\ref{volc} how this statement has to be understood.

\subsection{Historical background}

Let us describe briefly the origin of our proposals. Twenty years ago, Witten showed in his pioneering article \cite{Witten89} that Wilson loop observables in a Chern-Simons theory with gauge group $G$ on $\mathbb{S}^3\setminus \mathfrak{K}$ where $\mathfrak{K}$ is a knot, compute knot invariants. 
Moreover, he proposed a correspondence between Chern-Simons theory on a $3$-manifold $\mathfrak{M}$ and topological string theory on $T^*\mathfrak{M}$ \cite{Witteng}, which has been developed later on \cite{GopaV}, \cite{MarcosRev}. More recently, Bouchard, Klemm, Mari\~{n}o and Pasquetti \cite{BKMP} suggested\footnote{This conjecture has been proved lately \cite{BKMPproof}.} that amplitudes in topological string theory can be computed from the axiomatics of the "topological recursion" developed in \cite{EOFg}. Putting these two ideas together, Dijkgraaf, Fuji and Manabe \cite{DiFuji1}, \cite{DiFuji2}, proposed that the topological recursion's wave function, applied to the $\mathrm{SL}_2(\mathbb{C})$-character variety of the knot, coincides with $\mathcal{J}_{\hbar}^{\mathrm{CS}}(u)$. However, they kept only the "perturbative part" of the topological recursion's wave function, and their conjectured formula did not match $\mathcal{J}_{\hbar}^{\mathrm{CS}}(u)$ or $\mathcal{J}_{\hbar}^{\mathrm{H}}(u)$ for the figure-eight knot. They could fix this mismatching problem by introducing additional \textit{ad hoc} constants to all orders. Here, we propose a formula using the thoroughly non-perturbative wave function introduced in \cite{Ecv}, \cite{EMhol}, which should successfully match $\mathcal{J}_{\hbar}^{\mathrm{CS}}(u)$ without having to introduce additional terms.

\subsection{Short presentation}

Let us give a flavor of our construction (all terms will be defined in the body of the article). The geometric component $\mathfrak{A}(m,l) = 0$ of the A-polynomial of $\mathfrak{K}$ has a smooth model which is a compact Riemann surface $\mathcal{C}_0$ of genus $g$. It is endowed with a point $p_c$ corresponding to the complete hyperbolic metric on $\mathbb{S}_3\setminus\mathfrak{K}$, and a neighborhood $\tilde{\mathcal{U}} \subseteq \mathcal{C}_0$ of $p_c$ in bijection with a neighborhood $\mathcal{U}\subseteq \mathbb{C}$ of $\mathrm{i}\pi$ which parametrizes deformations of the hyperbolic metric of $\mathbb{S}_3\setminus\mathfrak{K}$. Let $p_u$ the unique point in $\tilde{\mathcal{U}}$ such that $m(p_u) = e^{u}$. We denote $\iota$ the involution of $\mathcal{C}_0$ sending $(m,l)$ to $(1/m,1/l)$. In particular we have $p_{-u} = \iota(p_u)$. In the following, $p$ denotes a point of the curve, and in the comparison with the asymptotics of the colored Jones near $u = {\rm i}\pi$, one wishes to specialize at $p = p_u$. Let  $(\mathcal{A},\mathcal{B})$ be a symplectic basis of homology on $\mathcal{C}_0$. We construct a formal asymptotic series with leading coefficient:
\beq
\widetilde{\jmath}_{-1}(p) = \int_{o}^{p} \ln l\,\dd \ln m,
\eeq
and for $\chi \geq 1$,
\beq
\label{fora}\widetilde{\jmath}_{\chi}(u) = \frac{1}{2} \sum_{\ell = 1}^{\chi} \sum_{\substack{2h_j - 2 +d_j+n_j> 0 \\ \sum_j 2h_j - 2 + d_j + n_j = \chi}} \Big(\bigotimes_{j = 1}^{\ell} \frac{\int_{\bullet}\cdots\int_{\bullet} \omega_{n_j}^{h_j,(d_j)}}{(2\mathrm{i}\pi)^{d_j}\,d_j!\,n_j!}\Big)\cdot U_{\ell,\mathbf{d},\bullet}.
\eeq
The notation $\int_{\bullet}$ is used for $\int^{p}_{o} + \int^{\iota(p)}_{\iota(o)}$ for some basepoint $o$, $\omega_n^h$ are the differentials forms computed by the topological recursion for the spectral curve $(\mathcal{C}_0,\ln m,\ln l)$ with a Bergman kernel normalized on $\mathcal{A}$-cycles, and $U_{\ell,\mathbf{d},\bullet}$ is $(0,|\mathbf{d}|)$ tensor which is a sum of terms of the form:
\beq
\bigotimes_{i = 1}^{s}\frac{\nabla^{\otimes(\sum_{j \in J_i} d_j)}\vartheta_{\bullet}\bigl[{}^{\mu}_{\nu}\bigr]}{\vartheta_{\bullet}\bigl[{}^{\mu}_{\nu}\bigr]},
\eeq
where $J_1,\ldots,J_s$ denote a partition of $\{1,\ldots,\ell\}$ is $s$ subsets. $\vartheta\bigl[{}^{\mu}_{\nu}\bigr]$ denotes the theta function with characteristics $\mu,\nu \in \mathbb{C}^g$ associated to the matrix of periods of $\mathcal{C}_0$ for the chosen basis of homology. The notations $\vartheta_{\bullet}\bigl[{}^{\mu}_{\nu}\bigr]$ means that we specialize its argument to $\mathbf{w} = \mathbf{w}_{\bullet} \equiv \int_{\bullet} \dd\ab + \zeta_{\hbar}$, where $\dd\ab$ is the vector of holomorphic differentials dual to the $\mathcal{A}$-cycles, $\zeta_{\hbar} \in \mathbb{C}^g$ is a constant defined in Eqn.~\ref{reca}, and $\nabla$ the gradient acting on the argument $\mathbf{w}$. Besides, the undotted version of theta means that we specialize to $\mathbf{w} = \zeta_{\hbar}$.

Formula~\ref{fora} depends on a choice of basepoint $o$ and of a characteristics $\mu,\nu \in \mathbb{C}^g$. If we change the homology basis $(\mathcal{A},\mathcal{B})$, we merely obtain the same quantities for a different characteristics $\mu,\nu$. The dependence of $\jmath_{-1}(u)$ of the choice of branches for the logarithms will be discussed later. For instance, the first coefficient is given by:
\beq
\begin{split}
\label{exampleJ1}
2\,\widetilde{\jmath}_1(p) 
&= \int^{\bullet} \om_1^1 + \frac{1}{6} \int_{\bullet}\int_{\bullet}\int_{\bullet} \om_3^0 + \Big(\frac{\nabla \vartheta_\bullet\bigl[{}^{\mu}_{\nu}\bigr]}{2\mathrm{i}\pi\,\vartheta_\bullet\bigl[{}^{\mu}_{\nu}\bigr]}-\frac{\nabla \vartheta\bigl[{}^{\mu}_{\nu}\bigr]}{2\mathrm{i}\pi\,\vartheta\bigl[{}^{\mu}_{\nu}\bigr]}\Big)\oint_{\bcycle} \om_1^1 \\ 
& + \frac{1}{2}\,\frac{\nabla \vartheta_\bullet\bigl[{}^{\mu}_{\nu}\bigr]}{2\mathrm{i}\pi\,\vartheta_\bullet\bigl[{}^{\mu}_{\nu}\bigr]}\,\oint_{\bcycle} \int_{\bullet} \int_{\bullet} \om_3^0 + \frac{1}{2}\,\frac{\nabla^{\otimes 2} \vartheta_\bullet\bigl[{}^{\mu}_{\nu}\bigr]}{(2\mathrm{i}\pi)^2\,\vartheta_\bullet\bigl[{}^{\mu}_{\nu}\bigr]}\,\oint_{\bcycle} \oint_{\bcycle} \int_{\bullet}\om_3^0  \\
& + \frac{1}{6}\Big(\frac{\nabla^{\otimes 3} \vartheta_\bullet\bigl[{}^{\mu}_{\nu}\bigr]}{(2\mathrm{i}\pi)^3\,\vartheta_\bullet\bigl[{}^{\mu}_{\nu}\bigr]} - \frac{\nabla^{\otimes 3} \vartheta\bigl[{}^{\mu}_{\nu}\bigr]}{(2\mathrm{i}\pi)^3\,\vartheta\bigl[{}^{\mu}_{\nu}\bigr]}\Big)\oint_{\bcycle} \oint_{\bcycle} \oint_{\bcycle} \om_3^0.
\end{split}
\eeq
In general, $\zeta_{\hbar}$ depend on $\hbar$ in a non trivial way, so our series is not a priori a power series in $\hbar$. It is however a well-defined formal asymptotic series: as we will see, $j_{\chi}(u)$ is actually a function of $\hbar$, which does not have a power series expansion in powers of $\hbar$. But A-polynomials of $3$-manifolds are very special polynomials: for K-theoretical reasons, $\zeta_{\hbar}$ is constant along sequences $\hbar = \mathrm{i}\pi/k$ where $k$ ranges over the integers. Hence, $j_{\chi}(u)$ specialized to such subsequences of $\hbar$, is indeed a function of $u$ only. We also point out that another huge simplification occurs for a certain class of knots (containing the figure-eight knot). Let us denote $\iota_*$, the linear involution induced by $\iota$ on the homology of $\mathcal{C}_0$. When $\iota_* = -\mathrm{id}$, we have actually $\zeta_{\hbar} = 0$. Thus, our series is always a power series in $\hbar$ (without restriction) in this case.

The proposal of Dijkgraaf, Fuji and Manabe \cite{DiFuji2} is tantamount to setting $U_{\ell,\mathbf d,\bullet} \equiv \delta_{\ell,1}\,\delta_{\mathbf d,\mathbf 0}$, and thus miss the theta functions. For the figure-eight knot, as we indicated, $U_{\ell,\mathbf{d},\bullet}$ contributes as constants in $\widetilde{j}_{\chi}$ for $\chi \geq 1$, and their value explain the renormalizations observed by these authors. To summarize, they are due the fact that the geometric component of the character variety is not simply connected. 

The leading coefficient $\widetilde{\jmath}_{-1}(p_u)$ is known to be related to the complexified volume of $M$ for a family of uncomplete hyperbolic metrics parametrized by $u$. Within our conjecture, the other coefficients $\widetilde{\jmath}_{\chi}(u)$ also acquires a geometric meaning, as primitives of certain meromorphic $1$-forms on the $\mathrm{SL}_2(\mathbb{C})$ character variety. The computation of the coefficients with our method is less efficient than making an ansatz like Eqn.~\ref{ququq}, pluging into the A-hat recursion relation and solving for the coefficients \cite{GZLD}, \cite{Shengmao}. However, it underlines the relevance of the geometry of the character variety itself for asymptotics of knot invariants, and also suggests unexpected links between knot theory and other topics in mathematical physics (Virasoro constraints, integrable systems, intersection theory on the moduli space, non-perturbative effects, etc.), via the topological recursion. It also provides a natural framework to discuss the arithmetic properties of perturbative knot invariants, at least when $\iota_* = -\mathrm{id}$.

\subsection{Outline}

We first review the notions of geometry of the character variety needed to present our construction (Section~\ref{S1}), and the axiomatics of the topological recursion with the definition of the correlators, the partition function and the kernels (Section~\ref{S2} and \ref{S3}). We state precisely our conjecture concerning the asymptotic expansion of the Jones polynomial in Section~\ref{S4}, and check it to first orders for the figure-eight knot and the manifold $L^2R$. Our intuition comes from two other aspects of the topological recursion, namely its relation to integrable systems \cite{BEInt} and to matrix integrals \cite{AM90}, \cite{E1MM}, \cite{CEdiag}. We give some heuristic motivations in Section~\ref{S7}, by examining the relation of our approach with computation of torus knots invariants from the topological recursion presented in \cite{BEMknots}. This section is however independent of the remaining of the text. In appendix \ref{appdiag}, we propose a diagrammatic way to write $\jmath_{\chi}$, which may help reading the formulas, but requires more notations.

\section{$\mathrm{SL}_2(\mathbb{C})$-character variety and algebraic geometry on A-spectral curves}
\label{S1}
We review standard facts on the $\mathrm{SL}_2(\mathbb{C})$-character variety of $3$-manifolds, especially of hyperbolic $3$-manifolds with $1$-cusp. A work of reference is \cite{CCGLS}, where most of the facts presented here are rigorously stated and proved. In Sections~\ref{algcurves}-\ref{bp}, we focus on the (irreducible components of the) character variety seen as a compact algebraic curves. In order to prepare the presentation of the topological recursion, we describe some algebraic geometry of the character variety, with the notion of branchpoints, symplectic basis of cycles, theta functions, Bergman kernel, etc.

\subsection{A-polynomial and spectral curve}

Let $\mathfrak{M}$ be a $3$-manifold with one cusp. If $R$ a $\mathrm{SL}_2(\mathbb{C})$ representation of $\pi_1(M)$ and $(\mathfrak{m},\mathfrak{l})$ a basis of $\pi_1(\partial M) \simeq \mathbb{Z}^2$, $R[\mathfrak{m}]$ and $R[\mathfrak{l}]$ can be written in Jordan form:
\beq
R[\mathfrak{m}] = \left(\begin{array}{ll} m & \,\star \\ 0 & m^{-1}\end{array}\right),\qquad R[\mathfrak{l}] = \left(\begin{array}{ll} l & \,\star \\ 0 & l^{-1}\end{array}\right),
\eeq
up to a global conjugation. When $\mathfrak{M}$ is a knot complement in $\mathbb{S}_3$, the choice of $\mathfrak{l}$ and $\mathfrak{m}$ is canonically the longitude and the meridian around the knot. In general, we continue to call the (arbitrarily chosen) $(\mathfrak{m},\mathfrak{l})$ \emph{meridian} and \emph{longitude}.

The locus of possible eigenvalues $(m,l) \in \mathbb{C}^*\times\mathbb{C}^*$, also called \emph{character variety}, has been studied in detail in \cite{CCGLS}. It is the union of points curves. In particular, the union of the $1$-dimensional components is non empty and coincide with the zero locus of a polynomial with integer coefficients: $\mathfrak{A}(m,l) = 0$. The latter is uniquely defined up to normalization and is called the A-polynomial of $\mathfrak{M}$. The A-polynomial is topological invariant of $3$-manifolds endowed with a choice of basis of $\pi_1(\partial M)$, and it contains a lot of geometric information about $\mathfrak{M}$.

The A-polynomial has many properties, and we shall highlight those we need along the way. The first one is that, since $(m,l)$ and $(1/m,1/l)$ describe the same representation up to conjugation, the A-polynomial is quasi-reciprocal: there exists integers $a,b$ and a sign $\varepsilon$ such that $\varepsilon\,m^al^b\,\mathfrak{A}(1/m,1/l) = \mathfrak{A}(l,m)$.
To simplify, we assume throughout the paper that $\mathfrak{M}$ is a knot complement in a homology sphere, although most of the ideas can be extended to arbitrary $3$-manifolds. In particular, this assumption implies \cite{CCGLS} that the A-polynomial is actually even in $m$. We take this property into account by defining $x = m^2$ and $y = l$. The 1-form:
\beq
\label{phidef1}\phi = \ln l\,\dd \ln m
\eeq
is related to a notion of volume and will play an important role.

The A-polynomial might not be irreducible. We denote generically $A(x,y)$ one of its irreducible factor, which is a polynomial with integer coefficients considered in the variables $x$ and $y$. We use the generic name \emph{component} to refer to the subvariety $\mathcal{C}_0$ defined by $A(x,y) = 0$ in $\mathbb{C}^{\times}\times\mathbb{C}^{\times}$. There always exists reducible representations with $l = 1$ and $m \neq 0$, so one of the irreducible factor is $(l - 1)$, and it defines the \emph{abelian component}. The A-polynomial of the unknot is precisely $(l - 1)$, but in general there are several non-abelian components. In a given non-abelian component $\mathcal{C}_0$, there always exists points corresponding to reducible representations, i.e. $\mathcal{Z}_0 = \mathcal{C}_0\cap \{(x = 1,y = 1),(x = 1,y = -1)\} \neq \emptyset$. $\mathcal{Z}_0$ is actually the set of singular points of $\mathcal{C}_0$. After a birational transformation with integer coefficients and poles at the singular values of $x$, we can always find a smooth algebraic curve $\mathcal{C}$ which models $\mathcal{C}_0$. $x$ and $y$ are then meromorphic functions of $\mathcal{C}$. We refer to the triple $(\mathcal{C},x,y)$ as the \emph{spectral curve} of the component we considered. The examples treated in Section~\ref{S6} illustrate the method to arrive unambiguously to the spectral curve.

\subsection{Properties of the A-polynomial}
\label{PropA}
As a polynomial, the A-polynomial of a $3$-manifold is very special: it satisfies the Boutroux condition and a quantization condition. These two properties hold for any $3$-manifold (and any component of its A-polynomial). They come from a property in K-theory, which is proved in \cite[p59]{CCGLS}, and were clarified in \cite{LiWang} Before coming to the K-theoretic point of view, let us describe these properties. 

\subsubsection*{Boutroux condition}

We have a \emph{Boutroux property}: for any closed cycle $\Gamma \subseteq \mathcal{C}\setminus\mathcal{Z}_0$,
 \beq
\label{Bou}\oint_{\Gamma}\mathrm{Im}\,\phi = 0.
\eeq
For hyperbolic $3$-manifolds, this is related to the existence of a function giving the hyperbolic volume. The Boutroux condition has been underlined in \cite{KapaB} for plane curves of the form $\mathrm{Pol}(x,y) = 0$ endowed with the $1$-form $\phi = y\,\dd x$. It appears naturally in the asymptotic study of matrix integrals, (bi)orthogonal polynomials and Painlev\'e transcendants, and is related to a choice of steepest descent integration contours to apply a saddle-point analysis \cite{BeMo}, \cite{Bertolacut}. Actually, Hikami observed \cite{Hikami} that the A-polynomial can be obtained as the saddle-point condition in integrals of product of quantum dilogarithm constructed from triangulations and related to knot invariants. So, it is not surprising to meet a Boutroux property here.

\subsubsection*{Quantization condition}

The real periods of $\phi$ are quantized: there exists a positive integer $\varsigma$ such that, for any closed cycle $\Gamma \subseteq \mathcal{C}\setminus\mathcal{Z}_0$ with base point $p_0$ such that $\ln m(p_0) = 0,{\rm i}\pi$,
\beq
\label{Quan}\oint_{\Gamma} \mathrm{Re}\,\phi \in \frac{2\pi^2}{\varsigma}\cdot\mathbb{Z}.
\eeq
This condition has first been pointed out by Gukov \cite{GukovV} in his formulation of the generalized volume conjecture, as a necessary condition for the $\mathrm{SL}_2(\mathbb{C})$-Chern-Simons theory to be quantizable. In our framework also, Eqn.~\ref{Quan} implies the existence of a expansion in powers of $\hbar$ for certain quantities. We explain the mechanism in Section~\ref{BAspinor}.



\subsection{Triangulations and hyperbolic structures on $3$-manifolds}
\label{HypTri}

The A-polynomial of a hyperbolic $3$-manifold $\mathfrak{M}$ is closely related to the deformations of the hyperbolic structure on $\mathfrak{M}$. Since all our examples are taken from hyperbolic manifolds, we review this relation and follow the foundational work of W.~Thurston \cite{ThurstonNotes} and Neumann and Zagier \cite{ZN}. By definition, a $3$-oriented manifold $\mathfrak{M}$ is \emph{hyperbolic} if it can be endowed with a smooth, complete hyperbolic metrics with finite volume. There exists an infinite number of hyperbolic knots, i.e. knots whose complement in the ambient space is hyperbolic.

Mostow rigidity theorem then states that the metrics in the definition above is unique. $\mathfrak{M}$ is either compact, or has $c$ cusps. Thurston explains that, modulo Dehn surgery on the cusps, $\mathfrak{M}$ can be decomposed in a set of ideal tetrahedra glued face to face. Ideal means that all vertices of the triangulations are on the cusps. We imagine that the Dehn surgery has already been performed and start with a triangulable manifold $\mathfrak{M}$. It is then the interior of an oriented, bordered compact $3$-manifold, whose boundary consists in $c$ tori. So, its Euler characteristics is $0$, and counting reveals that the number of tetrahedra $N_T$ equals the number of distinct edges in the triangulation. And, by construction, the number of vertices is $c$, the number of cusps.

In a ideal tetrahedron $\mathfrak{T}$, let us choose an oriented edge $e$ pointing towards a vertex $\circ$. If we intersect $\mathfrak{T}$ by a horosphere centered at $\circ$, we obtain a triangle whose sum of angles is $\pi$. It is thus similar to some euclidean triangle $T(z_e)$ with vertices $0$, $1$ and $z_e$. We choose a representative for which the image of $z_e$ in the tetrahedron belongs to $e$, and such that $\mathrm{Im}\,z_{e} > 0$. $z_e$ is called a \emph{shape parameter}, and we may define in a unique way logarithmic shape parameters $\zeta_e = \ln z_e$, which are more natural to express geometric conditions.

For a given vertex with incident edges $e_1,e_2,e_3$ in cyclic order, the shape parameters are $z_{e_1}$, $1 - z_{e_1}^{-1}$ and $(1 - z_{e_1})^{-1}$. As a manifestation of the angle sum condition around a Euclidean condition, we have:
\beq
\zeta_{e_1} + \zeta_{e_2} + \zeta_{e_3} = \mathrm{i}\pi
\eeq
and in particular, the product of the shape parameters around a vertex is always $-1$. The shape parameter of an edge in opposite orientation is $z_{-e} = -z_{e}^{-1}$. Opposite edges in the tetrahedron have the same shape parameter. Thus, the triangulation depends a priori on $N_T$ shape parameters.

An oriented edge in the ideal triangulation of $\mathfrak{M}$ correspond to the identification of a collection of distinct oriented edges $(e_j)_j$ of the tetrahedra. Since $\mathfrak{M}$ is smooth along edges, we have $N_T$ gluing conditions, which are in general redundant:
\beq
\label{gluing}\sum_{j} \zeta_{e_j} = 2\mathrm{i}\pi,
\eeq
The data of $N_T$ shape parameters satisfying Eqn.~\ref{gluing} fixes a hyperbolic metrics with finite volume for $\mathfrak{M}$, which in general becomes singular at the vertices of the triangulation. The completion $\mathfrak{M}_{\mathbf{z}}$ of $\mathfrak{M}$ with respect to this metrics is a topological space, which may differs from $\mathfrak{M}$ by addition of set of points at the cusps. $\mathfrak{M}_{\mathbf{z}}$ happens to be a genuine hyperbolic manifold iff for any vertex $\bullet \in \{1,\ldots,c\}$ in the triangulation:
\beq
\label{completeg}\alpha_{\bullet} \equiv \sum_{e_\bullet} \zeta_{e_\bullet} = 2\mathrm{i}\pi.
\eeq
where $\{e_\bullet\}$ is the set of oriented edges of tetrahedra whose image in the triangulation points towards $\bullet$. It is shown in \cite{ThurstonNotes}, \cite{ZN}, that the set of solutions of \ref{gluing}-\ref{completeg} is discrete. Moreover, in the neighborhood of a solution (i.e. of a manifold $\mathfrak{M}_{\mathbf{z}_0}$), the cusp anomalies $\alpha_\bullet$ are local coordinates for the set of solutions of \ref{gluing}.

In a triangulated hyperbolic $3$-manifold $\mathfrak{M}_{\mathbf{z}_0}$, there is a natural $\mathrm{PSL}_2(\mathbb{C})$ representation, namely the holonomy representation. If we assume only $c = 1$ cusp, let us choose two closed paths $\gamma_{\mathfrak{m}},\gamma_{\mathfrak{l}} \subseteq \mathfrak{N}$ which are representatives of a meridian and a longitude. Then, the holonomy eigenvalues $(m,l)$ arise such that $m^2$ (resp. $l^2$) is the product of the shape parameters of the oriented edges crossed by $\gamma_{\mathfrak{m}}$ (resp. $\gamma_{\mathfrak{l}}$). The holonomy representation can be lifted \cite[p71]{CCGLS} to a $\mathrm{SL}_2(\mathbb{C})$ representation. The lift is not unique because of a choice of square root, but we always have $l_c = -1$ \cite{Calegari}, and we can choose $m_c = -1$. Now, the hyperbolic structure of $\mathfrak{M}$ has a $1$-parameter deformation in the neighborhood of $\mathfrak{M}_{\mathbf{z}_0}$, and $(m,l)$ can be defined in a unique way as continuous (in fact, holomorphic) functions along the deformation. Also, the locus $\mathcal{U}$ of $(m,l)$ achieved by the deformation is included in some $1$-dimensional component of the $\mathrm{SL}_2(\mathbb{C})$-character variety: the \emph{geometric component} $\mathcal{C}_0^{\mathrm{geom}}$. In other words, the deformation selects an irreducible factor of the A-polynomial, as well as a point $p_c$ on the spectral curve $\mathcal{C}^{\mathrm{geom}}$. $p_c$ is uniquely defined by the initial value $(m_c,l_c) = (-1,-1)$ and the infinitesimal deformation around. Logarithmic variables on $\mathcal{C}^{\mathrm{geom}}$ are also very useful. We define $(u,v)$ as analytic functions on $\mathcal{C}$ assuming the initial value $(\mathrm{i}\pi,\mathrm{i}\pi)$ at $p_c$, and such that $(m,l) = (e^{u},e^{v})$. The branches of the logarithm in Eqn.~\ref{phidef1} can be unambiguously chosen as:
\beq
\phi = v\,\dd u.
\eeq

For a countable set of points $p$ in $\mathcal{C}^{\mathrm{geom}}$, the space $\mathfrak{M}_{\mathbf{z}}$ is not as wild
as in the generic case. Indeed, if there exists coprime integers $q,q'$ such that $qu + q'v = \mathrm{i}\pi$, $\mathfrak{M}_{\mathbf{z}}$ is a manifold that differ from $\mathfrak{M}$ by adjonction a geodesic circle at the cusp, which is obtained by performing a $(q,q')$ Dehn surgery on $\mathfrak{M}$.

 \subsection{Volume and Chern-Simons invariant}
 \label{VCS}
By a standard computation, the volume of an ideal tetrahedron with shape parameter $z$, endowed with its complete hyperbolic metrics, is given by the Bloch-Wigner dilogarithm $D(z)$, which is a continuous function defined on $\mathbb{C}\setminus\{0,1\}$ as:
\beq
\label{BlochW}D(z) = \mathrm{Im}\,\mathrm{Li}_2(z) + \mathrm{arg}(1 - z)\ln |z|.
\eeq
The hyperbolic volume of $\mathfrak{M}_{\mathbf{z}}$ is thus:
\beq
\label{vold}\mathsf{Vol}(\mathfrak{M}_{\mathbf{z}}) =  \sum_{e} D(z_{e}),
\eeq
and the functional relations satisfied by the dilogarithm ensure that it does not depend on the triangulation.

Another invariant of hyperbolic $3$-manifolds is the \emph{Chern-Simons invariant}. For compact manifolds, it was introduced in \cite{ChernS} and belongs to $\mathbb{R}/(2\pi^2\mathbb{Z})$. For manifolds $\mathfrak{M}_{\mathbf{z}}$ obtained by Dehn surgery on a hyperbolic manifold $\mathfrak{M}$, this definition was generalized in \cite{Meyerhoff} and the invariant belongs to $\mathbb{R}/\pi^2\mathbb{Z}$. Its definition in terms of a triangulation involves $\sum_{e} \mathrm{Re}\,\mathrm{Li}_2(z_e)$ plus a tricky part described in \cite{NeumannTri}. Notice that, a priori, the Chern-Simons invariant $\mathsf{CS}(\mathfrak{M}_{\mathbf{z}})$ only make sense when $\mathfrak{M}_{\mathbf{z}}$ is a true manifold.

From a differential geometry standpoint, \cite{ZN} and independently \cite[Theorem 2]{Yoshida} proved that both invariants can be extracted from the function on $\mathcal{C}^{\mathrm{geom}}$:
\beq
\Phi(p) \equiv \int_{o}^p \phi = \int_o^p v\,\dd u,\qquad \dd\Phi(p) = \frac{{\rm i}}{2}\,\dd\big(\mathsf{Vol}_a + i\mathsf{CS}_a).
\eeq
For a point $p \in \tilde{U}\subseteq \mathcal{C}^{\mathrm{geom}}$, the volume of $\mathfrak{M}_{\mathbf{z}(p)}$ is directly related to the imaginary part of $\Phi(p)$:
\beq
\mathsf{Vol}(\mathfrak{M}_{\mathbf{z}(p)}) - \mathsf{Vol}(\mathfrak{M}_{\mathbf{z}(p_c)}) = \mathsf{Vol}_a(p) - 2\,\mathrm{Re}\,u\,\mathrm{Im}\,v,
\eeq
and thanks to the Boutroux condition, it does not depend of the path from $p_c$ to $p$. If we assume that $\mathfrak{M}_{\mathbf{z}(p)}$ is a manifold obtained by $(q,q')$ Dehn surgery, the real part is related to the Chern-Simons invariant. The formula involves the conjugate integers $(r,r')$ such that $qr' - q'r = 1$:
\beq
\label{Csphi}\mathsf{CS}(\mathfrak{M}_{\mathbf{z}(p)}) - \mathsf{CS}(\mathfrak{M}_{\mathbf{z}(p_0)}) = \mathsf{CS}_a(p) + \pi(r\,\mathrm{Im}\,u + r'\,\mathrm{Im}\,v),
\eeq
and thanks to the quantization condition, it does not depend modulo $2\pi^2\mathbb{Z}/\varsigma$ of the choice of path from $p_c$ to $p$. In this article, we call $\mathsf{Vol}_a$ the \emph{analytic volume}, and $\mathsf{CS}_a$ the \emph{analytic Chern-Simons term}.

\begin{remark}\label{AAAI} Even if $\mathfrak{M}$ is not hyperbolic, the primitive of the $1$-form $\ln l\,\dd\ln m$ defined over (one of the component of) the $\mathrm{SL}_2(\mathbb{C})$ character varieties defines a notion of complexified volume, whose imaginary part is closely related to the notion of volume of a representation.
\end{remark}


It is enlightening to understand the volume, the Chern-Simons invariant and the properties raised in \S~\ref{PropA} from the point of view of K-theory. This is the matter of the next two paragraphs.

\subsection{Bloch group and hyperbolic geometry}
\label{BlochH}
Let $\mathbb{K}$ be a number field or a function field. To fix notations, $\mathbb{K}^{\times}$ is the multiplicative group of invertible elements of $\mathbb{K}$, and $\mathbb{K}^+$ is just $\mathbb{K}$ considered as an additive group. For an abelian group $\mathbb{G}$, the exterior product $\mathsf{\Lambda}^2_{\mathbb{Z}}\mathbb{G}$ is the $\mathbb{Z}$-module generated by the antisymmetric elements $x\wedge y$ for $x,y \in \mathbb{G}$, modulo the relations of compatibility with the group law $(n\cdot x)\wedge y = n(x\wedge y)$. When $S$ is a set, $\mathbb{Z}\cdot S$ is the free $\mathbb{Z}$-module with basis the elements of $S$.

The \emph{pre-Bloch group} \cite{Bloch} $\mathrm{P}(\mathbb{K})$ is the quotient of $\mathbb{Z}\cdot(\mathbb{K}^{\times}\setminus\{1\})$ by the relations $[z] + [1 - z] = 0$ and $[z] + [1/z] = 0$ for any $z \in \mathbb{K}^{\times}\setminus\{0\}$, and the five term relations:
\beq
[z] + [z'] + [1 - zz'] + \big[\frac{1 - z}{1 - zz'}\big] + \big[\frac{1 - z'}{1 - zz'}\big] = 0
\eeq
for any $z,z' \in \mathbb{K}^{\times}\setminus\{1\}$ such that $zz' \neq 1$. Those combinations appear precisely in the functional relations of the function $D(z)$ of \ref{BlochW}. Indeed, $D$ induces a well-defined function $\mathrm{D}\,:\,\mathrm{P}(\mathbb{K}) \rightarrow \mathbb{R}$ if we interprete $\mathrm{D}(\xi = \sum_j [z_j]) = \sum_j D(z_j)$.

For a hyperbolic manifold $\mathfrak{M}$ with a triangulation, a point $p$ in $\mathcal{C}^{\mathrm{geom}}$ determines shape parameters $\mathbf{z}(p) = (z_e(p))_e$ for the triangulation. We can apply the above construction to a field $\mathbb{K}$ where the functions $\mathbf{z}$ live. It is in general an extension of the field $\mathbb{C}(\mathcal{C}^{\mathrm{geom}})$, of finite degree that we denote $d$. The element
\beq
\label{PBlochele}\xi_{\mathbf{z}} = \sum_e [z_e] \in \mathrm{P}(\mathbb{K})
\eeq
is actually independent of the triangulation. Also, the volume is a well-defined function on $\mathcal{C}^{\mathrm{geom}}$,  given by $\mathrm{D}(\xi_{\mathbf{z}})$.

Up to now, the introduction of the pre-Bloch group has served merely as a rephrasing of \S~\ref{VCS}. Neumann and Yang \cite{YN} took a step further to reach the Chern-Simons invariant. We introduce the Rogers dilogarithm, which is a multivalued analytic function on $\mathbb{C}^{\times}\setminus\{1\}$:
\beq
R(z) = \mathrm{Li}_2(z) + \frac{\ln z\,\ln(1 - z)}{2}.
\eeq
Some computations shows that the diagram below\footnote{The factor of $2$ is convenient for applications to knot theory in the homology spheres.} is well-defined and commutative:
\begin{center}
\begin{minipage}[l]{0.3\linewidth}
\hspace{2cm}\begin{tikzpicture}[node distance=2cm,auto]
\node (P) {$\mathrm{P}(\mathbb{C})$} ;
\node (BP) [below of=P] {$\mathsf{\Lambda}^2_{\mathbb{Z}} \mathbb{C}^+$} ;
\node (RP) [right of=P] {$\mathsf{\Lambda}^2_{\mathbb{Z}} \mathbb{C}^{\times}$} ;
\node (LP) [left of =P] {$\mathrm{B}(\mathbb{C})$} ;
\node (LBP) [left of =BP] {$\mathbb{C}/\mathbb{Q}$} ;
\draw[->,densely dotted,font=\scriptsize] (LP) to node {$\subseteq$} (P) ;
\draw[->] (P) to node {$\mu$} (RP) ;
\draw[->,densely dotted] (LP) to node [swap] {$\hat{\rho}$} (LBP) ;
\draw[->,densely dotted] (BP) to node {} (LBP) ;
\draw[->] (P) to node [swap] {$\rho$} (BP) ;
\draw[->] (BP) to node [swap] {$\mathrm{e}$} (RP) ;
\end{tikzpicture}
\end{minipage} \hfill \begin{minipage}[l]{0.6\linewidth}
\vspace{-0.1cm}
\beq
\begin{split}
\mu([z]) & = 2\,z\wedge (1 - z), \nonumber \\
\rho([z]) & = \frac{\ln z}{2\mathrm{i}\pi}\wedge\frac{\ln(1 - z)}{2\mathrm{i}\pi} + 1\wedge\frac{R(z)}{2\pi^2}, \nonumber \\
\mathrm{e}(\zeta\wedge\zeta') & = 2\,e^{2\mathrm{i}\pi\zeta}\wedge e^{2\mathrm{i}\pi\zeta'} \nonumber
\end{split}
\eeq
\end{minipage}
\end{center}
The \emph{Bloch group} of $\mathbb{C}$ by definition $\mathrm{B}(\mathbb{C}) = \mathrm{ker}\,\mu$, and we have $\rho(\mathrm{B}(\mathbb{C})) \subseteq \mathrm{Ker}\,\mathrm{e}$. As a matter of fact, $\zeta \mapsto 1 \wedge \zeta \in \mathsf{\Lambda}^2_{\mathbb{Z}}\mathbb{C}^+$ induces an isomorphism between $\mathbb{C}/\mathbb{Q}$ and $\mathrm{Ker}\,\mathrm{e}$. Thus, there is a map $\hat{\rho}$ from the Bloch group to $\mathbb{C}/\mathbb{Q}$. Coming back to hyperbolic geometry: since two edges carry the same shape parameter in each tetrahedron, the element $\xi_{\mathbf{z}}$ defined in \ref{PBlochele} actually sits in $\mathrm{B}(\mathbb{C}) \subseteq \mathrm{P}(\mathbb{C})$. When $\mathfrak{M}_{\mathbf{z}}$ is a manifold, it was proved in \cite{YN2} that $\hat{\rho}$ gives the irrational part of the Chern-Simons invariant:
\beq
\mathsf{CS}(\mathfrak{M}_{\mathbf{z}}) = -2\pi^2\,\mathrm{Re}\,\hat{\rho}(\xi_{\mathbf{z}}) \,\,\mathrm{mod}\, 8\pi^2\mathbb{Q}.
\eeq

\subsection{K-theory viewpoint}
\label{Tame}

We now review the interpretation of the Boutroux and quantization condition in the context of K-theory, and its relations to hyperbolic geometry.

\subsubsection*{Symbols}

After a classical result of Matsumoto \cite[\S 11]{MilnorK}, the second K-group $\mathrm{K}_2(\mathbb{K})$ of a field $\mathbb{K}$ is isomorphic to $\mathsf{\Lambda}^2_{\mathbb{Z}}\mathbb{K}^{\times}$ modulo the relations $z \wedge (1 - z) = 0$. In other words, $\mathrm{K}_2(\mathbb{K}) = \mathrm{coker}\,\mu/2$, where $\mu$ is the morphism introduced in \S~\ref{BlochH}. The elements of $\mathrm{K}_2(\mathbb{K})$ are usually called \emph{symbols}, and denoted $\{z_1,z_2\}$. When $\mathcal{C}$ is a component of an A-polynomial of a $3$-manifold, a theorem \cite[p. 61]{CCGLS} shows the existence of a integer $\varsigma$, that we choose minimal, such that
\beq
\label{torsi}2\varsigma\cdot\{m,l\} = 0 \in \mathrm{K}_2(\mathbb{C}(\mathcal{C})).
\eeq

\subsubsection*{Regulators}

If $z_1,z_2 \in \mathbb{C}(\mathcal{C})$, and $\mathcal{Z}$ denotes the set of zeroes and poles of $z_1,z_2$, the \emph{regulator map} is defined as:
\beq
\begin{aligned}
r[z_1,z_2]\, :\, H_1(\mathcal{C}\setminus\mathcal{Z},\mathbb{Z}) & \longrightarrow \mathbb{C}^{\times} \\
 \gamma & \longmapsto \exp\Big[\frac{1}{2\mathrm{i}\pi}\Big(\oint_{\gamma} \ln z_1\,\dd\ln z_2 - \ln z_1(o)\oint_{\gamma}\dd\ln z_2\Big)\Big].
 \end{aligned}
\eeq
$o$ is a basepoint in $\gamma$ and given a choice of branch of $\ln z_1$ and $\ln z_2$ at $o$, the logarithms are analytically continued starting from $o$ along $\gamma$. One can show that this definition does not depend on $o$, on the initial choice of branches for the logarithm, and of the representative $z_1,z_2$ of the symbol $\{z_1,z_2\}$. Hence, there exists a map:
\beq
r\,:\,\mathrm{K}_2(\mathbb{K}) \rightarrow  \varinjlim_{\mathcal{Z}\,\,\mathrm{finite}} \mathrm{Hom}(H_1(\mathcal{C}\setminus\mathcal{Z}),\mathbb{C}^{\times}).
\eeq
If $\{z_1,z_2\}$ is $2\varsigma$-torsion (as in Eqn.~\ref{torsi}), we see that $r[z_1,z_2](\gamma)$ is a $2\varsigma^{\mathrm{th}}$-root of unity for all closed cycles $\gamma$. We deduce that, for any closed cycle $\gamma$ with basepoint $o$ such that $\ln z_1(o) = 0,{\rm i}\pi$ and the integral is well-defined,
\beq
\label{bba}\oint_{\gamma} \ln z_1\,\dd \ln z_2 \in \frac{2\pi^2}{\varsigma}\,\mathbb{Z}.
\eeq
This line of reasoning has been written explicitly in \cite{LiWang}. This can be applied to $\{m,l\}$ for a component of an $A$-polynomial, and justifies the Boutroux and the quantization condition of \S~\ref{PropA}.

\subsubsection*{Tame symbol and Boutroux condition}

Given an algebraic curve $\mathcal{C}$ with two functions $z_1$, $z_2$ defined on it, it might not be easy to check if $\{z_1,z_2\}$ is torsion. However, it is elementary to check if there is a local obstruction to being torsion, i.e. if Eqn.~\ref{bba} holds for all contractible, closed cycles $\gamma$ in $\mathcal{C}$. We focus in this paragraph only on the imaginary part of Eqn.~\ref{bba}, which gives rise to the Boutroux condition, and discuss its relation with the tame condition. The reason is that the Boutroux condition already has interesting consequences for the Baker-Akhiezer kernel (\S~\ref{BAspinor}) and thus the construction of Section~\ref{S3}.

This is formalized as follows. To any $z_1,z_2 \in \mathbb{K}^{\times}$, we can associate the \emph{regulator form}, which is the $1$-form:
\beq
\begin{split}
\eta[z_1,z_2] & = \ln |z_1|(\dd\,\mathrm{arg}\,z_2) - \ln |z_2|(\dd\,\mathrm{arg}\,z_1) \\
& = \mathrm{Im}(\ln z_1\,\dd\ln z_2) - \dd(\mathrm{arg}\,z_1\ln|z_2|).
\end{split}
\eeq
For any point $p \in \mathcal{C}_0$, let $T_p\,:\,\mathrm{K}_2(\mathbb{K}) \rightarrow \mathbb{C}^{\times}$ be the map defined by:
\beq
\begin{split}
& \label{saia}T_p(\{z_1,z_2\}) = (-1)^{\mathrm{ord}_p z_1 \cdot\,\mathrm{ord}_p z_2}\,z_1(p)^{\mathrm{ord}_p z_2}\,z_2(p)^{-\mathrm{ord}_p z_1} \\
& = e^{\mathrm{i}\pi\,(\Res_{p} \dd\ln z_1)(\Res_p \dd \ln z_2) + \Res_p [(\dd \ln z_2)\ln z_1 - (\dd \ln z_1)\ln z_2]}.
\end{split}
\eeq
This expression is indeed independent of the representative of $\{z_1,z_2\}$. It is also independent of the branches of the logarithms and of the basepoint to define the integral over a small circle around $p$. A computation shows:
\beq
\Res_p \eta[z_1,z_2] = -i\ln\big|T_p(\{z_1,z_2\})\big|,
\eeq
so $T_p$ is closely related to the regulator map $r[z_1,z_2]$ evaluated on a small circle around $p$. For a given curve, $z_1$ and $z_2$ only have a finite number of zeroes and poles, so $T_p(\{z_1,z_2\}) = 1$ except at a finite number of points. Notice that the Riemann bilinear identity applied to the meromorphic $1$-forms $\frac{\dd z_1}{z_1}$ and $\frac{\dd z_2}{z_2}$ implies $\prod_{p \in \mathcal{C}_0} T_p(\{z_1,z_2\}) = 1$. We say that $\{z_1,z_2\}$ is a \emph{weakly tame} symbol if $|T_p(\{z_1,z_2\})| = 1$ for all $p$, i.e. we define the subgroup:
\beq
\mathrm{K}_2^{\mathrm{w-tame}}(\mathcal{C}) = \bigcap_{p \in \mathbb{C}(\mathcal{C})} \mathrm{ker}\,|T_p|.
\eeq
It is very easy to check if an element of $K_2(\mathbb{C}(\mathcal{C}))$ is weakly tame of not, given Eqn.~\ref{saia}, and this provides a local obstruction for the Boutroux condition, and a fortiori for being torsion. Moreover, if there exists an integer $\varsigma_0$ such that $T_p(\{z_1,z_2\})$ is a $2\varsigma_0^{\mathrm{th}}$-root of unity for all $p \in \mathcal{C}$, and if $\{z_1,z_2\}$ is torsion, $\varsigma_0$ must divide the order of torsion. The \emph{tame group} itself is defined as:
\beq
\mathrm{K}_2^{\mathrm{tame}}(\mathcal{C}) = \bigcap_{p \in \mathbb{C}(\mathcal{C})} \mathrm{ker}\,T_p.
\eeq
$\eta[z_1,z_2]$ is always closed, since $\eta[z_1,z_2] = \mathrm{Im}\big(\dd\ln z_1\wedge\dd\ln z_2) = 0$. It is in general not exact, but $\eta[z,1 - z] = \dd D(z)$. So, we can illustrate this discussion in the context of hyperbolic $3$-manifolds. The shape parameters $(z_e)_e$ sit in an extension $\mathbb{K}$ of $\mathbb{C}(\mathcal{C}^{\mathrm{geom}})$ of some degree $\varsigma$, we have at our disposal the element $\xi \in \mathrm{B}(\mathbb{K})$ (see Eqn.~\ref{PBlochele}) and the symbol $\{m^2,l\} = \sum_e \{z_e,1 - z_e\}$ is by construction zero in $\mathrm{K}_2(\mathbb{K})$. By coming back to $\mathrm{K}_2(\mathbb{C}(\mathcal{C}^{\mathrm{geom}}))$, one only obtains that $\varsigma_0\cdot\{m^2,l\}$, so $\{m,l\}$ is $2\varsigma$-torsion. Hence $\{m,l\}$ is weakly tame in a trivial way.

\subsection{Arithmetics and cusp field}

We now come to aspects of the A-polynomial which are relevant to the arithmeticity properties of the perturbative invariants of $3$-manifolds. We aim at preparing for a clarification of the arithmetic nature of the invariants defined from the topological recursion in Section~\ref{toporeca}, especially when applied to A-polynomials. Unless precised otherwise, we work in the remaining of this section with any of the irreducible factor of the A-polynomial which is not of the form $(lm^{a} \pm 1)$, and the corresponding spectral curve $(\mathcal{C},u,v)$ is endowed with a marked point $p_c \in \mathcal{C}$ such that $m^2(p_c) = l^2(p_c) = 1$.

We already stated that $m(p_c),l(p_c)$ is a singular point for $A$. More precisely, $A(m,l) \propto (l - l(p_c))^a\,C\big(\frac{l - l(p_c)}{m - m(p_c)}\big)$ in the neighborhood of this singularity. $C$ is a polynomial with integer coefficients, called the \emph{cusp polynomial}. Although it contains less information than the A-polynomial, it retains some geometric significance and is closely related to the C-polynomial studied by Zhang \cite{CZhang}. We also introduce the \emph{cusp field} $\mathbb{F}$, which is the splitting field of the cusp polynomial. In particular, at the vicinity of $p_c$ in $\mathcal{C}$, we have $(l + 1)\sim \gamma(m + 1)$ where $\gamma$ is a root of $C$, thus an element of the cusp field.

There are several notions of fields associated to a hyperbolic $3$-manifold $\mathfrak{M}$. In presence of an ideal triangulation of $\mathfrak{M}$, the \emph{tetrahedron field} is the field generated by the shape parameters of the tetrahedra. From another point of view, $\mathfrak{M}$ can be realized as quotients $\mathbb{H}_3/\Gamma$ where $\Gamma$ is a discrete subgroup of $\mathrm{PSL}_2(\mathbb{C})$ of finite covolume. One can define the \emph{invariant trace field}, which is the field generated by the trace of squares of elements of $\Gamma$. It is clear that:
\beq
\mathrm{cusp}\,\mathrm{field}\,\,\,\subseteq\,\,\,\mathrm{invariant}\,\mathrm{trace}\,\mathrm{field}\,\,\,\subseteq\,\,\,\mathrm{tetrahedron}\,\mathrm{field}.
\eeq
There are examples of hyperbolic knot complements where the cusp field is strictly smaller than the tetrahedron field \cite{RN}. The numbers produced from the topological recursion will naturally live in the cusp field $\mathbb{F}$.

\begin{remark}
When $\mathcal{C}_0$ is the geometric component of an A-polynomial of a triangulated $3$-manifold, \cite{Abhijit} ensures that the shape parameters are rational functions of $l$ and $m$. Hence, the order of torsion of $\{m,l\}$ is $2$ (i.e. $\varsigma = 1$), and the invariant trace field coincides with the tetrahedron field. 
\end{remark}

\subsection{Definition of A-spectral curves}
\label{algcurves}
Since we will often use this setting, we give the name \emph{A-spectral curve} (over a field $\mathbb{K}$) to the data of:
\begin{itemize}
\item a curve $\mathcal{C}$ defined by an equation of the form $\mathrm{Pol}(m,l) = 0$ (with coefficients in $\mathbb{A}$), such that $\{m,l\}$ is $2\varsigma$-torsion in $\mathrm{K}_2(\mathcal{C}(\mathbb{C}))$ for some minimal integer $\varsigma$. We assume that $\mathrm{Pol}(m,l)$ is irreducible and not proportional to $lm^b \pm 1$ for some integer $b$.
\item a compact Riemann surface $\mathcal{C}_0$ which is a smooth model for $\mathcal{C}$, and a marked point $p_c \in \mathcal{C}_0$ such that $l(p_c)^2 = m(p_c)^2 = 1$.
\item two functions $u = \ln m$ and $v = \ln l $ on $\mathcal{C}_0$, and the differential form $\phi = v\,\dd u$.
\item we add the technical assumption that the zeroes of $v$ are simple.
\end{itemize}
One may wonder if all A-spectral curves over $\mathbb{K}$ arise as components of the A-polynomial of some $3$-manifold. The answer does not seem to be known. $\mathrm{K}_2^{\mathrm{tame}}(\mathcal{C})$ (and a fortiori $\mathrm{K}_2^{\mathrm{w.tame}}(\mathcal{C})$) for a compact Riemann surface $\mathcal{C}_0$ of genus $g \geq 1$ defined over $\mathbb{Q}$ is in general not trivial. Part of a conjecture of Beilinson predicts that a certain subgroup of $\mathrm{K}_2^{\mathrm{tame}}(\mathcal{C})$ has rank $g$. Yet, non zero tame symbols are not easy to exhibit, see for instance \cite{DokJeuZ} where elements in the tame group of some hyperelliptic curves are constructed.

\subsection{Algebraic geometry on the spectral curve}
\label{algcurve}

We now come to the study of algebraic geometry on the spectral curve $(\mathcal{C},u,v)$

\subsubsection*{Topology, cycles, and holomorphic $1$-forms}

The curve $\mathcal{C}_0$ defines a compact Riemann surface of a certain genus $g$, which does not depends on the smooth model $\mathcal{C}_0$ for $\mathcal{C}$. Actually, the genus can be computed from the polynomial $A(m,l)$ as the dimension $g$ of the space of holomorphic forms, i.e. rational expressions $h(m^2,l)\dd m$ which are nowhere singular. Let $(\mathcal{A}_j,\mathcal{B}_j)_j$ be a symplectic basis of homology cycles:
\beq
\label{symplb}\forall j,j' \in \{1,\ldots,g\},\qquad \mathcal{A}_{j}\cap\mathcal{A}_{j'} = 0\qquad \mathcal{B}_j\cap\mathcal{B}_{j'} = 0\qquad \mathcal{A}_j\cap\mathcal{B}_{j'} = \delta_{j,j'}.
\eeq
For the moment, we choose an arbitrary basis, and we will have to consider later how objects depend on the basis, i.e to describe the action of the modular group $\mathrm{Sp}_{2g}(\mathbb{Z})$. There is a notion of dual basis of holomorphic forms $(\dd \mathrm{a}_j)_j$, characterized by:
\beq
\forall j \in \{1,\ldots,g\},\qquad\qquad  \oint_{\mathcal{A}_{j'}}\dd \mathrm{a}_j = \delta_{j,j'}.
\eeq
Then, the \emph{period matrix} is defined as:
\beq
\oint_{\mathcal{B}_{j'}} \dd \mathrm{a}_j = \tau_{j,j'},
\eeq
and a classical result states that it is symmetric with positive definite imaginary part. We choose an arbitrary base point $o$, for example $o = p_c$, and introduce the \emph{Abel map}:
\beq
\begin{array}{lcccl} \ab & : & \mathcal{C}_0 & \longrightarrow & \mathbb{J} = \mathbb{C}^g/(\mathbb{Z}^g \oplus \tau\mathbb{Z}^g) \\
& & p & \longmapsto & (\int_o^p \dd \mathrm{a}_1,\ldots,\int_o^p \dd \mathrm{a}_g)\quad {\rm mod}\,\,\mathbb{Z}^g \oplus \tau\mathbb{Z}^g. \end{array}
\eeq
When $g = 1$, $\mathcal{C}_0$ is an elliptic curve and $\ab$ is an isomorphism. When $g \geq 2$, this is only an immersion.

\subsubsection*{Theta functions and characteristics}

For any $g \times g$ matrix $\tau$ which is symmetric with positive definite imaginary part, we can define the theta function:
\beq
\forall \mathbf{w} \in \mathbb{C}^g,\qquad \theta(\mathbf{w}|\tau) = \sum_{\mathbf{p} \in \mathbb{Z}^{g}} e^{\mathrm{i}\pi\mathbf{p}\cdot\tau\cdot \mathbf{p} + 2\mathrm{i}\pi\mathbf{w}\cdot\mathbf{p}}.
\eeq
Where there is no confusion, we omit to write the dependance in $\tau$. $\theta$ is an even, quasi periodic function with respect to the lattice $\mathbb{Z}^g \oplus \tau\mathbb{Z}^g$:
\beq
\label{funct}\theta(\mathbf{w} + \mathbf{m} + \tau\cdot\mathbf{n}) = e^{-\mathrm{i}\pi(2\mathbf{n}\cdot\mathbf{w} + \mathbf{n}\cdot\tau\cdot\mathbf{n})}\,\theta(\mathbf{w}).
\eeq
We define a gradient $\nabla$ acting implicitly on the variable $\mathbf{w}$, and a gradient $D$ acting on the variable $\tau$:
\beq
\nabla\theta = \Big(\frac{\partial\theta}{\partial w_1},\ldots,\frac{\partial\theta}{\partial w_g}\Big)\qquad\qquad D\theta = \Big(4\mathrm{i}\pi\,\frac{\partial\theta}{\partial\tau_{i,j}}\Big)_{1 \leq i,j \leq g}.
\eeq
The theta function is solution to the heat equation:
\beq
4\mathrm{i}\pi\partial_{\tau_{j,j'}} \theta = \partial_{w_j}\partial_{w_{j'}} \theta
\qquad \quad {\rm i.e.}\quad
D\vartheta = \nabla^{\otimes 2}\vartheta.
\eeq
Throughout the article, we are going to use tensor notations, and indicate with a $\cdot$ the contraction of indices. We consider $\nabla\theta$ and $D\theta$, and more generally $\nabla^{\otimes l}\theta$ (resp. $D^{\otimes l}\theta$) as a $l$-linear form (resp. a $2l$-linear form), i.e. a $[0,l]$ tensor. For example, if $T$ is a $[l,0]$ tensor, we may write:
\beq
\nabla^{\otimes l} \theta\cdot T = \sum_{j_1 = 1}^g \cdots \sum_{j_l = 1}^g \frac{\partial^l \theta}{\partial w_{j_1}\cdots \partial w_{j_l}}\,T_{j_1,\ldots,j_l}.
\eeq

A \emph{half-characteristics} is a vector $\mathbf{c} = \frac{1}{2}(\mathbf{n} + \tau\cdot\mathbf{m})$ where $\mathbf{n},\mathbf{m} \in \mathbb{Z}^g$. It is said \emph{odd} or \emph{even} depending on the parity of the scalar product $\mathbf{n}\cdot\mathbf{m}$. Eqn.~\ref{funct} implies that $\theta(\mathbf{c}|\tau)$ and its even-order derivatives vanish at odd half-characteristics, while the odd-order derivatives of $\theta(\mathbf{w})e^{\mathrm{i}\pi\mu\cdot\mathbf{w}}$ vanish at even half-characteristics of the form $\frac{1}{2}(\mathbf{n} + \tau\cdot\mathbf{m})$. There is a notation for theta functions whose argument is shifted by a half-characteristics $\mathbf{c} = \nu + \tau\cdot\mu$,
\beq
\begin{split}
\vartheta\bigl[{}^{\mu}_{\nu}\bigr](\mathbf{w}) & = \sum_{\mathbf{p} \in \mathbf{Z}^g} e^{\mathrm{i}\pi(\mathbf{p} + \mu)\cdot\tau\cdot(\mathbf{p} + \mu) + 2\mathrm{i}\pi(\mathbf{w} + \nu)\cdot(\mathbf{p} + \mu)} \\
& = e^{\mathrm{i}\pi\mu\cdot\tau\cdot\mu + 2\mathrm{i}\pi\mu\cdot\nu + 2\mathrm{i}\pi\mathbf{w}\cdot\mu}\,\theta(\mathbf{w} + \nu + \tau\cdot\mu). 
\end{split}
\eeq
Notice that we still have:
\beq
\nabla^{\otimes 2}\vartheta\bigl[{}^{\mu}_{\nu}\bigr] = D\vartheta\bigl[{}^{\mu}_{\nu}\bigr].
\eeq

\subsubsection*{Bergman kernel}

For us, a \emph{Bergman kernel} is a symmetric $(1,1)$ form $B(p_1,p_2)$ on $\mathcal{C}_0\times\mathcal{C}_0$ which has no residues and has no singularities except for a double pole with leading coefficient $1$ on the diagonal, i.e. in a local coordinate $\lambda$:
\beq
B(p_1,p_2) \mathop{=}_{p_1 \rightarrow p_2} \frac{\dd\lambda(p_1)\otimes \dd\lambda(p_2)}{\big(\lambda(p_1) - \lambda(p_2)\big)^2} + O(1).
\eeq
If we pick up a symplectic basis of homology $(\mathcal{A},\mathcal{B})$, there is a unique Bergman kernel $B(p_1,p_2)$ which is normalized on the $\mathcal{A}$-cycles:
\beq
\label{norma}\forall j \in \{1,\ldots,g\},\qquad\qquad \oint_{\mathcal{A}_j} B(p_1,\cdot) = 0.
\eeq
Moreover, $B$ is symmetric in $p_1$ and $p_2$ and the basis of holomorphic form is retrieved by:
\beq
\label{norma2}\forall j \in \{1,\ldots,g\},\qquad\qquad \oint_{\mathcal{B}_j} B(p_1,\cdot) = 2\mathrm{i}\pi\,\dd \mathrm{a}_j.
\eeq
Any other Bergman kernel takes the form:
\beq
B_{\kappa}(p_1,p_2) = B(p_1,p_2) + 2\mathrm{i}\pi\,\dd\ab(p_1)\cdot\kappa\cdot\dd\ab^{t}(p_2),
\eeq
where $\kappa$ is a symmetric $g \times g$ matrix of complex numbers and ${}^t$ denotes the transposition. As a matter of fact, $B_{\kappa}$ satisfies the relations \ref{norma} and \ref{norma2} if we replace $(\mathcal{A},\mathcal{B})$ by a symplectic basis of generalized cycles $(\mathcal{A}^{\kappa},\mathcal{B}^{\kappa})$ defined by:
\beq
\mathcal{A}^{\kappa} = \mathcal{A} - \kappa\mathcal{B}^{\kappa},\qquad \mathcal{B}^{\kappa} = \mathcal{B} - \tau\mathcal{A}.
\eeq
In this formula, $\mathcal{A}$ and $\mathcal{B}$ should be interpreted as column vectors with $g$ rows.

The Bergman kernel normalized of the $\mathcal{A}$-cycles can always be expressed in terms of theta functions:
\beq
B_0(p_1,p_2) = \dd_{p_1}\dd_{p_2} \ln\theta(\ab(p_1) - \ab(p_2) + \mathbf{c}|\tau),
\eeq
where $\mathbf{c}$ is any non singular odd half-characteristics. Non-singular means that the right hand does not vanish identically when $p_1,p_2 \in \mathcal{C}_0$, and such characteristics exist \cite{TataII}. Yet, this formula is not very useful for computations when $g \geq 2$. In practice, one may start from the equation $A(m^2,l) = 0$ defining $\mathcal{C}$ and $\mathcal{C}_0$, and find "by hand" a Bergman kernel and a basis of holomorphic forms expressed as rational expressions in $m^2$ and $l$ with rational coefficients. Both methods are illustrated for genus $1$ curves in Section~\ref{Bergman}.

\subsubsection*{Prime form}

Let $\mathbf{c}$ be a non singular odd half-characteristics. We introduce a holomorphic $1$-form:
\beq
\dd h_{\mathbf{c}}(p) = \nabla\theta(\mathbf{c})\cdot\dd\ab(p).
\eeq
It is such that its $2g-2$ zeroes are all double. Then, the \emph{prime form} \cite{TataII} $E(p_1,p_2)$ is a $(-1/2,-1/2)$ form defined on the universal cover of $\mathcal{C}_0\times\mathcal{C}_0$:
\beq
E(p_1,p_2) = \frac{\theta(\ab(p_1) - \ab(p_2) + \mathbf{c})}{\sqrt{\dd h_{\mathbf{c}}(p_1)}\otimes\sqrt{\dd h_{\mathbf{c}}(p_2)}}.
\eeq
It is antisymmetric in $p_1$ and $p_2$, it has has a zero iff $p_1 = p_2$ in $\mathcal{C}_0$, and in a local coordinate $\lambda$:
\beq
E(p_1,p_2) \mathop{=}_{p_1 \rightarrow p_2} \frac{\lambda(p_1) - \lambda(p_2)}{\sqrt{\dd \lambda(p_1)}\otimes\sqrt{\dd\lambda(p_2)}} + O\big((\lambda(p_1) - \lambda(p_2))^3\big).
\eeq
The prime form appears in this article through the formulas:
\beq
\begin{split}
& \exp\Big[\frac{1}{2}\int_{p_2}^{p_1}\int_{p_2}^{p_1}\Big(B_0(p,p') - \frac{\dd u(p)\,\dd u(p')}{(u(p) - u(p'))^2}\Big)\Big] = \frac{u(p_2) - u(p_1)}{E(p_1,p_2)\,\sqrt{\dd u(p_1)\,\dd u(p_2)}}, \\
& \exp\Big[\int_{p_2}^{p_1}\int_{p_4}^{p_3} B_0(p,p') \Big] = \frac{E(p_1,p_3)\,E(p_2,p_4)}{E(p_1,p_4)\,E(p_2,p_3)}.
\end{split}
\eeq

\subsubsection*{Modular transformations}

The group $\mathrm{Sp}_{2g}(\mathbb{Z})$ acts on those objects by transformation of the symplectic basis of homology cycles. Let $\gamma$ be an element of $\mathrm{Sp}_{2g}(\mathbb{Z})$.
\begin{itemize}
\item The cycles $\mathcal{A}$ and $\mathcal{B}$, interpreted as column vectors with $g$ rows, transform by definition as:
\beq
\label{cyclemod}\left(\begin{array}{c} {}^{\gamma}\mathcal{A} \\ {}^{\gamma}\mathcal{B} \end{array}\right) = \mathbf{M}_{\gamma}\cdot\left(\begin{array}{c} \mathcal{A} \\ \mathcal{B} \end{array}\right),\qquad \mathbf{M}_{\gamma} = \left(\begin{array}{cc} d & c \\ b & a\end{array}\right),
\eeq
where $a,b,c,d$ are $g \times g$ integer matrices. The new basis $({}^{\gamma}\mathcal{A},{}^{\gamma}\mathcal{B})$ is symplectic (see Eqn.~\ref{symplb}) iff ${}^{t}b\,d$ and ${}^{t}c\,a$ are symmetric and ${}^{t}a\,d - {}^{t}c\,b = 1$. These are indeed the condition under which $\mathbf{M}_{\gamma}$ belongs to $\mathrm{Sp}_{2g}(\mathbb{Z})$.
\item The dual basis of holomorphic forms, interpreted as a row vector with $g$ columns, transforms as a modular weigth $-1$ vector:
\beq
{}^{\gamma}\dd\ab(p) = {}^{t}(c\tau + d)^{-1}\,\dd\ab(p).
\eeq
\item The matrix of periods transforms as:
\beq
{}^{\gamma} \tau = (a\tau + b)(c\tau + d)^{-1}.
\eeq
Using the relations defining $\mathrm{Sp}_{2g}$, one can check:
\beq
\label{egalite1}(c\tau + d)^t(a\tau + b)(c\tau + d)^{-1} = (a\tau + b)^t,
\eeq
so that ${}^{\gamma}\tau$ is indeed symmetric. We have denoted $M^{t}$, the transposed of a matrix $M$.
\item The Bergman kernel $B_{\kappa\,;\,\tau}$ defined from the chosen basis of cycles (we have stressed the dependance in $\tau$), transforms as:
\beq
\begin{split}
{}^{\gamma}(B_{\kappa\,;\,\tau})(p_1,p_2) & = B_{{}^{\gamma}\kappa\,;\,{}^{\gamma}\tau}(p_1,p_2) ,\nonumber \\
\label{trans} {}^{\gamma}\kappa & = c(c\tau + d)^{t} + (c\tau + d)\kappa(c\tau + d)^{t}.
\end{split}
\eeq
\item The generalized cycles $(\mathcal{A}^{\kappa},\mathcal{B}^{\kappa})$ on which $B_{\kappa}$ is normalized are modular expression of weight $1$:
\beq
{}^\gamma\mathcal{A}^{\kappa} = (c\tau + d)\mathcal{A}^{\kappa},\qquad {}^\gamma\mathcal{B}^{\kappa} = [(c\tau + d)^{-1}]^t\,\mathcal{B}^{\kappa}.
\eeq
We have used the relation $[a - (a\tau + b)(c\tau + d)^{-1}c] = [(c\tau + d)^{-1}]^{t}$ which can be deduced from Eqn.~\ref{egalite1}.
\item The theta function transforms as:
\beq
\label{thetatrans}\theta(\mathbf{w} + \Delta_{\gamma}|{}^{\gamma}\tau) = \Xi_{\gamma}\,\sqrt{\mathrm{det}(c\tau + d)}\,e^{-\mathbf{w}\cdot\tau\cdot\mathbf{w}}\,\theta\big((c\tau + d)\mathbf{w}|\tau\big),
\eeq
where $\Delta_{\gamma}$ is the half-characteristics $\Delta_{\gamma} = \frac{1}{2}(\mathrm{diag}(ab^{t}) + \mathrm{diag}(cd^{t})\tau)$ and $\Xi_{\gamma}$ a eighth root of unity. 
\end{itemize}

\subsection{Baker-Akhiezer spinors}
\label{BAspinor}
Given a $1$-form $\omega$ on $\mathcal{C}_0$, a complex number $\hc\in\mathbb C^{\times}$, and vectors $\mu,\nu \in \mathbb{C}^g/\mathbb{Z}^g$, we set:
\beq
\begin{split}
\label{psiBA}\psi_{\mathrm{BA}}(p_1,p_2) & = \frac{\exp\Big(\frac{1}{\hc}\int_{p_2}^{p_1} \omega\Big)}{E(p_1,p_2)}\,\frac{\vartheta\bigl[{}^{\mu}_{\nu}\bigr]\big(\ab(p_1) - \ab(p_2) + \zeta_{\hc}\big)}{\vartheta\bigl[{}^{\mu}_{\nu}\bigr]\big(\zeta_{\hc})} \\
 & =  \sqrt{\nabla\theta(\mathbf{c})\cdot\dd\ab(p_1)}\otimes\sqrt{\nabla\theta(\mathbf{c})\cdot\dd\ab(p_2)}\,\exp\Big(\int_{p_2}^{p_1} (\hc^{-1}\omega + 2\mathrm{i}\pi\mu\cdot\dd\ab)\Big) \\
 & \times \,\frac{\theta\big(\ab(p_1) - \ab(p_2) + \zeta + \nu + \tau\cdot\mu\big)}{\theta(\ab(p_1) - \ab(p_2) + \mathbf{c})\,\theta(\zeta_{\hc} + \nu + \tau\cdot\mu)},
\end{split}
\eeq
with
\beq
\zeta_{\hc} = \mathrm{frac}\Big[\frac{1}{2\mathrm{i}\pi\hc}\oint_{\mathcal{B}}\omega\Big] - \tau\cdot\mathrm{frac}\Big[\frac{1}{2\mathrm{i}\pi\hc}\oint_{\mathcal{A}} \omega\Big].
\eeq
For a vector $\mathbf{w} \in \mathbb{C}^g$, we have denoted $\mathrm{frac}[\mathbf{w}]$ the vector of $[0,1[^g$ which is equal to $\mathbf{w}$ modulo $\mathbb{Z}^g$. $\psi_{\mathrm{BA}}$ is called a \emph{Baker-Akhiezer spinor}, it is a $(1/2,1/2)$-form defined a priori on the universal cover of $\mathcal{C}_0\times\mathcal{C}_0$, since we have:
\beq
\frac{\psi_{\mathrm{BA}}(p_1 + \mathbf{m}\mathcal{A} + \mathbf{n}\mathcal{B},p_2)}{\psi_{\mathrm{BA}}(p_1,p_2)} = e^{(2\mathrm{i}\pi\mu + \hc^{-1}\,\oint_{\mathcal{A}} \omega)\cdot\mathbf{m} + (\zeta + 2\mathrm{i}\pi(\nu - \mathbf{c}) + \hc^{-1}\,\oint_{\mathcal{B}} \omega)\cdot\mathbf{n}}.
\eeq
It is regular apart from a simple pole when $p_1 = p_2$:
\beq
\label{p1}\psi_{\mathrm{BA}}(p_1,p_2) \mathop{\sim}_{p_1 \rightarrow p_2} \frac{\sqrt{\dd\lambda(p_1)}\otimes\sqrt{\dd\lambda(p_2)}}{\lambda(p_1) - \lambda(p_2)},
\eeq
and has an essential singularity when $p_1$ or $p_2$ reach a singularity of $\omega$, of the form:
\beq
\label{p2}\psi_{\mathrm{BA}}(p_1,p_2) \propto \exp\Big(\frac{1}{\hc}\int_{p_2}^{p_1} \omega\Big).
\eeq
Baker-Akhiezer functions have been introduced in \cite{Kri77} to write down some explicit solutions of the KP hierarchy. They can be obtained from the Baker-Akhiezer spinor when $\omega$ is a meromorphic $1$-form, and by sending $p_2$ to a pole of $\omega$ with an appropriate regularization (see for instance \cite{BEInt}). Modular transformations act on $\psi_{\mathrm{BA}}$ only by a change of the vectors $\mu,\nu$. We have introduced a normalization constant $\hc\in\mathbb C^{\times}$, to be adjusted later. In general, the ratio involving $\vartheta\bigl[{}^{\mu}_{\nu}\bigr]$ does not have a limit, neither has a power series expansion when $\hc \rightarrow 0$.

But we can say more if we assume the Boutroux and the quantization condition, i.e. that there exists $\varsigma \in \mathbb{N}^*$ such that, for all closed cycles $\Gamma$:
\beq
\Big(\mathrm{Im}\,\oint_{\Gamma} \omega\Big) = 0\qquad\qquad \Big(\mathrm{Re}\,\oint_{\Gamma} \omega\Big) \in \frac{2\pi^2}{\varsigma}\cdot\mathbb{Z}.
\eeq
Let us denote $\mathbf{s}_A$ and $\mathbf{s}_B$ integer vectors such that:
\beq
\oint_{\mathcal{A}} \omega = \frac{2\pi^2}{\varsigma}\,\mathbf{s}_A\qquad\qquad\oint_{\mathcal{B}} \omega = \frac{2\pi^2}{\varsigma}\,\mathbf{s}_B.
\eeq
It is then natural to consider values of $\hc^{-1}$ belonging to arithmetic subsequences on the imaginary axis:
\beq
\hc = \frac{\mathrm{i}\pi}{k}\qquad\qquad k \in \varsigma\cdot\mathbb{Z} + r.
\eeq
Indeed, we find:
\beq
\zeta = \mathrm{frac}\Big[\frac{r\mathbf{s}_B}{\varsigma}\Big] - \tau\cdot\mathrm{frac}\Big[\frac{r\mathbf{s}_A}{\varsigma}\Big],
\eeq
so that the argument of the theta functions only depend on $r = k\,\,\mathrm{mod}\,\,\varsigma$. We have:
\beq
\label{ays}\frac{2\pi}{k}\,\ln|\psi_{\mathrm{BA}}(p_1,p_2)| \mathop{\sim}_{\substack{k \rightarrow \infty \\ k \in \varsigma\cdot\mathbb{Z} + r}} \mathrm{Im}\Big(\int_{p_2}^{p_1} 2\,\omega\Big),
\eeq
and the Boutroux condition also ensure that $\mathrm{Im}\big(\int_{p_2}^{p_1} \omega\big)$ does not depend on the path of integration between $p_1$ and $p_2$. For a hyperbolic $3$-manifold, if we choose $\omega = v\,\dd u$, the right hand side is $\mathsf{Vol}_a(p_1) - \mathsf{Vol}_a(p_2)$ and this asymptotics is exactly the one involved in the generalized the volume conjecture (see \S~\ref{volc})

\subsection{Branchpoints and local involution}
\label{bp}
In this article, we reserve the name \emph{ramification points} to points in $\mathcal{C}_0$ which are zeroes of $\dd u = \dd\ln m$. The value of $m$ at a ramification point is called a \emph{branchpoint}. We use generically the letter $a$ to denote a ramification point. Since $\mathcal{C}$ is defined by a polynomial equation $A(e^u,e^v) = 0$, we must have $m(a) \neq 0,\infty$. When $a$ is a simple zero of $\dd\ln m$, we call it a \emph{simple ramification point}, and we can define at least in a neighborhood $\mathcal{U}_a \subseteq \mathcal{C}_0$ of $a$ the \emph{local involution} $p \mapsto \overline{p}$:
\beq
p,\overline{p} \in \mathcal{U}_a,\qquad m(p) = m(\overline{p})\,\,\,\mathrm{and}\,\,\,p \neq \overline{p}.
\eeq
Since $A$ is quasi-reciprocal and has real coefficients, the involution $\iota\,:\,(l,m) \mapsto (1/m,1/l)$ and the complex conjugation $*$ act on the set of coordinates $(m(a),l(a))$ of the ramification points, and decompose it into orbits with $2$ elements (for an $a$ such that $(l(a),m(a))$ is real or unitary) or $4$ elements (in general). \emph{Amphichiral} knot complements admit an orientation reversing automorphism, so that $\iota_{\mathfrak{m}}(m,l) = (1/m,l)$ by $\iota_{\mathfrak{l}}(m,l) = (m,1/l)$ are separately symmetries of their A-polynomial. Then at the level of spectral curves, the set of ramification points can be decomposed further into orbits of $2$, $4$ or $8$ elements.

\section{Topological recursion}
\label{S2}
The \emph{topological recursion} associates, to any spectral curve $(\mathcal{C}_0,u,v)$, a family of symmetric $(1,\ldots,1)$ forms $\omega_n^{h}(p_1,\ldots,p_n)$ on $\mathcal{C}_0^n$ ($n \in \mathbb{N}^*$, $h \in \mathbb{N}$) and a family of numbers $F_h$ ($h \in \mathbb{N}$). These objects have many properties, we shall only mention those we use without proofs. We refer to \cite{EORev} for a detailed review of the topological recursion. The fact that, here or in topological strings, one encounters spectral curves of the form $\mathrm{Pol}(e^u,e^v) = 0$ rather than $\mathrm{Pol}(u,v) = 0$, does not make a big difference in the formalism.

We assume that all ramification points are simple. This is satisfied for most of the A-polynomials we have studied (see Figs.~\ref{Fig34}-\ref{Fig35}). The topological recursion can also be defined when some ramification points are not simple \cite{PratsF}, \cite{Bouchm}, but we do not address this issue here.

\subsection{Definitions}
\label{toporeca}

Let $(\mathcal{C}_0,u,v)$ be a spectral curve endowed with a basis of cycles $(\mathcal{A},\mathcal{B})$. Hence, there is a privileged Bergman kernal $B(p_1,p_2)$. T. To shorten notations, we write $\dd z_1\cdots\dd z_n$ instead of $\dd z_1\otimes\cdots\otimes\dd z_n$ for a $(1,\ldots,1)$-form.

\subsubsection*{Recursion kernel}

We introduce the recursion kernel:
\beq
\label{reck}K(p_0,p) = \frac{-\frac{1}{2}\int_{\overline{p}}^p B(\cdot,p_0)}{\big(v(p) - v(\overline{p})\big)\dd u(p)}.
\eeq
$K(p_0,p)$ is a $1$-form with respect to $p_0$ globally defined on $\mathcal{C}_0$, and a $(-1)$-form with respect to $p$ which is defined locally around each ramification point.

\subsubsection*{Differential forms}

\begin{figure}[h!]
\begin{center}
\includegraphics[width=1.0\textwidth]{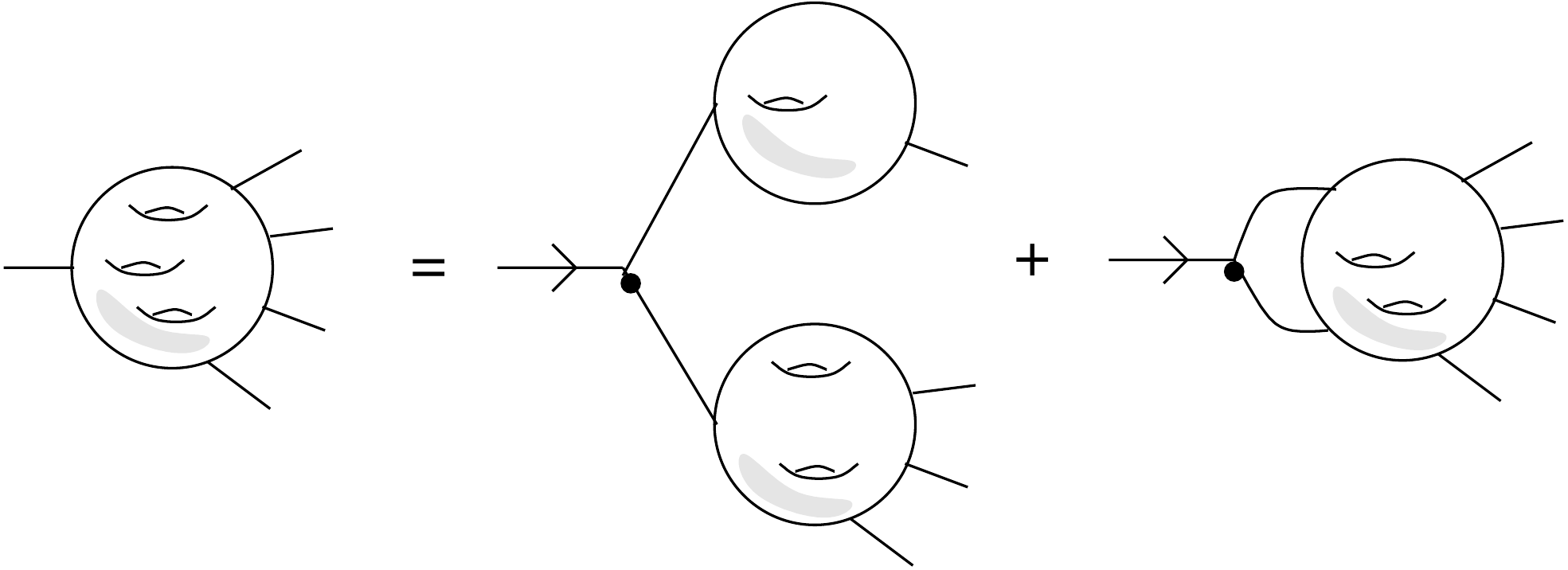}
\end{center}
\caption{Diagrammatic representation of the topological recursion which defines $\omega_n^h$. Each $\omega_n^h$ is represented as a "surface" with $h$ handles and $n$ punctures, i.e. with Euler characteristics $\chi=2-2h-n$. The diagrammatic representation of the topological recursion, is that one can compute $\om_n^h$ with Euler characteristics $\chi$ in terms of $\om_{n'}^{h'}$ with $\chi'=2-2h'-n'>\chi$, by "removing a pair of pants" from the corresponding surface. \label{toprec}}
\end{figure}

We define:
\beq
\omega_1^{0}(p) = v(p)\dd u(p),\qquad\qquad \omega_2^{0}(p_1,p_2) = B(p_1,p_2),
\eeq
and recursively:
\beq
\label{defwng}\omega_n^{h}(p_0,p_I) = \sum_{a} \Res_{p \rightarrow a} K(p_0,p)\Big[\omega_{n + 1}^{h - 1}(p,\overline{p},p_I) + \sum_{h',J}^{'} \omega_{|J| + 1}^{h'}(p,p_J)\omega_{n - |J|}^{h - h'}(\overline{p},p_{I\setminus J})\Big].
\eeq
In the left hand side, $I = \{1,\ldots,n - 1\}$ and $p_I$ in a $(n - 1)$-uple of points of $\mathcal{C}_0$. For any $J \subseteq I$, $p_J$ is the uple of points indexed by the subset $J$. In the right hand side, we take the residues at all ramification points, and the $\sum'$ in the right hand side ranges over $h' \in \{0,\ldots,h\}$ and all splitting of variables $J \subseteq I$, excluding $(J,h') = (\emptyset,0)$ and $(I,h)$. The formula above is a recursion on the \emph{level} $\chi = 2h - 2 + n$. $\omega_n^{h}$ has a diagrammatic interpretation (Fig. \ref{toprec}), it can be written as a sum over graphs with $n$ external legs, $h$ handles, and thus Euler characteristics $-\chi$. However, the weights of the graphs are non local, they involve stacks of $2g + 2 - n$ residues where the ordering matters.

Although Eqn.~\ref{defwng} seems to give a special role to the variable $p_0$, one can prove (e.g. from the diagrammatic representation) that $\omega_n^{h}(p_0,\ldots,p_{n - 1})$ is symmetric in $p_0,\ldots,p_{n - 1}$. Except maybe $\omega_1^0$, the $\omega_n^{h}$ are meromorphic $(1,\ldots,1)$-forms on $\mathcal{C}_0^n$, which have no residues and have poles only at the ramification points.

We illustrate the computation at level $1$. To write down the residues it is convenient to choose a local coordinate at each ramification point, for instance $\lambda_a(p) = \sqrt{m(p) - m(a)} = \sqrt{e^{u(p)} - e^{u(a)}}$, which has the advantage that $\lambda_a(\overline{p}) = -\lambda_a(p)$. If $r$ is a function or $R$ is a $1$-form, we denote:
\beq
\partial^j r(\underline{a}) = \frac{\partial^j r(p)}{\partial \lambda_a(p)^j}\Big|_{p = a},\qquad\qquad \partial^j R(\underline{a}) = \left.\frac{\partial^j\big(\frac{R(p)}{\dd\lambda_a(p)}\big)}{\partial \lambda_a(p)^j}\right|_{p = a}.
\eeq
Then, we find:
\beq
\begin{split}
\omega_3^{0}(p_0,p_1,p_2) & = \sum_{a} \Res_{z \rightarrow a} K(p_0,p)\big[B(p,p_1)\,B(p_2,\overline{p}) + B(\overline{p},p_1)\,B(p,p_2)\big] \\
\label{w30}& = \sum_a \frac{B(p_0,\underline{a})B(p_1,\underline{a})B(p_2,\underline{a})}{2\,v'(\underline{a})}.
\end{split}
\eeq
To write down $\omega_1^{1}$, we need to expand:
\beq
\frac{B(p,\overline{p})}{\big(\dd\lambda_a(p)\big)^2} \mathop{=}_{p \rightarrow a} -\frac{1}{4\lambda_a^2(p)} + S_B(a) + o(1).
\eeq
Then, we have:
\beq
\begin{split}
\omega_1^{1}(p_0) & = \sum_a \Res_{p \rightarrow a} K(p_0,p)B(p,\overline{p}) \\
& = \sum_a \frac{S_B(a)}{2}\,B(p_0,\underline{a}) + \frac{1}{48}\big(v'''(\underline{a})\,B(p_0,\underline{a}) - v'(\underline{a})\,\partial_2^2B(p_0,\underline{a})\big).
\end{split}
\eeq

\subsubsection*{Stable free energies}

We have already met the abelian function:
\beq
\Phi(p) = \int_{o}^p \omega_1^{0} = \int_{o}^p v\,\dd u.
\eeq
For $h \geq 2$, we define:
\beq
F_h = \frac{1}{2h - 2}\,\sum_a \Res_{p \rightarrow a} \Phi(p)\,\omega_1^{h}(p).
\eeq
Since $\omega_1^h$ has no residues, $F_h$ does not depend on the basepoint $o$. The numbers $F_h$ are called the \emph{stable free energies} of the spectral curve. We are not going to give an explicit definition of the \emph{unstable free energies} $F_0$ and $F_1$. Actually, for the computation of the BA kernels and later the asymptotics of the colored Jones polynomial, it is not necessary to know how to compute the free energies, we only need one of their key property called special geometry (see Eqn.~\ref{specgeom}). So, we just state that there exists $F_0$ and $F_1$ satisfying Eqn.~\ref{specgeom}, it is in fact a way to define them.

\subsection{Deformation of spectral curves}

By abuse of notations, we write $F_h = \omega_{n = 0}^{h}$, i.e. we consider $\omega_n^{h}$ for all $n,h\in \mathbb{N}$. Unless specified, the properties mentioned below also hold for the unstable free energies. Special geometry expresses the variation of $\omega_n^{h}$ when $\phi = v\,\dd u$ is deformed by addition of a meromorphic $1$-form $\Omega$. By form-cycle duality on $\mathcal{C}_0$, to any meromorphic 1--form $\Omega$ we can associate a cycle $\Omega^*$ and a germ of holomorphic function on $\Omega^*$ denoted $\Lambda_{\Omega}$, such that:
\beq
\Omega(p) = \int_{\Omega^*} \Lambda_{\Omega}(\cdot)\,B(\cdot,p).
\eeq
Then, for a smooth family of spectral curves $\mathcal{S}_{\alpha} = (\mathcal{C}_0,u_{\alpha},v_{\alpha})$ such that:
\beq
(v_{\alpha} - v)\dd u - (u_{\alpha} - u)\dd v \mathop{\sim}_{\alpha \rightarrow 0} \alpha\,\Omega.
\eeq
we have:
\beq
\label{specgeom} \left. \frac{\partial}{\partial_{\alpha}} \,\omega_n^{h}[\mathcal{S}_{\alpha}](p_I)\right|_{\alpha = 0} = \int_{\Omega^*} \Lambda_{\Omega}(\cdot)\,\omega_{n + 1}^{h}[\mathcal{S}_{\alpha = 0}](\cdot,p_I).
\eeq
Notice that from the expression of $\omega_3^0$ in Eqn.~\ref{w30}, one retrieves as a special case the analog of Rauch variational formula \cite{Rauchvar} for the variation of the Bergman kernel $\omega_2^0 = B$ along any meromorphic deformation.

In this article, deformations by holomorphic $1$-forms and by $1$-forms with simple poles will play a special role.

\subsubsection*{Variations of filling fractions}

The \emph{filling fractions} are defined by:
\beq
\epsilon_j = \frac{1}{2\mathrm{i}\pi}\oint_{\mathcal{A}_j} v\,\dd u.
\eeq
Performing a variation of filling fractions amounts to add to $v\,\dd u$ a holomorphic $1$-form, i.e. use the deformation:
\beq
\label{holdef}\Omega(p) = 2\mathrm{i}\pi\,\dd \mathrm{a}_j(p) = \int_{\mathcal{B}_j} B(\cdot,p).
\eeq
We denote $\omega_{n}^{h,(l)}$ the $[0,l]$-tensor of $l^{\mathrm{th}}$ derivatives of $\omega_n^h$ with respect to the filling fractions, and according to Eqn.~\ref{specgeom}:
\beq
\omega_n^{h,(l)}(p_I) = \underbrace{\oint_{\mathcal{B}}\cdots\oint_{\mathcal{B}}}_{l\,\,\mathrm{times}} \omega_{n + l}^{h}(\cdots,p_I).
\eeq
In particular, the tensor of second derivatives of $F_0=\omega_0^0$ is the matrix of periods:
\beq
F_0'' = \oint_{\mathcal{B}}\oint_{\mathcal{B}} B = 2\mathrm{i}\pi\,\tau.
\eeq

\subsubsection*{Deformation by simple poles}

Given a couple of distinct points $(p_1,p_2)$, we denote:
\beq
\label{Sfdef}\mathop{\dd S}_{p_2,p_1}(p) = \int_{p_2}^{p_1} B(\cdot,p).
\eeq
This $1$-form is characterized by a simple pole at $p = p_1$ (resp. $p = p_2$) with residue $1$ (resp. $-1$), no other singularities, and vanishing $\mathcal{A}$-cycle integrals. If we perform an infinitesimal deformation with $\Omega(p) = \dd S_{p_2,p_1}(p)$, we obtain according to Eqn.~\ref{specgeom}:
\beq
\delta \omega_n^h(p_I) = \int_{p_2}^{p_1} \omega_{n + 1}^h(\cdot,p_I).
\eeq

\subsection{Symplectic invariance}
\label{sympeq}
The topological recursion also has nice properties under global transformations of the spectral curve $(\mathcal{C}_0,u,v)$. To simplify, we consider in this paragraph $(n,h) \neq (0,0),(0,1),(1,0)$, and just mention that the properties below are slightly modified for those cases.

It is very easy to prove from the definitions:
\begin{property}
If $v\,\dd u$ is replaced by $\alpha\,v\,\dd u$ for some $\alpha \in \mathbb{C}^{\times}$, $\omega_n^h$ is replaced by $\alpha^{2 - 2h - n}\,\omega_n^h$. In particular, the stable free energies $F_h$ are unchanged when $(u,v)$ is replaced by $(-u,v)$ or $(u,-v)$.
\end{property}
\begin{property}
If $(u,v)$ is replaced by $(u,v + f(u))$ with $f$ at least a germ of holomorphic function in the neighborhood of the values $u(a)$, $\omega_n^h$ are unchanged.
\end{property}
According to the first property, replacing $m = e^u$ and $l = e^v$ by some of their powers i.e. use $(\pm m^a, \pm l^b)$ instead of $(m,l)$, only affect the $\omega_n^h$ by a scaling factor. The second property tells us that the $\omega_n^h$ are the same if we change the signs of $m$ and $l$, or even replace\footnote{This last operation is very useful to lower the degree of $m$ in A-polynomials. For instance, the A-polynomial of the $\mathrm{Pretzel}(-2,3,7)$
\beq
\mathfrak{A}(l,m) = m^{110}l^6 - m^{90}(m^2 - 1)^2l^5 - m^{72}(2m^2 + 1)l^8 + m^{36}(m^2 + 2)l^4 + m^{16}(m^2 - 1)^2l - 1
\nonumber
\eeq
looks simpler if we use the variable $\ell = lm^8$:
\beq
\mathfrak{A}(l,m) = m^{14}\ell^6 - m^{10}(m^2 - 1)\ell^5 - m^8(2m^2 + 1)\ell^4 + m^4(m^2 + 2)\ell^2 + (m^2 - 1)^2\ell - 1
\nonumber
\eeq} $l$ by $lm^a$ for some power $a$. There is conjecturally a third property concerning the exchange of $u$ and $v$:
\begin{property}
\label{P3} If $(u,v)$ is replaced by $(v,-u)$, the $F_h$ are unchanged, and for $n \geq 1$, the cohomology class of $\omega_n^h$ is multiplied by the sign $(-1)^n$.
\end{property}
This has only been proved \cite{EO2MM} when $u$ and $v$ are meromorphic function on the curve $\mathcal{C}_0$, that is for spectral curves defined by an equation $\mathrm{Pol}(u,v) = 0$. This invariance of the free energies under this exchange has meaningful consequences in random matrix theory and enumerative geometry (see \cite[\S 10.4.1]{EOFg} for an example). Here and in topological strings, we rather have to consider spectral curves of the form $\mathrm{Pol}(e^u,e^v) = 0$. We believe that Property~\ref{P3} survives in this context with a few extra assumptions, although this has not been established yet. For example, within "remodeling the B-model", it implies the framing independence of the closed topological string sector.

In other words, if Property~\ref{P3} holds, the $F_g$, and cohomology classes of the $\omega_n^h$ up to a sign, are invariant under all the transformations which preserve the symbol $\dd u\wedge \dd v$. This suggests to consider the $F_h$ and the $\omega_n^h$ up to a sign as "symplectic invariants" of the function field $\mathbb{K} = \mathbb{C}(\mathcal{C})$. We have seen in \S~\ref{BlochH} that the real part of the primitive of $\omega_1^0$ essentially coincide with the Bloch regulator of the symbol $\{m,l\}$. It would be interesting to investigate the possible meaning of the topological recursion in terms of K-theory of $\mathbb{K}$.

\subsection{Deformation of the Bergman kernel}

Instead of $B(p_1,p_2)$, we could have used in the definitions \ref{reck} and \ref{defwng} another Bergman kernel:
\beq
B_{\kappa}(p_1,p_2) = B(p_1,p_2) + 2\mathrm{i}\pi\,\ab(p_1)\cdot\kappa\cdot\dd\ab^{t}(p_2).
\eeq
We denote $\omega_{n|\kappa}^h$ the corresponding objects. They are polynomials of degree $3h - 3 + n$ in $\kappa$, and it is not difficult to prove:
\beq
\begin{split}\label{kappavar} \frac{\partial\omega_{n|\kappa}^h(p_I)}{\partial\kappa} & = \frac{1}{2}\,\frac{1}{2\mathrm{i}\pi}\Big(\oint_{\mathcal{B}_{\kappa}}\oint_{\mathcal{B}_{\kappa}} \omega_{n + 2|\kappa}^{h - 1}(\cdot,\cdot,p_I) \\
& + \sum_{\substack{J \subseteq I \\ 0 \leq h' \leq h}} \oint_{\mathcal{B}_{\kappa}} \omega_{|J| + 1|\kappa}^h(\cdot,p_J)\oint_{\mathcal{B}_{\kappa}} \omega_{n - |J| + 1|\kappa}^{h - h'}(\cdot,p_{I\setminus J})\Big).
\end{split}
\eeq
The special geometry (Eqn.~\ref{specgeom}) for meromorphic deformations normalized on the $\mathcal{A}$-cycles still holds for $\omega_{n|\kappa}^h$ at any fixed $\kappa$. However, variations of $\kappa$ and filling fractions are mixed, since the holomorphic forms $\dd a_j$ in Eqn.~\ref{holdef} are defined from $B = B_{\kappa = 0}$ and not $B_{\kappa}$. The appropriate formula can be found in \cite{EOFg}, it is closely related to "holomorphic anomaly equations" \cite{BCOV}, but it will not be used in this article.

\subsection{Effect of an involution}
\label{effeinv}
The A-polynomial comes with an involution $\iota\,:\,(m,l) = (1/m,1/l)$. It induces an involutive linear map $\iota_{*}$ on the space of holomorphic $1$-forms on $\mathcal{C}_0$. The $g$ eigenvalues of $\iota_*$ are thus $\pm 1$. By integration, it induces an involutive isomorphism of the Jacobian of the curve, that we denote $\underline{\iota}_*$. The number of $+1$ eigenvalues is the genus of the quotient curve $\mathcal{C}_0/\iota$.

The case $\iota_* = \varepsilon\,\mathrm{id}$ is of particular interest. When $\varepsilon = 1$, $\underline{\iota}_*$ is a translation by a half-period, and when $\varepsilon = -1$, $\underline{\iota}_*$ is a central symmetry. In these two situations, all admissible Bergman kernels:
\beq
B_{\kappa}(z_1,z_2) = \mathrm{d}_{z_1}\mathrm{d}_{z_2} \ln \theta\big(\mathbf{a}(z_1) - \mathbf{a}(z_2) + \mathbf{c}\big) + 2\mathrm{i}\pi\,\mathrm{d}\mathbf{a}(z_1)\cdot\kappa\cdot\mathrm{d}\mathbf{a}(z_2)
\eeq
are invariant under $\iota$, and so is the recursion kernel $K_{\kappa}(z_0,z)$. Since the set of ramification points is stable under $\iota$, we can recast the residue formula by choosing a representative $a'$ in each pair $\{a,\iota(a)\}$ of ramification points:
\beq
\begin{split}
\omega_{n|\kappa}^h(p_0,p_I) & = \sum_{a'} \Res_{p \rightarrow a'} K_{\kappa}(p_0,p)\,E_{n|\kappa}^{h}(p,z_I) + \Res_{p \rightarrow \iota(a')} K_{\kappa}(p_0,p)\,E_{n|\kappa}^h(p,p_I) \\
& = \sum_{a'} \Res_{p \rightarrow a'} \big[ K_{\kappa}(p_0,p)\,E_{n|\kappa}^h(p,p_I) + K_{\kappa}(p_0,\iota(p))\,E_{n|\kappa}^h(\iota(p),p_I)\big] \\
& = \sum_{a'} \Res_{p \rightarrow a'} \big[K_{\kappa}(p_0,p)\,E_{n|\kappa}^h(p,p_I) + K_{\kappa}(\iota(p_0),p)\,E_{n|\kappa}^h(\iota(p),p_I)\big].
\end{split}
\eeq
By recursion on $2g - 2 + n \leq 0$, we infer that $E_{n|\kappa}^h(\iota(p),p_I) = E_{n|\kappa}^h(p,\iota(p_I))$ and $\omega_{n|\kappa}(p_0,p_I) = \omega_{n|\kappa}^h(\iota(p_0),\iota(p_I))$.

This result has an interesting corollary when $\iota_* = -\mathrm{id}$: by duality, $\iota_*\mathcal{B}_{\kappa} = -\mathcal{B}_{\kappa}$, hence
\begin{property}
If $\iota_* = -\mathrm{id}$,
\beq
\underbrace{\oint_{\mathcal{B}_{\kappa}}\cdots\oint_{\mathcal{B}_{\kappa}}}_{d\,\,\mathrm{times}} \omega_{n|\kappa}^h(p_1,\ldots,p_{n - d},\cdot) = 0\qquad\mathrm{when}\,\,d\,\,\mathrm{is}\,\,\mathrm{odd}.
\eeq
\end{property}
and in the case $(n,h) = (1,0)$, since $\ln\,\dd\ln m$ is always invariant under $\iota$, we have:
\begin{property} \label{psoq}If $\iota_* = -\mathrm{id}$, for any closed cycle $\Gamma \subseteq \mathcal{C}_0$,
\beq
\oint_{\Gamma} v\,\dd u = 0
\eeq
\end{property}
As one can see in Figs.~\ref{Fig34} and \ref{Fig35}, $\iota_{*} = -\mathrm{id}$ is neither rare nor the rule for complement of hyperbolic knots. We observe however that the genus of the quotient $\mathcal{C}_0/\iota$ is low compared to the genus of $\mathcal{C}_0$: the "simplest" knot we found for which the quotient has not genus $0$ is $\mathbf{8}_{21}$. The geometrical significance of these observations from the point of view of knot theory is unclear.

\section{Non-perturbative topological recursion}
\label{S3}
The \emph{perturbative partition} function is usually defined as:
\beq
Z_{\mathrm{pert},\hc} = \exp\Big(\sum_{h \geq 0} \hc^{2h - 2} F_h\Big),
\eeq
where $F_h$ are the free energies. However, the genuine partition function of a quantum field theory (like the Chern-Simons theory or topological string theory) should have properties that $Z_{\mathrm{pert},\hc}$ does not satisfy. For instance, it should be independent of the classical solution chosen to quantize the theory (background independence), and it should have modular properties (e.g. S-duality) whenever this makes sense.

From the topological recursion applied to a spectral curve $(\mathcal{C}_0,u,v)$, and theta functions, we are going to define a non-perturbative partition function $\mathcal{T}_{\hc}$ which implements such properties. Modular transformations correspond here to change of symplectic basis of cycles on $\mathcal{C}_0$. Then, one can define non-perturbative "wave functions". To keep a precise vocabulary, we shall introduce quantities
$$\psi_{\hc}^{[\mathrm{n}|\mathrm{n}]}(p_1,p_2 ; \ldots ; p_{2n - 1},p_{2n}),$$
that we call $\mathrm{n}|\mathrm{n}$-kernels, which depend on $2\mathrm{n}$ points on the curve. In particular, the leading order of $\psi_{\hc}^{[1|1]}(p_1,p_2)$ when $\hc \rightarrow 0$ will be given by the Baker-Akhiezer spinor. We prefer to use a new letter $\hbar$ for the formal parameter. We shall find later that in the application we consider, it must be identified to $\hbar$ defined in terms of the parameter $q = e^{2\hbar}$ in which the colored Jones polynomial is a Laurent polynomial, but this identification might be different when considering other problems.

\subsection{Definitions}
\label{dedede}
We use the notations of \S~\ref{algcurve}. We take as data a spectral curve $(\mathcal{C}_0,u,v)$ endowed with a basis of cycles, we choose two vectors $\mu,\nu \in \mathbb{C}^g$ and we set:
\beq
\label{reca}\tau = \frac{F_0''}{2\mathrm{i}\pi}\qquad\qquad \zeta = \mathrm{frac}\Big[\frac{1}{2\mathrm{i}\pi\hc}\oint_{\mathcal{A}}v\,\dd u\Big] - \tau\cdot\mathrm{frac}\Big[\frac{1}{2\mathrm{i}\pi\hc}\oint_{\mathcal{B}}v\,\dd u\Big].
\eeq
We give the definitions, which we comment in \S~\ref{defcom}.

\subsubsection*{Partition function}

The \emph{non-perturbative partition function} is by definition:
\beq
\begin{split}
\label{taudef}\mathcal{T}_{\hc} & = \exp\Big(\sum_{h \geq 0} \hc^{2h - 2}\,F_h\Big) \\
& \times \Big\{\sum_{r \geq 0} \frac{1}{r!} \sum_{\substack{h_j \geq 0,\,\,d_j \geq 1 \\ 2h_j - 2 + d_j > 0}} \hspace{-10pt} \hc^{\sum_j 2h_j - 2+ d_j}\,\bigotimes_{j = 1}^r \frac{F^{(d_j)}_{h_j}\cdot\nabla^{\otimes d_j}}{(2\mathrm{i}\pi)^{d_j}\,d_j!}\Big\}\vartheta\bigl[{}^{\mu}_{\nu}\bigr](\zeta_{\hc}|\tau).
\end{split}
\eeq
We may isolate its leading behavior by writing
\beq
\label{taudefr}\mathcal{T}_{\hc} = e^{\hc^{-2}F_0 + F_1}\,\vartheta\bigl[{}^{\mu}_{\nu}\bigr](\zeta_{\hc}|\tau)\,\hat{\mathcal{T}}_{\hc},
\eeq
where now $\lim_{\hc \rightarrow 0} \hat{\mathcal{T}}_{\hc}  = 1$. We consider this expression as a formal asymptotic series with parameter $\hc \rightarrow 0$. The coefficient of $\hc^{\chi}$ in general depend on $\hc$, but does not have a power series expansion in $\hc$. Thus, it is meaningful to speak of the $\chi^{\mathrm{th}}$-order term in the expansion, keeping in mind that this coefficient may also depend on $\hc$.

\subsubsection*{($1|1$)- Kernel}
\label{spinker}

In integrable systems, the Sato formula expresses the wave function as Schlesinger transforms of the tau function, which in our language correspond to adding a $1$-form with simple poles to $\phi = v\,\dd u$. Actually, we prefer to work with the \emph{kernel} $\psi_{\hc}(p_1,p_2)$ which is a function on $\mathcal{C}_0\times\mathcal{C}_0$, defined as:
\beq
\label{psidefa}\psi_{\hc}(p_1,p_2) = \frac{\mathcal{T}_{\hc}\big[v\dd u \longrightarrow v\,\dd u + \hc\,\mathop{\dd S}_{p_2,p_1}\big]}{\mathcal{T}_{\hc}\big[v\,\dd u\big]},
\eeq
where $\dd S$ was defined in Eqn.~\ref{Sfdef}. We introduce shortcut notations:
\beq
\label{bulletdeft}\vartheta = \vartheta\bigl[{}^{\mu}_{\nu}\bigr](\zeta_{\hc}|\tau),\qquad\qquad \vartheta_{\bullet} = \vartheta\bigl[{}^{\mu}_{\nu}\bigr]( \zeta_{\hc} + \ab(p_1) - \ab(p_2)|\tau).
\eeq
$\psi_{\hc}(p_1,p_2)$ can be computed thanks to special geometry:
\beq
\begin{split}
\label{psidef}& \psi_{\hc}(p_1,p_2) = \frac{\exp\Big(\frac{1}{\hc}\int_{p_2}^{p_1} v\,\dd u\Big)}{E(p_1,p_2)\big(\dd u(p_1)\,\dd u(p_2)\big)^{1/2}} \\
& \times \frac{\Big\{\sum_{r \geq 0} \frac{1}{r!} \sum_{\substack{h_j,n_j \geq 0,\,\,d_j \geq 1 \\ 2h_j - 2 + d_j + n_j > 0}} \hc^{\sum_j 2h_j - 2 + d_j + n_j} \bigotimes_{j = 1}^r \frac{\int_{p_2}^{p_1}\cdots\int_{p_2}^{p_1}\omega_{n_j}^{h_j,(d_j)}\cdot\nabla^{\otimes d_j}}{(2\mathrm{i}\pi)^{d_j}\,d_j!\,n_j!} \Big\}\vartheta_{\bullet}}{\Big\{\sum_{r \geq 0} \frac{1}{r!} \sum_{\substack{h_j \geq 0,\,\,d_j \geq 1 \\ 2h_j - 2 + d_j > 0}} \hc^{\sum_j 2h_j - 2 + d_j} \bigotimes_{j = 1}^r \frac{F_{h_j}^{(d_j)}\cdot\nabla^{\otimes d_j}}{(2\mathrm{i}\pi)^{d_j}\,d_j!} \Big\}\vartheta}.
\end{split}
\eeq
In the second line of Eqn.~\ref{psidef}, all the $n_j$ variables in $\omega_{n_j}^{h_j,(d_j)}$ are integrated over $\int_{p_2}^{p_1}$, and recall from special geometry that:
\beq
\int_{p_2}^{p_1}\cdots\int_{p_2}^{p_1} \omega_{n}^{h,(d)} = \int_{p_2}^{p_1}\cdots\int_{p_2}^{p_1} \underbrace{\oint_{\mathcal{B}}\cdots\oint_{\mathcal{B}}}_{d\,\,\mathrm{times}}\omega_{n + d}^{h}.
\eeq
Again, Eqn.~\ref{psidef} should be understood as a formal asymptotic series with parameter $\hc \rightarrow 0$. It can be shown \cite{BEInt} that $\psi_{\hc}(p_1,p_2)$ does not change when $p_1$ or $p_2$ goes around an $\mathcal{A}$ or a $\mathcal{B}$ cycle. Since $\psi$ is the ratio of two partition function, the exponential involving the free energies $F_h$ in the numerator of the first line of Eqn.~\ref{psidef} cancels with the same factor present in the denominator. As we claimed earlier, only the expression of $\omega_n^h$ is needed to compute the kernel, not the expression of the free energies. We may isolate its leading behavior:
\beq
\psi_{\hc}(p_1,p_2) = \frac{\psi_{\mathrm{BA}}(p_1,p_2)}{\big(\dd u(p_1)\,\dd u(p_2)\big)^{1/2}}\,\hat{\psi}_{\hc}(p_1,p_2),
\eeq
where now $\lim_{\hc \rightarrow 0} \hat{\psi}_{\hc}(p_1,p_2) = 1$.

\subsubsection*{$n|n$-kernels}

If we perform $n$ successive Schlesinger transformations, we are led to define the \emph{$\mathrm{n}|\mathrm{n}$- kernels}:
\beq
\label{dkea}\psi_{\hc}^{[\mathrm{n}|\mathrm{n}]}(p_1,o_1\,;\,\cdots\,;\,p_{\mathrm{n}},o_{\mathrm{n}}) = \frac{\mathcal{T}_{\hc}\big[v\,\dd u \rightarrow v\,\dd u + \sum_{k = 1}^\mathrm{n} \hc\,\dd S_{o_{k},p_{k}}\big]}{\mathcal{T}_{\hc}\big[v\,\dd u\big]},
\eeq
which are functions on $\mathcal{C}^{2\mathrm{n}}_0$. Eqn.~\ref{psidef} has a straightforward generalization:
\beq\label{nnk}\begin{split}
& \psi_{\hc}^{[\mathrm{n}|\mathrm{n}]}(p_1,o_1\,;\,\ldots\,;\,p_{\mathrm{n}},o_{\mathrm{n}})  \\
& = \prod_{1 \leq k < k' \leq \mathrm{n}} \frac{E(p_{k},p_{k'})\,E(o_{k},o_{k'})}{E(p_{k},o_{k'})\,E(o_{k},p_{k'})}\,\times\,\frac{\exp\Big(\frac{1}{\hc}\int_{\bullet} v\,\dd u\Big)}{\prod_{k = 1}^{\mathrm{n}}E(p_{k},o_{k})\big(\dd u(p_{k})\,\dd u(o_{k})\big)^{1/2}}  \\
& \times \frac{\Big\{\sum_{r \geq 0} \frac{1}{r!} \sum_{\substack{h_j,n_j \geq 0,\,\,d_j \geq 1 \\ 2h_j - 2 + d_j + n_j > 0}} \hc^{\sum_j 2h_j - 2 + d_j + n_j} \bigotimes_{j = 1}^r \frac{\int_{\bullet}\cdots\int_{\bullet}\omega_{n_j}^{h_j,(d_j)}\cdot\nabla^{\otimes d_j}}{(2\mathrm{i}\pi)^{d_j}\,d_j!\,n_j!} \Big\}\vartheta_{\bullet}}{\Big\{\sum_{r \geq 0} \frac{1}{r!} \sum_{\substack{h_j \geq 0,\,\,d_j \geq 1 \\ 2h_j - 2 + d_j > 0}} \hc^{\sum_j 2h_j - 2 + d_j} \bigotimes_{j = 1}^r \frac{F_{h_j}^{(d_j)}\cdot\nabla^{\otimes d_j}}{(2\mathrm{i}\pi)^{d_j}\,d_j!} \Big\}\vartheta}.
\end{split}
\eeq
In this context, $\int_{\bullet} = \sum_{i = 1}^{{\rm n}} \int_{o_i}^{p_i}$ and $\vartheta_{\bullet}$ stands for:
\beq
\vartheta_{\bullet} = \vartheta\bigl[{}^{\mu}_{\nu}\bigr]\Big(\zeta_{\hc} + \sum_{k = 1}^{\mathrm{n}} \ab(p_{k}) - \ab(o_{k})\Big|\tau\Big).
\eeq

\subsubsection*{Diagrammatic representation} In Appendix~\ref{appdiag}, we explain that the formulae for the non perturbative partition function (Eqn.~\ref{taudef}) and the $\mathrm{n}|\mathrm{n}$ kernels (Eqn.~\ref{nnk}) can be represented as a sum over (maybe disconnected) diagrams. To a given order in $\hc$, there is only a finite sum of allowed diagrams. With this formalism, it is easy to reexponentiate the series above, i.e. to compute the asymptotic series for $\ln \mathcal{T}_{\hc}$ or $\ln \hat{\psi}_{\hc}$: they can be written as sum over connected diagrams.

\subsubsection*{Special properties for A-spectral curves}

When the symbol $\{e^u,e^v\}$ is $2\varsigma$-torsion in $\mathrm{K}_2(\mathbb{C}(\mathcal{C}))$, the spectral curve satisfies the Boutroux condition and the quantization condition. So, when $k$ is an integer going to infinity along arithmetic subsequences of step $\varsigma$ and:
\beq
\hc = \frac{\mathrm{i}\pi}{k},
\eeq
the non-perturbative partition function and the $\mathrm{n}|\mathrm{n}$-kernels do have an expansion in powers of $\hc$, and besides, the factors $\exp\big(\frac{1}{\hc}\int_{o}^p v\,\dd u\big)$ do not depend on the path from $o$ to $p$.
This follows from the discussion of \S~\ref{BAspinor}, and especially from the fact that the argument of the theta functions are independent of $\hc$ along such subsequences. If one consider $\hc \rightarrow 0$ in a generic way, or if the symbol $\{e^u,e^v\}$ is not torsion and $\zeta_{\hc}$ is not zero for other reasons, the asymptotic expansion features fast oscillations at all orders in $\hc$ arising from the theta functions and their derivatives.

\subsection{Remarks}
\label{defcom}
Eqn.~\ref{taudef} for the non-perturbative partition function was first derived in \cite{Ecv} as a heuristic formula to compute the asymptotics of matrix integrals, $N = \hc^{-1}$ playing the role of the matrix size. In \cite{EMhol} it was proved that it has order by order in powers of $\hc^{-1}$ a property of background indepedence, and that it transform like a theta function of characteristics $[\mu,\nu]$ under modular transformation. Actually, Eqn.~\ref{taudef} is the result of summing all perturbative partition functions over filling fractions shifted by integer multiples of $\hc$. This operation looks very much like the Whitham averaging in integrable systems, and we conjectured (and checked to the first non trivial order) in \cite{BEInt} that $\mathcal{T}_{\hc}$ is indeed a formal tau function of an integrable system whose times are moduli of the spectral curves. In that article, we also introduced a spinor version of the kernel $\psi_{\hc}(p_1,p_2)$ (Eqn.~\ref{psidefa}), in order to build a wave function in the language of integrable systems. "Wave function" is a generic name for any complex-valued solution of a linear ODE's or difference equation. $\Psi^{\alpha}(m) \equiv \psi_{\hc}(p^{\alpha}(m),p_0)$ should be considered as the asymptotic series of a wave function, where $p_0$ is a point hold fixed in $\mathcal{C}_0$. Different branches $p^{\alpha}(m)$ give rise to wave functions with dominant asymptotic behavior in different sectors. Typically, a wave function is a linear combination of $\Psi^{\alpha}(m)$, and thus its asymptotics is subject to the Stokes phenomenon when one goes from one sector to the other. The advantage of introducing the kernels is that the Stokes phenomenon is described by a single object $\psi_{\hc}(p,p_0)$ with $p \in \mathcal{C}_0$, through the branching structure of the covering $\mathrm{m}\,:\,\mathcal{C}_0 \rightarrow \widehat{\mathbb{C}}$.

One can in principle derive the difference equation satisfied by $\psi_{\hc}(p,p_0)$ order by order in $\hc$, and we expect it to have an expansion in powers of $\hc$, no matter if the spectral curve satisfy the Boutroux and the quantization condition. However, a general  expression for the resummated difference operator annihilating the wave function (and its $n|n$ counterpart) just from the data of the spectral curve is not available. Recently, Gukov and Su\l{}kowski \cite{GSq} have pointed out that, adding some assumption on the form of the answer, allows to reconstruct the full ODE or difference operator from the knowledge of the first orders. In particular in the context of hyperbolic geometry, the $A$-hat polynomial \cite{AJconj} is expected to appear as one of those operators (cf. \S~\ref{AJpar}). The $A$-hat polynomial is known in closed form for many knots, and  Dimofte \cite{Dimofte} has discussed a procedure to construct the $\hat{\mathrm{A}}$-polynomial from the A-polynomial. Those observations might give hints towards a general theory for the reconstruction of an exact integrable system whose tau function has precisely an asymptotic given to all order by Eqn.~\ref{taudef} in the limit $\hc \rightarrow 0$.

\subsection{Rewriting in terms of modular quantities}
\label{modre}
It was proved in \cite{EMhol} that $\mathcal{T}_{\hc}$ has modular properties. Since the Bergman kernel $B_0$ is not modular invariant, the $\omega_n^h$ are not either modular. Similarly, although the theta function is modular, its derivative are not.
So, the modular properties in the expression \ref{taudef} are not manifest.

From Eqn.~\ref{trans}, one sees that the deformed Bergman kernel $B_{\kappa}$ is modular invariant if we have chosen $\kappa = \underline{\kappa}(\tau)$ as a function of $\tau$ which is quasimodular of weight $2$, namely:
\beq
\underline{\kappa}((a\tau + b)(c\tau + d)^{-1}) = c(c\tau + d)^{t} + (c\tau + d)\cdot\underline{\kappa}(\tau)\cdot(c\tau + d)^{t}.
\eeq
If this is the case, it is straightforward to deduce from the topological recursions formula that $\omega_{n|\underline{\kappa}(\tau)}^h$ is modular invariant when $2h - 2 + n \geq 0$. It is often easier to compute modular objects than non-modular ones, so imagine that we have computed the $\omega_{n|\underline{\kappa}(\tau)}^h$. We would like to write $\mathcal{T}_{\hc}$ only in terms of $\omega_{n|\underline{\kappa}(\tau)}^h$. This can be done using Eqn.~\ref{kappavar} to express $\omega_{n}^h \equiv \omega_{n|\kappa = 0}^h$ in terms of $\omega_{n|\kappa}^h$. The result (valid for any $\kappa$) is:
\beq
\mathcal{T}_{\hc} = \exp\Big(\sum_{h \geq 0} \hc^{2h - 2}\,F_{h|\kappa}\Big)\,\vartheta \times \Big\{\sum_{r \geq 0} \frac{1}{r!} \sum_{\substack{h_j \geq 0,\,d_j \geq 1 \\ 2h_j - 2 + d_j > 0}} \prod_{j = 1}^{r} \frac{F_{h_j|\kappa}^{(d_j)}\cdot T_{d_j|\kappa}}{(2\mathrm{i}\pi)^{d_j}\,d_j!}\Big\},
\eeq
where:
\beq
F_{h|\kappa}^{(d)} = \oint_{\mathcal{B}_{\kappa}}\cdots\oint_{\mathcal{B}_{\kappa}} \omega_{d|\kappa}^{h},
\eeq
and:
\beq
\label{astana}T_{d|\kappa} = \sum_{d' = 0}^{\lfloor d/2 \rfloor} \frac{d!(-1)^{d'}(2\mathrm{i}\pi)^{d'}}{2^{d'}\,d'!}\,\kappa^{\otimes d'}\,\otimes \frac{\nabla^{\otimes(d - 2d')}\vartheta}{\vartheta}.
\eeq
Certain linear combinations of derivatives of theta are modular, and the $T_{d|\underline{\kappa}(\tau)}$ precisely provide such combinations. In fact, the proof that $\mathcal{T}_{\hc}$ is modular given in \cite{EMhol} amounts to prove that $T_{d|\underline{\kappa}(\tau)}$ are modular. In the context of elliptic curves, we shall see in \S~\ref{ellipticgen} that it is natural to choose $\underline{\kappa}(\tau)$ proportional to $E_2(\tau)$, and for this choice, $T_{2d|\underline{\kappa}(\tau)}$ is related to the $d^{\mathrm{th}}$-order Serre derivative of theta functions.

\subsection{Effect of an involution}
\label{Evenorder}
When the genus of the quotient $\mathcal{C}_0/\iota$ is zero, only the terms with even $d_j$ remain in the partition function (Eqn.~\ref{taudef}) and the kernel (Eqn.~\ref{psidef}). In this paragraph, we assume it is the case. The conclusion of \S~\ref{effeinv} was that only even order derivatives of theta functions appear in the formulas, since the odd order derivatives are contracted with zero. Then, we may trade $\nabla^{\otimes 2}$ for a derivative with respect to the period matrix:
\beq
\nabla^{\otimes 2} = 4\mathrm{i}\pi\partial_{\tau} = D,
\eeq
Besides, from Property~\ref{psoq} we learn that $\zeta_{\hc} = 0$ \emph{for any $\hc$}. Then, the non-perturbative partition function and the non-perturbative $\mathrm{n}|\mathrm{n}$ kernels happen to be formal power series in $\hc$. And, in order to compute them, we only have to compute derivatives of Thetanullwerten with respect to the matrix of periods.  For instance, the partition function reads:
\beq\begin{split}
\label{PA}\mathcal{T}_{\hc} & = \exp\Big(\sum_{h \geq 0} \hc^{2h - 2}\,F_h\Big)  \\
&  \times \Big\{\sum_{r \geq 0} \frac{1}{r!} \sum_{\substack{h_j \geq 0,\,d_j' \geq 1 \\ 2h_j - 2 + 2d_j' > 0}} \hc^{\sum_j 2h_j - 2 + 2d_j'}\,\bigotimes_{j = 1}^{r} \frac{F_{h_j}^{(2d_j')}\cdot D^{\otimes d_j'}}{(2\mathrm{i}\pi)^{2d_j'}(2d_j')!}\Big\}\vartheta 
\end{split}\eeq

On the other hand, if we compute the \emph{perturbative partition function} with the Bergman kernel $B_{\kappa}$, we find with help of Eqn.~\ref{kappavar}:
\beq\begin{split}
\label{PB} Z_{\mathrm{pert},\hc|\kappa} & \equiv \exp\Big(\sum_{h \geq 0} \hc^{2h - 2}\,F_{h|\kappa}\Big) \\
& = \exp\Big(\sum_{h \geq 0} \hc^{2h - 2}\,F_h\Big)\Big\{\sum_{r \geq 0} \frac{1}{r!}\hspace{-5pt}\sum_{\substack{h_j \geq 0,\,d_j' \geq 1 \\ 2h_j - 2 + 2d_j' > 0}} \hspace{-10pt}\hc^{\sum_j 2h_j - 2 + 2d_j'}\,\prod_{j = 1}^{r} \frac{F_{h_j}^{(2d_j')}\cdot \kappa^{\otimes d_j'}}{d_j'!\,2^{d_j'}}\Big\}.
\end{split}\eeq
This expression is very similar to the non-perturbative partition function computed with the Bergman kernel $B_0$. More precisely:
\beq
\label{sqjua}\mathcal{T}_{\hc} = Z_{\mathrm{pert},\hc|\kappa} \quad \mathrm{where}\quad \frac{\kappa^{\otimes d'}}{2^{d'}(2\mathrm{i}\pi)^{d'}\,d'!} \rightarrow \frac{1}{(2\mathrm{i}\pi)^{2d'}\,(2d')!}\,\frac{D^{\otimes d'} \vartheta}{\vartheta}. \nonumber
\eeq
The analogy carries at the level of the kernels. For instance, the \emph{perturbative kernel} computed with $B_{\kappa}$ is defined as:
\beq
\psi_{\mathrm{pert},\hc|\kappa}(p_1,p_2) = \exp\Big(\sum_{n \geq 1}\frac{1}{n!} \sum_{h \geq 0} \hc^{2h - 2 + n}\int_{p_2}^{p_1}\cdots\int_{p_2}^{p_1} \omega_{n|\kappa}^{h}\Big),
\eeq
and we observe that:
\beq
\label{sajuq}\psi_{\hc}(p_1,p_2) = \frac{\vartheta_{\bullet}}{\vartheta}\,\psi_{\mathrm{pert},\hc|\kappa}(p_1,p_2)\quad \mathrm{where} \quad \frac{\kappa^{\otimes d'}}{2^{d'}(2\mathrm{i}\pi)^{d'}\,d'!} \rightarrow \frac{1}{(2\mathrm{i}\pi)^{2d'}\,(2d')!}\,\frac{D^{\otimes d'} \vartheta}{\vartheta}.
\eeq

Examples of knots for which $\iota_* = -\mathrm{id}$ can be read off Figs.~\ref{Fig34}-\ref{Fig35}. For instance, it happens for the figure eight-knot and the manifold $L^2R$. These two examples have be studied in  \cite{DiFuji2}, where it was proposed that asymptotics of the colored Jones polynomial could be computed from $\psi_{\mathrm{pert},\hc|\kappa}$, at the price of \textit{ad hoc} renormalizations of $\kappa^{\otimes d'}$ to all orders. This phenomenon is explained by Eqns.~\ref{sqjua} and \ref{sajuq}, and this explanation is verified on examples in Section~\ref{S6}.

\section{Application to knot invariants}
\label{S4}
Our main conjecture is formulated in \S~\ref{conmi}. We first explain the background of Chern-Simons theory and facts about volume conjectures, which allow a better understanding of the identification of parameters and of the complicated Stokes phenomenon when we consider functions in the variable $u$.

\subsection{Generalities on Chern-Simons theory}
\label{CStheo}

\subsubsection*{With compact gauge group}

The partition function of Chern-Simons theory of compact gauge group $G$ (and corresponding Lie algebra $\mathfrak{g}$) in a closed $3$-manifold $\overline{\mathfrak{M}}$ is formally the path integral over $\mathfrak{g}$-connections $\mathcal{A}$ on $\overline{\mathfrak{M}}$, of the Chern-Simons action:
\beq\begin{split}
\label{Cscompact}Z_{\mathrm{CS}\,;\,G}(\overline{\mathfrak{M}}) & = \int [\mathcal{D}\mathcal{A}]\,e^{-\frac{1}{\hbar}\,S_{\mathrm{CS}}[\mathcal{A}]}  \\
S_{\mathrm{CS}}[\mathcal{A}] & = \int_{\overline{\mathfrak{M}}} \frac{1}{4}\Big(\mathcal{A}\wedge\dd\mathcal{A} + \frac{2}{3}\mathcal{A}\wedge\mathcal{A}\wedge\mathcal{A}\Big).
\end{split}\eeq
It depends on the Planck constant $\hbar$.

A way to define properly this integral is to choose a saddle point $\mathcal{A}^{\mathrm{cl}}$ of the action $S_{\mathrm{CS}}$, and perform an expansion around $\mathcal{A}_{\mathrm{cl}}$ as usual in perturbative quantum field theory. By construction of Chern-Simons theory, the saddle points (also called "classical solutions") are flat connections on $\overline{\mathfrak{M}}$, i.e. those satisfying $\dd\mathcal{A}^{\mathrm{cl}} + \mathcal{A}^{\mathrm{cl}}\wedge\mathcal{A}^{\mathrm{cl}} = 0$. However, there are in general many equivalence classes of flat connections, and one wishes the genuine partition function to be a sum over all classes of the perturbative partition functions, with some coefficients $\alpha_{\mathrm{cl}}$:
\beq
Z_{\mathrm{CS}\,;\,G}(\overline{\mathfrak{M}}) = \sum_{\mathrm{cl}} \alpha_{\mathrm{cl}}\,Z_{G}^{\mathrm{cl}}(\overline{\mathfrak{M}}).
\eeq
This sum is finite when $\hbar$ assumes a value\footnote{There exists two conventions for the Planck constant in Chern-Simons theory: either one puts $\hbar$ \cite{Dimofte} in the denominator in the exponential, or $2\hbar$ \cite{GZLD}, \cite{DiFuji2}. We adopt the second convention, where $\hbar = 2{\rm i}\pi/\mathrm{integer}$.} of the form:
\beq
\hbar = \frac{{\rm i}\pi}{K + H^{\vee}},
\eeq
where $K$ is an integer called \emph{level} and $H^{\vee}$ is the dual Coxeter number of $G$. There is actually a rigorous definition of the Wilson lines for these values \cite{Tura}, \cite{TuraRe}.

\subsubsection*{With complex gauge group}

The complexification comes in two steps. We shall be sketchy here and refer to \cite{CSQ} for details. Firstly, one considers a Chern-Simons theory with complex gauge group $G_{\mathbb{C}}$, whose Lie algebra is obtained from $\mathfrak{g}$ by Weyl's unitary trick, and with the new action $S_{\mathrm{CS}}[\mathcal{A}] + S_{\mathrm{CS}}[A^*]$. The partition function then admits a decomposition in perturbative blocks, defined by expansion around a $\mathfrak{g}_{\mathbb{C}}$-valued flat connexion:
  \beq\begin{split}
\label{decoq} Z_{G_{\mathbb{C}}} & = \sum_{\mathrm{cl}} \alpha_{\mathrm{cl}}\,Z_{G_{\mathbb{C}}}^{\mathrm{cl}} =  \sum_{\mathrm{cl}} \alpha_{\mathrm{cl}}\,Z_{G}^{\mathrm{cl}}\,(Z_{G}^{\mathrm{cl}})
\end{split}
\eeq
The partition function is real when $\hbar^*$ and $\mathcal{A}^*$ are the complex conjugates of $\hbar$ and $\mathcal{A}$. Secondly, at the level of the perturbative blocks, one consider a complexified version of the theory by assuming $\mathcal{A}$ and $\mathcal{A}^*$ independent $\mathfrak{g}_{\mathbb{C}}$-valued connections. The blocks then have a factorization $Z_{G_{\mathbb{C}}}^{\mathrm{cl}} = \Phi_{G}^{\mathrm{cl}}\Phi_{G}^{*,\mathrm{cl}}$. The $\Phi_{G}^{\mathrm{cl}}$ are called \emph{holomorphic blocks}, and they will play an important r\^{o}le in the following. By construction, they have an expansion in power series of $\hbar$, whose coefficients can be computed as well-defined sums over Feynman diagrams.

\subsubsection*{Wilson loops and colored Jones polynomial}

The most important observables in Chern-Simons theory are the \emph{Wilson loops}: given an oriented loop $\mathfrak{K}$ in $\overline{\mathfrak{M}}$, and a representation $R$ of $G$, they are defined as
\beq
\mathcal{W}_{G,R}(\mathfrak{K},\hbar) = \Big\langle \mathrm{Tr}_{R}\,\mathrm{P}\exp\Big(\oint_{\mathfrak{K}} \mathcal{A}\Big)\Big\rangle,
\eeq
where $\mathrm{P}$ is the ordering operation along the loop. $\mathcal{L}$ can be considered as a knot drawn in $\overline{\mathfrak{M}}$, and in fact the Wilson loop is a partition function for the knot complement $\overline{\mathfrak{M}}\setminus\mathfrak{K}$, where the classical solutions are now flat connections on $\overline{\mathfrak{M}}\setminus\mathfrak{K}$ with a meridian holonomy prescribed by $R$. To be precise, if $\rho=\frac{1}{2}\sum_{\alpha>0} \alpha$ is the vector of Weyl's constants, $\Lambda_R = (\lambda_j)_j$ is the highest weight associated to $R$, we must identify the holonomy eigenvalues to $e^{\hbar(\rho_j + \lambda_j)} = e^{u_j}$.

A foundational result is that the Wilson loops define knot invariants. When $G = \mathrm{SU(2)}$ and $R$ is the spin $\frac{N - 2}{2}$ representation, which has dimension $N$ and is represented by the Young diagram
\beq
\lambda_{N - 1} = \underbrace{\tableau{3}\cdots\tableau{2}}_{N - 1\,\,\mathrm{boxes}},
\eeq
the Wilson loop is related to the colored Jones polynomial $J_N(\mathfrak{K},q)$, with identifications:
\beq
\label{Jonesdef} q = e^{2\hbar},\qquad N\hbar = u,\qquad J_N(\mathfrak{K},q) = \frac{\mathcal{W}_{\mathrm{SU}(2),\lambda_{N - 1}}(\mathfrak{K},\hbar)}{\mathcal{W}_{\mathrm{SU}(2),\lambda_{N - 1}}(\varbigcirc,\hbar)}.
\eeq
The denominator accounts for the normalization of the Jones polynomial, which is $1$ for the unknot in $\mathbb{S}_3$, denoted $\varbigcirc$. The Wilson loop of the unknot is itself given by:
\beq
\mathcal{W}_{\mathrm{SU}(2),\lambda_{N - 1}}(\varbigcirc,\hbar) = \frac{q^{N/2} - q^{-N/2}}{q^{1/2} - q^{-1/2}} = \frac{\mathrm{sh}\,u}{\mathrm{sh}\,\hbar}.
\eeq

\subsection{The volume conjectures}
\label{volc}
Initially, the Jones polynomial $J_2(\mathfrak{K},q)$ has been defined in \cite{Jonesini} and its colored version $J_N(\mathfrak{K},q)$ in \cite{Tura}, \cite{TuraRe}, in the context of quantum groups. $J_N$ is a Laurent polynomial in $q$ with integer coefficients. The number $J_N(\mathfrak{K},q = e^{2\mathrm{i}\pi/N})$ is usually called the Kashaev invariant, and the original volume conjecture is:
\begin{conjecture}\cite{Kashaev}
\label{Kconj}For any hyperbolic knot $\mathfrak{K}$ in $\mathbb{S}_3$:
\beq
\frac{2\pi}{N}\,\lim_{N \rightarrow \infty} \ln |J_N(\mathfrak{K},q=e^{2\mathrm{i}\pi/N})| = \mathsf{Vol}(\mathbb{S}_3\setminus\mathfrak{K}).
\eeq
\end{conjecture}
It was later enhanced by Gukov \cite{GukovV} to include hyperbolic deformations of $\mathbb{S}_3\setminus\mathfrak{K}$, and subleading terms:
\begin{conjecture} \cite{GukMur}
\label{sauq}For any knot $\mathfrak{K}$ in $\mathbb{S}_3$, and $u$ in certain open domains $\mathcal{U}^{(\alpha)} \subseteq \mathbb{C}$, in the regime $N \rightarrow \infty$, $\hbar \rightarrow 0$ while $N\hbar \rightarrow u$, the colored Jones has an asymptotic expansion of the form
\beq
\label{ooooo}J_{N}(q = e^{2\hbar},\mathfrak{K}) \sim \hbar^{\delta^{(\alpha)}/2}\exp\Big(\sum_{\chi \geq -1} \hbar^{\chi}\jmath_{\chi}^{(\alpha)}(u)\Big)
\eeq
\begin{itemize}
\item The leading order is a complexified volume $\jmath_{-1}^{(\alpha)}(u) = \int^{p_u^{(\alpha)}} \ln l\,\dd\ln m$, where $p_u^{(\alpha)}$ is a point in some component of the $A$-polynomial such that $m(p_u^{(\alpha)}) = e^{u}$ (see Remark~\ref{AAAI} when $\mathfrak{K}$ is not hyperbolic),
\item $\delta^{(\alpha)}$ is an integer computed from cohomology, and $\jmath_0^{(\alpha)}(u)$ is related to the Ray-Singer torsion.
\item $\jmath_{\chi}(u)$ for $\chi \geq 1$ are the coefficients in the $\hbar$-expansion of a certain holomorphic block $\Phi^{(\alpha)}(u)$ for $\mathrm{SL}_2(\mathbb{C})$ Chern-Simons theory on $\mathfrak{M}$ with boundary condition specified by $u$.
\end{itemize}
\end{conjecture}
The statement about the leading order is called the \emph{generalized volume conjecture} (GVC). The range of validity in $u$ in not obvious, because of resonances and Stokes phenomena, that we attempt to describe in the next paragraph. The Kashaev invariant is retrieved for $u = {\rm i}\pi$. We first recall two rigorous results about the leading order of the GVC. The first one is due to Garoufalidis and L\^{e}:
\begin{theorem} \cite{GarouLe}
\label{volsmall} For any knot $\mathfrak{K}$, if $u$ is nonnegative and small enough, then
\beq
\lim_{\substack{N \rightarrow \infty \\ u = N\hbar}} \ln J_N(\mathfrak{K},q = e^{2\hbar}) = 0,
\eeq
i.e. the GVC holds with the choice of the abelian component.
\end{theorem}
The second is due to Murakami, who studied the figure-eight knot starting from a closed formula available in this case.
\begin{theorem} \cite{Murakami1}
\label{Mura1} For the figure-eight knot, the expression $\hbar\,\ln J_N(\mathbf{4}_1,q = e^{2\hbar})$ has the following behavior in the regime $N \rightarrow \infty$, $\hbar \rightarrow 0$ while $N\hbar = u$:
\begin{itemize}
\item When $u$ is real and $|u| \leq \frac{1}{2}\ln\Big(\frac{3 + \sqrt{5}}{2}\Big)$, or $u \in [0,\frac{{\rm i}\pi}{6}[$, the $\lim$ is $0$ (GVC for the abelian component).
\item When $u = {\rm i}\pi$ or $u \in \big]\frac{5{\rm i}\pi}{6},\frac{7{\rm i}\pi}{6}\big[ \notin {\rm i}\pi\mathbb{Q}$, the $\lim$ is given by the GVC for the geometric component.
\item When $u = \frac{{\rm i}\pi P}{Q}$ with $P,Q$ coprime integers and $P/Q \in \big]5/6,7/6[$, the $\lim$ is $0$ when $N \rightarrow \infty$ along multiples of $Q$, whereas the $\lim$ is given by the GVC for the geometric component if $N \rightarrow \infty$ avoiding multiples of $Q$.
\end{itemize}
\end{theorem}
We recall that the $A$-polynomial of the figure-eight knot has two components, one abelian $(l - 1)$ and one geometric, which intersect at $m^2 = -1$ and $m^2 = \frac{3 \pm \sqrt{5}}{2}$. One recognizes in the latter a value of $u$ at which a transition between components occur for the GVC to be valid according to Theorem~\ref{Mura1}. The example of the figure-eight is special in two ways. Firstly, its branchpoints are located at $m^2 = e^{\pm 2{\rm i}\pi/3}$ and $m^2 = \frac{3 \pm \sqrt{5}}{2}$, so two of them coincide with the intersection points. So, we do not see a change of branch within a single component at $\sqrt{3 + \sqrt{5}}{2}$ but actually a transition to the abelian component, and the behavior around the other branchpoints is beyond the range of validity of Theorem~\ref{Mura1}. Secondly, $\mathsf{CS}_{a}(p)$ vanishes along the path from $p_c$ to $p$ in the geometric component such that:
\beq
m^2(p) \in \big[\frac{3 - \sqrt{5}}{2},\frac{3 + \sqrt{5}}{2}\big]\cup \big\{ e^{i\varphi} \quad \phi \in [-2\mathrm{i}\pi/3,2\mathrm{i}\pi/3]\big\}.
\eeq
so that Theorem~\ref{Mura1} is only sensitive to the volume, not to the Chern-Simons part.

\subsection{$\hat{A}$-polynomial, AJ conjecture and Stokes phenomenon}
\label{AJpar}

Garoufalidis and L\^{e} \cite{AhatGLe} showed that $J_N(\mathfrak{K},q)$ always satisfy some recurrence relation on $N$. At the level of the analytic continuation, this turns into the existence of an operator $\widehat{\mathfrak{A}}_{\mathfrak{K}} \in \mathbb{Z}[e^{\frac{\hbar}{2} \partial_u},e^{u},e^{\hbar}]$ so that:
\beq
\label{ahtl}\widehat{\mathfrak{A}}_{\mathfrak{K}}\cdot J_{u/\hbar}(\mathfrak{K},q = e^{2\hbar}) = 0.
\eeq
The AJ conjecture \cite{AJconj} states that the limit $\hbar \rightarrow 0$ of $\widehat{\mathfrak{A}}$ coincides withwith the $A$-polynomial of $\mathfrak{K}$ up to a factor which is a polynomial in $e^{u}$, i.e. the $A$-polynomial is the semiclassical spectral curve associated to the difference equation Eqn.~\ref{ahtl}. It has been proved recently in \cite{AJLe} for hyperbolic knots satisfying some technical assumptions and for which the A-polynomial has only a single irreducible factor apart from $(l - 1)$.

If we treated Eqn.~\ref{ahtl} like an ODE, the leading asymptotic of the colored Jones when $\hbar \rightarrow 0$ would be given naively by a WKB analysis, namely:
\beq
\label{absee}J_{u/\hbar}(\mathfrak{K},q = e^{2\hbar}) \sim \exp\Big(\frac{1}{\hbar}\,\int^{p_u} \ln l\,\dd \ln m\Big),
\eeq
where $l$ and $m$ satisfies $\lim_{\hbar \rightarrow 0} \hat{A}_{\hbar}(m,l) = 0$, and $p_u$ is a point on this curve such that $\ln m(p_u) = u$. At the heuristic level, it explains the appearance of the complexified volume in the leading asymptotics of the colored Jones polynomial, by combining the AJ conjecture and Neumann-Zagier results reviewed in \S~\ref{VCS}. Going a step further, we could imagine to introduce a infinite set of times and embed Eqn.~\ref{ahtl} (at least perturbatively in the new times) in a system of compatible ODE's, for which we know how to associate quantities satisfying loop equations \cite{BEloop}, \cite{BETW}, \cite{BBEloop}. Those loop equations have many solutions, and the non-perturbative topological recursion applied to the semiclassical spectral curve provide distinguished solutions as formal asymptotic series in $\hbar$ \cite{BEInt}. This naive approach can be seen as a vague intuition why it is sensible to compare objects computed from the topological recursion to the asymptotics of solutions of the $\widehat{\mathfrak{A}}$ recursion relation, which we attempt to do in \S~\ref{conmi}.

Difference equation are of discrete nature, and if we treat it like an ODE we may miss resonance phenomena, which here occur when $q$ is a root of unity. On top of that, we have to take into account the usual Stokes phenomenon, hidden in the specification of the point $p_u$ on the semiclassical spectral curve which projects to $\ln m(p_u) = u$. This choice comes in three part:
\begin{itemize}
\item to which component $\mathcal{C}^{(\alpha)}$ of the $A$-polynomial should $p_u$ belong ?
\item in which sheet of the covering $m^2\,:\mathcal{C}^{(\alpha)} \rightarrow \mathbb{C}$ should $p_u$ belong ?
\item which determination of the logarithms in $\jmath_{-1}(u)$ should be chosen ?
\end{itemize}
We call the data of such a triple $\mathcal{T}^{(\alpha)} = (\mathcal{C}^{(\alpha)},p^{(\alpha)},\ln)$ a \emph{determination}. Although we can consider the RHS intrinsically as a function of a point $p$ in the the universal covering of the $\mathrm{SL}_2(\mathbb{C})$ character variety (defined component by component), it is a non trivial issue to predict for which determination it can be matched to the asymptotics of the LHS which is a function of $u$. The transition between different determinations occurs across \emph{Stokes curves} in the $u$-complex plane. Although the values of $u$ at which several components intersect, and branchcut structures of the coverings $m^2\,:\,\mathcal{C} \rightarrow \mathbb{C}$ represented in the $u$-plane obviously play a role, we do not know of a unambiguous algorithm which would give, for any knot, the determination corresponding to each domain and the correct pattern of Stokes curves which separate them. For second order differential equation (i.e. for semiclassical spectral curves having a single component, the form $y^2 = \mathrm{Pol}(x)$), the algorithm yielding the Stokes curves is known \cite{Bertolacut}, but it is not obvious to generalize this construction to curves of the form $\mathrm{Pol}(e^{x},e^{y}) = 0$ and having several components. The only reliable facts are that, for hyperbolic knots, one has to choose:
\begin{itemize}
\item for $u$ close to ${\rm i}\pi$: the determination corresponding to the geometric branch of the geometric component (see \S~\ref{HypTri}). We call it the \emph{geometric determination}.
\item and for $u$ nonnegative and close to $0$, the determination corresponding to the abelian component, so that $\jmath_{-1}(u) \equiv 0$.
\end{itemize}

\subsection{Main conjectures}
\label{conmi}

Let $\mathfrak{M}$ is a hyperbolic $3$-manifold with $1$-cusp, and let us consider an A-spectral curve $(\mathcal{C}_0,u,v)$ coming from an irreducible component of the $A$-polynomial of $\mathfrak{M}$. We would like to consider the asymptotics series constructed from the $2|2$-kernel introduced in \S~\ref{dedede}:
\beq
\mathcal{J}_{\hbar}^{\mathrm{n.p.TR}}(p) = \big(\psi_{\hbar}^{[2|2]}(p,o;\iota(p),\iota(o))\big)^{1/2}
\eeq
depending on a choice of basepoint $o$ and a characteristics $\mu,\nu \in \mathbb{C}^g$. We identified the formal parameter $\hc$ to $\hbar$. We recall that $\iota$ is the involution $(m,l) \rightarrow (1/m,1/l)$ defined on $\mathcal{C}_0$. Let us recapitulate its properties:
\begin{itemize}
\item $\mathcal{J}_{\hbar}^{\mathrm{n.p.TR}}(p)$ is defined as a formal asymptotic series:
\beq
\mathcal{J}^{\mathrm{n.p.TR}}_{\hbar}(p) = \exp\Big(\sum_{\chi \geq -1} \hbar^{\chi}\,\widetilde{\jmath}_{\chi}(p)\Big).
\eeq
\item The leading order is the complexified volume up to a constant:
\beq
\widetilde{j}_{-1}(p) = \frac{1}{2}\int_{o,\iota(o)}^{p,\iota(p)} v\,\dd u = \int_{o}^p v\,\dd u = \frac{i}{2}\big(\mathsf{Vol}_a(p) + i\mathsf{CS}_a(p)\big).
\eeq
\item For any $\chi \geq 1$,  $\widetilde{\jmath}_{\chi}(p)$ is a meromorphic function of $p \in \mathcal{C}_0$, which is either independent of $\hbar$, or is a function of $\hbar$ which does not have a power series expansion when $\hbar \rightarrow 0$. We give in \S~\ref{firstf} its expression up to $\chi = 3$.
\item If $2\varsigma$ is the order of torsion of the symbol $\{m,l\}$ in $\mathrm{K}_2(\mathbb{C}(\mathcal{C}_0))$, for any $\chi \geq 1$, $\widetilde{\jmath}_{\chi}(p)$, seen as a function of $\hbar$, assumes a constant value on the subsequences $\hbar = \frac{{\rm i}\pi}{k}$ where $k$ is a integer with fixed congruence modulo $\varsigma$.
\item When $\iota_{*} = -\mathrm{id}$, for any $\chi \geq 1$, $\widetilde{\jmath}_{\chi}(p)$ is independent of $\hbar$.
\end{itemize}

\begin{conjecture}
\label{ahconj}There exists a choice of $o$ and $\mu,\nu$ such that $\mathcal{J}_{\hbar}^{\mathrm{n.p.TR}}(p)$ is annihilated by the $\widehat{\mathfrak{A}}$-operator.
\end{conjecture}

We also attempt to formulate a stronger version of the conjecture to identify this series with the all-order asymptotics of the colored Jones polynomial:

\begin{conjecture}
\label{qiq}If $\mathfrak{M}$ is the complement of a prime\footnote{A knot $\mathfrak{K}$ is \emph{composite} if ${}^{c}\mathfrak{K}$ can be written as the connected sum of two knot complements. Else, $\mathfrak{K}$ is said \emph{prime}. If $\mathfrak{K} = \mathfrak{K}_1\amalg\mathfrak{K}_2$ is obtained by such a connected sum, it is known that $J_N(\mathfrak{K},q) = J_N(\mathfrak{K}_1,q)\times J_N(\mathfrak{K}_2,q)$ and that $\mathfrak{A}_{\mathfrak{K}_1}(m,l)\mathfrak{A}_{\mathfrak{K}_2}(m,l)/(l - 1)$ divides $\mathfrak{A}_{\mathfrak{K}}(m,l)$ \cite[Proposition 4.3]{RemCL}. Thus, it is straightforward to adapt our proposal to composite knots.} hyperbolic knot, with a choice of determination as in the GVC (Conjecture~\ref{sauq}) and keeping the same notations, we have the all-order asymptotic expansion:
\beq
J_N(\mathfrak{K},q = e^{2\hbar}) \sim C_{\hbar}\,B(u)\,\mathcal{J}_{\hbar}^{\mathrm{n.p.TR}}(p^{(\alpha)}_u)
\eeq
for a constant $C_{\hbar}$ independent of $u$, and a prefactor $B(u)$ independent of $\hbar$. In other words, for any $\chi \geq 1$, the $\jmath_{\chi}^{(\alpha)}(u)$ of Eqn.~\ref{ooooo} coincide with $\widetilde{\jmath}_{\chi}(p^{(\alpha)}(u))$ up to a constant independent of $\hbar$ and $u$.
\end{conjecture}

\subsection{First few terms}
\label{firstf}
In the comparison to the colored Jones polynomial, there is always an issue of normalization, which is reflected in the prefactors $C_{\hbar}$ and $B(u)$ that we do not attempt to predict. Thus, the definition of $\widetilde{\jmath}_0$ is irrelevant here, and we refer to \cite{DiFuji1} for some discussion on the computation of the constant term $\jmath_0(u)$ in the GVC in terms of algebraic geometry on the A-polynomial.

We now write down the general expression for $\widetilde{\jmath}_{\chi}(p)$ for $\chi = 1,2,3$, in terms of modular quantities (see Section~\ref{S3} and in particular \S~\ref{modre}). We first introduce the $[l,0]$ tensors:
\beq
G_{n|\kappa}^{h,(d)}(p) = \frac{1}{n!}\,\frac{1}{(2\mathrm{i}\pi)^d\,d!}\,\underbrace{\int_{o,\iota(o)}^{p,\iota(p)}\cdots\int_{o,\iota(o)}^{p,\iota(p)}}_{n\,\,\mathrm{times}}\underbrace{\oint_{\mathcal{B}_{\kappa}}\cdots\oint_{\mathcal{B}_{\kappa}}}_{d\,\,\mathrm{times}} \omega_{n + d|\kappa}^{h}.
\eeq
and $G_{n|\kappa}^{h}(p) \equiv G_{n|\kappa}^{h,(0)}(p)$. Then, we have:
\beq\begin{split}
2\,\widetilde{\jmath}_1(p) & = G_{1|\kt}^{1,(0)}(p) + G_{3|\kt}^{0,(0)}(p) + {\blue T_{1,\bullet|\kt}\,G_{2|\kt}^{0,(1)}(p)} + T_{2,\bullet|\kt}\,G_{1|\kt}^{0,(2)}(p)  \\
&  {\blue + (T_{3,\bullet|\kt} - T_{3|\kt})G_{0|\kt}^{0,(3)}}  \\
2\,\widetilde{\jmath}_2(p) & =  G_{2|\kt}^{1,(0)}(p) {+ \blue T_{1,\bullet|\kt}\,G_{1|\kt}^{1,(1)}(p) + (T_{2,\bullet|\kt} - T_{2|\kt})G_{0|\kt}^{1,(2)}} + G_{4|\kt}^{0,(0)}(p) \\
& {\blue + T_{1,\bullet|\kt}\,G_{3|\kt}^{0,(1)}(p)} + T_{2,\bullet|\kt}\,G_{2|\kt}^{0,(2)}(p) + {\blue T_{3,\bullet|\kt}\,G_{1|\kt}^{0,(3)}(p)}  \\
& {\blue + (T_{4,\bullet|\kt} - T_{4|\kt})G_{0|\kt}^{0,(4)} + \frac{1}{2}(V_{\bullet|\kt}^{(1,1)} - T_{1|\kt}^2)\big(G_{0|\kt}^{1,(1)}\big)^2} {\blue + V_{\bullet|\kt}^{(1,1)}\,G_{0|\kt}^{0,(1)}\,G_{2|\kt}^{0,(1)}(p)} \\
&  {\blue + V_{\bullet|\kt}^{(1,2)}\,G_{0|\kt}^{0,(1)}\,G_{1|\kt}^{0,(2)}(p)} {\blue + (V_{\bullet|\kt}^{(1,3)} - T_{1|\kt}T_{3|\kt})G_{0|\kt}^{1,(1)}\,G_{0|\kt}^{0,(3)}}  \\
& {\blue + \frac{1}{2}\,V_{\bullet|\kt}^{(1,1)}\,\big(G_{2|\kt}^{0,(1)}(p)\big)^2} {\blue + V_{\bullet|\kt}^{(1,2)}\,G_{2|\kt}^{0,(1)}(p)\,G_{1|\kt}^{0,(2)}(p)}  \\
&  {\blue + V_{\bullet|\kt}^{(1,3)}\,G_{2|\kt}^{0,(1)}(p)\,G_{0|\kt}^{0,(3)}} + \frac{1}{2}\,V_{\bullet|\kt}^{(2,2)}\,\big(G_{1|\kt}^{0,(2)}(p)\big)^2 \\
& {\blue + V_{\bullet|\kt}^{(2,3)}\,G_{1|\kt}^{0,(2)}(p)\,G_{0|\kt}^{0,(3)}} {\blue + \frac{1}{2}(V_{\bullet|\kt}^{(3,3)} - T_{3|\kt}^{2})\big(G_{0|\kt}^{0,(3)}\big)^2}. \\
2\,\widetilde{\jmath}_3(p) & = G_{1|\kt}^{2,(0)}(p) {\blue + (T_{1,\bullet|\kt} - T_{1|\kt})G_{0|\kt}^{2,(1)}} +  G_{3|\kt}^{1,(0)}(p) \\
& {\blue + T_{1,\bullet|\kt}\,G_{2|\kt}^{1,(1)}(p)} + T_{2,\bullet|\kt}\,G_{1|\kt}^{1,(2)}(p)  {\blue + (T_{3,\bullet|\kt} - T_{3|\kt})G_{0|\kt}^{1,(3)}} \\
& +  G_{5|\kt}^{0,(0)}(p) {\blue + T_{1,\bullet|\kt}\,G_{4|\kt}^{0,(1)}} + T_{2,\bullet|\kt}\,G_{3|\kt}^{0,(2)}(p)  \\
& {\blue + T_{3,\bullet|\kt}\,G_{2|\kt}^{0,(3)}(p)} + T_{4,\bullet|\kt}\,G_{1|\kt}^{0,(4)}(p) \\
& {\blue + (T_{5,\bullet|\kt} - T_{5|\kt})G_{0|\kt}^{0,(5)}} {\blue + V_{\bullet|\kt}^{(1,1)}\,G_{0|\kt}^{1,(1)}\,G_{1|\kt}^{1,(1)}(p)}  \\
& {\blue + V_{\bullet|\kt}^{(1,2)}\,G_{0|\kt}^{1,(1)}\,G_{0|\kt}^{1,(2)} + V_{\bullet|\kt}^{(1,2)}\,G_{0|\kt}^{1,(1)}\,G_{3|\kt}^{0,(1)}(p)}  \\
& {\blue +V_{\bullet|\kt}^{(1,2)}\,G_{0|\kt}^{1,(1)}\,G_{2|\kt}^{0,(2)}(p)  + V_{\bullet|\kt}^{(1,3)}\,G_{0|\kt}^{1,(1)}\,G_{1|\kt}^{0,(3)}(p)}  \\
& {\blue + (V_{\bullet|\kt}^{(1,4)} - T_{1|\kt}T_{4|\kt})G_{0|\kt}^{1,(1)}\,G_{0|\kt}^{0,(4)} + V_{\bullet|\kt}^{(1,1)}\,G_{2|\kt}^{0,(1)}(p)\,G_{1|\kt}^{1,(1)}(p)}  \\
& {\blue + V_{\bullet|\kt}^{(1,1)}\,G_{2|\kt}^{0,(1)}(p)\,G_{0|\kt}^{1,(2)} + V_{\bullet|\kt}^{(1,1)}\,G_{2|\kt}^{0,(1)}(p)\,G_{3|\kt}^{0,(1)}(p)}  \\
& {\blue + V_{\bullet|\kt}^{(1,2)}\,G_{2|\kt}^{0,(1)}\,G_{2|\kt}^{0,(2)}(p) + V_{\bullet|\kt}^{(1,3)}\,G_{2|\kt}^{0,(1)}(p)\,G_{1|\kt}^{0,(3)}(p)} \\
& {\blue + (V_{\bullet|\kt}^{(1,4)} - T_{1|\kt}T_{4|\kt})G_{2|\kt}^{0,(1)}(p)\,G_{0|\kt}^{0,(4)} +V_{\bullet|\kt}^{(2,1)}\,G_{1|\kt}^{0,(2)}(p)\,G_{1|\kt}^{1,(1)}(p)}  \\
& + V_{\bullet|\kt}^{(2,2)}\,G_{1|\kt}^{0,(2)}(p)\,G_{0|\kt}^{1,(2)} {\blue + V_{\bullet|\kt}^{(2,1)}\,G_{1|\kt}^{0,(2)}(p)\,G_{3|\kt}^{0,(1)}(p)}  \\
& + V_{\bullet|\kt}^{(2,2)}\,G_{1|\kt}^{0,(2)}(p)\,G_{2|\kt}^{0,(2)}(p) {\blue + V_{\bullet|\kt}^{(2,3)}\,G_{1|\kt}^{0,(2)}(p)\,G_{1|\kt}^{0,(3)}(p)} \\
& + V_{\bullet|\kt}^{(2,4)}\,G_{1|\kt}^{0,(2)}(p)\,G_{0|\kt}^{0,(4)} {\blue + V_{\bullet|\kt}^{(3,1)}\,G_{0|\kt}^{0,(3)}\,G_{1|\kt}^{1,(1)}(p)}  \\
& {\blue + (V_{\bullet|\kt}^{(3,2)} - T_{3|\kt}T_{2|\kt})G_{0|\kt}^{0,(3)}\,G_{0|\kt}^{1,(2)} + V_{\bullet|\kt}^{(3,1)}\,G_{0|\kt}^{0,(3)}\,G_{3|\kt}^{0,(1)}(p)} \\
& {\blue +V_{\bullet|\kt}^{(3,2)}\,G_{0|\kt}^{0,(3)}\,G_{2|\kt}^{0,(2)}(p) + V_{\bullet|\kt}^{(3,3)}\,G_{0|\kt}^{0,(3)}\,G_{1|\kt}^{0,(3)}(p)}  \\
& {\blue + (V_{\bullet|\kt}^{(3,4)} - T_{3|\kt}T_{4|\kt})G_{0|\kt}^{0,(3)}\,G_{0|\kt}^{0,(4)} + \frac{1}{6}(V_{\bullet|\kt}^{(1,1,1)} - T_{1|\kt}^3)\big(G_{0|\kt}^{1,(1)}\big)^3}  \\
& {\blue + \frac{1}{2}\,V_{\bullet|\kt}^{(1,1,2)}\,\big(G_{0|\kt}^{1,(1)}\big)^2\,G_{2|\kt}^{0,(1)}(p) + \frac{1}{2}\,V_{\bullet|\kt}^{(1,1,2)}\,\big(G_{0|\kt}^{1,(1)}\big)^2\,G_{1|\kt}^{0,(2)}(p)}  \\
& {\blue + \frac{1}{2}(V_{\bullet|\kt}^{(1,1,3)} - T_{1|\kt}^2T_{3|\kt})\big(G_{0|\kt}^{1,(1)}\big)^2\,G_{0|\kt}^{0,(3)} + V_{\bullet|\kt}^{(1,1,2)}\,G_{0|\kt}^{1,(1)}\,G_{2|\kt}^{0,(1)}(p)\,G_{1|\kt}^{0,(2)}(p)}
\end{split}
\eeq
\beq
\begin{split}
\phantom{2\,\widetilde{\jmath}_3(p)} & \cdots {\blue + V_{\bullet|\kt}^{(1,1,3)}\,G_{0|\kt}^{1,(1)}\,G_{2|\kt}^{0,(1)}(p)\,G_{0|\kt}^{0,(3)} + \frac{1}{2}\,V_{\bullet|\kt}^{(1,2,2)}\,G_{0|\kt}^{1,(1)}\big(G_{1|\kt}^{0,(2)}(p)\big)^2} \\
& {\blue + V_{\bullet|\kt}^{(1,2,3)}\,G_{0|\kt}^{1,(1)}\,G_{1|\kt}^{0,(2)}(p)\,G_{0|\kt}^{0,(3)} + \frac{1}{2}(V_{\bullet|\kt}^{(1,3,3)} - T_{1|\kt}T_{3|\kt}^2)G_{0|\kt}^{1,(1)}\big(G_{0|\kt}^{0,(3)}\big)^2} \\
& {\blue + \frac{1}{6}\,V_{\bullet|\kt}^{(1,1,1)}\,\big(G_{2|\kt}^{0,(1)}(p)\big)^3 + \frac{1}{2}\,V_{\bullet|\kt}^{(1,1,2)}\,\big(G_{2|\kt}^{0,(1)}(p)\big)^2\,G_{1|\kt}^{0,(2)}(p)}  \\
& {\blue + \frac{1}{2}\,V_{\bullet|\kt}^{(1,1,3)}\,\big(G_{2|\kt}^{0,(1)}(p)\big)^2\,G_{0|\kt}^{0,(3)} + \frac{1}{2}\,V_{\bullet|\kt}^{(1,2,2)}\,G_{2|\kt}^{0,(1)}(p)\big(G_{1|\kt}^{0,(2)}(p)\big)^2}  \\
& {\blue + V_{\bullet|\kt}^{(1,2,3)}\,G_{2|\kt}^{0,(1)}(p)\,G_{1|\kt}^{0,(2)}(p)\,G_{0|\kt}^{0,(3)} + \frac{1}{2}\,V_{\bullet|\kt}^{(1,3,3)}\,G_{2|\kt}^{0,(1)}(p)\big(G_{0|\kt}^{0,(3)}\big)^2}  \\
& + \frac{1}{6}\,V_{\bullet|\kt}^{(2,2,2)}\,\big(G_{1|\kt}^{0,(2)}(p)\big)^3 {\blue + \frac{1}{6}\,V_{\bullet|\kt}^{(2,3,3)}\,\big(G_{1|\kt}^{0,(2)}(p)\big)^2\,G_{0|\kt}^{0,(3)}}  \\
& {\blue + \frac{1}{2}\,V_{\bullet|\kt}^{(2,3,3)}\,G_{1|\kt}^{0,(2)}(p)\,\big(G_{0|\kt}^{0,(3)}\big)^2 + \frac{1}{6}(V_{\bullet|\kt}^{(3,3,3)} - T_{3|\kt}^3)\big(G_{0|\kt}^{0,(3)}\big)^3}.
\end{split}\eeq
In the formulas above, the $G_{l|\kt}^{h,(d)}$ are contracted (from left to right) with the tensors $T_{d,\bullet|\kt}$ which were defined in Eqns.~\ref{astana} and their combinations:
\beq\begin{split}
V_{2,\bullet|\kt}^{(d_1,d_2)} & = T_{d_1 + d_2,\bullet|\kt} - T_{d_1,\bullet|\kt}T_{d_2,\bullet|\kt} \\
V_{2,\bullet|\kt}^{(d_1,d_2,d_3)} & = T_{d_1 + d_2 + d_3,\bullet|\kt} \\
& - (T_{d_1,\bullet|\kt}T_{d_2 + d_3,\bullet|\kt} + T_{d_2,\bullet|\kt}T_{d_1 + d_3,\bullet|\kt} + T_{d_3,\bullet|\kt}T_{d_1 + d_2,\bullet|\kt}) \\
& + 2T_{d_1,\bullet|\kt}T_{d_2,\bullet|\kt}T_{d_3,\bullet|\kt}
\end{split}
\eeq
They are combination of derivatives of theta functions evaluated at:
\beq
\label{secaq}\mathbf{w}_{\bullet} = \ab(p) - \ab(o) + \ab(\iota(p)) - \ab(\iota(o)) + \zeta_{\hbar},
\eeq
and the constant $\zeta$ is defined in Eqn.~\ref{reca}. When $\iota_* = -\mathrm{id}$, several simplifications occur: the blue terms vanish ; $\ab(p) - \ab(o) + \ab(\iota(p)) - \ab(\iota(o)) = 0$ and $\zeta_{\hbar} = 0$, the argument of the $\vartheta$ and $\vartheta_{\bullet}$ is always zero, i.e. $\mathbf{w}_{\bullet} = \mathbf{w} = 0$. In particular, this implies that, for any $\chi \geq 1$, $\widetilde{\jmath}_{\chi}(p)$ is independent of $\hbar$.

\subsection{Comments}

We check that Conjecture~\ref{qiq} holds for the figure-eight knot in \S~\ref{figure8} up to $o(\hbar^3)$, with $o$ chosen (in a certain sense, as explained later) at a branchpoint, $[\mu,\nu]$ the unique half-integer characteristics with reality properties, and $C_{\hbar} \equiv 1 + o(\hbar^3)$. This very natural choice of the normalization allows to retrieve the asymptotic expansion of the Kashaev invariant $J_N(\mathbf{4}_1,q = e^{2{\rm i}\pi/N})$ when $N \rightarrow \infty$, by specializing to $u = {\rm i}\pi$ and taking the geometric determination.

For the once punctured torus bundle $L^2R$ (a knot complement in lens space), we check in \S~\ref{L2R} up to $o(\hbar^2)$ that:
\beq
\mathcal{J}_{\hbar}^{\mathrm{H}}(u) \sim C_{\hbar}\,B(u)\,\mathcal{J}_{\hbar}^{\mathrm{n.p.TR}}(p^{(\alpha)}(u)),
\eeq
where $\mathcal{J}_{\hbar}^{\mathrm{H}}(u)$ is a Hikami-type integral associated to $L^2R$, and for the RHS, $o$ is chosen at a branchpoint, $[\mu,\nu]$ is the unique half-integer characteristics with reality properties, and we choose the geometric determination. However, the normalization is now $C_{\hbar} = 1 + \frac{\hbar^2}{32} + o(\hbar^2)$.

The free parameters in our conjecture are the basepoint $o$ for computing primitives, and the characteristics $[\mu,\nu]$ of the theta functions. Notice that different choices of $o$ affects $\widetilde{\jmath}_{\chi}(p)$ in a non trivial way, since it contains products of primitives. We have not found a general rule to specify neither $o$, nor $\mu,\nu$, and the choices might be also subjected to Stokes jumps regarding the identification to asymptotics of the Jones polynomial. In the examples treated in Section~\ref{S6}, the curve has $g = 1$ and we find natural choices for them. In general, we think that it must chosen among even half-integer characteristics, so $\binom{2g + 1}{g}$ possibilities are left. Recall that, for hyperelliptic curves, they are in bijection with partitions of the $2g + 2$ Weierstra{\ss} points in two sets of $g + 1$ elements. For A-spectral curves listed in Appendix~\ref{Fig1} that we found to be hyperelliptic\footnote{On top of curves of genus $g = 1,2$ which are necessarily hyperelliptic, we found that all genus $3$ curves of listed in Appendix~\ref{Fig1} are hyperelliptic, as well as $\mathbf{7}_2,\mathbf{9}_{10}^{(1)}, \mathbf{10}_{146}^{(2)}$ (genus $4$), $\mathbf{8}_1$ (genus $5$) and $\mathbf{9}_2$ (genus $7$). This list is not exhaustive within Fig.~\ref{Fig1}, because we could not obtain an answer from \textsc{maple} in reasonable time for curves of high degree.}, it turned out that they can be represented after birational transformations with rational coefficients $(m,l) \mapsto (X,Y)$, in the form:
\beq
Y^2 = S_1(X)S_2(X),
\eeq
where $S_1$ and $S_2$ are polynomials with integer coefficients and of the same degree $g + 1$, hence providing a canonical choice of even half-characteristics, for which $\big(\vartheta\bigl[{}^{\mu}_{\nu}\bigr](0)\big)^8$ computed by Thomae formula is an integer. This suggests that a deeper study of the $\mathrm{SL}_2(\mathbb{C})$ character variety could entirely fix the appropriate choice of $\mu,\nu$.

In such a conjecture, it is natural to identify the Planck constant $\hbar$ of Chern-Simons theory with the parameter $\hc$ of the non-perturbative partition functions of Section~\ref{S3}, since special properties arise on each side when $\hc$ and $\hbar$ assume values of the form ${\rm i}\pi/\mathrm{integer}$. In the framework of Chern-Simons theory, the Wilson line can be thought as a wave function, hence it is natural to compare them to kernels. The $2$-kernel $\psi^{[2|2]}_{\hc}(p_1,p_1' ; p_2,p_2')$ is symmetric by exchange of $(p_1,p_1')$ with $(p_2,p_2')$, so the right hand side is invariant under the involution $\iota$, which is also a property of the holomorphic blocks. We attempt to motivate\footnote{Dijkgraaf, Fuji and Manabe \cite{DiFuji1} also provided topological string arguments for the identification of parameters in Eqn.~\ref{Jonesdef} and the role of $\iota$.} further the precise form of the conjecture in Section~\ref{S7}. We shall see that, for torus knots, $\psi_{\hbar}^{[2|2]}$ \emph{without the power $1/2$} appears heuristically in the computation of the colored Jones polynomial. For torus knots, it is known \cite[Appendix B]{GukovV} that the Chern-Simons partition function $Z_{\mathrm{SL}_2(\mathbb{C})}$ coincide with $Z_{\mathrm{SU}(2)}$ up to a simple factor. For hyperbolic knots, we rather have Eqn.~\ref{decoq}, which incite to identify the holomorphic block with the analytic continuation of $\sqrt{Z_{\mathrm{SL}_2(\mathbb{C})}}$. This may account for the power $1/2$ in Conjecture~\ref{qiq}.

\section{Examples}
\label{S6}
From the point of view adopted in this article, the complexity of hyperbolic $3$-manifolds with $1$-cusp is measured by the complexity of the algebraic curve defined by the geometric component $\mathcal{C}^{\mathrm{geom}}_0$ of its A-polynomial: to compute $\mathcal{J}_{\hbar}^{\mathrm{n.p.TR}}(p)$, we need to compute explicitly meromorphic forms (and their primitives) on the curve, as well as values of theta functions and their derivatives. From the tables of A-polynomials of Culler \cite{Cullerweb}, \cite{CullerwebT}, we collected the genus of the A-polynomial components of various knots in Fig.~\ref{Fig1}.

The simplest non trivial class of manifolds correspond to those for which $\mathcal{C}^{\mathrm{geom}}_0$ is a genus $1$ curve, i.e. an elliptic curve. This happens for the geometric components of the figure $8$-knot and the manifold $L^2R$. The theta and theta derivatives values can be computed in a simple and efficient way thanks to the theory of modular forms (Section~\ref{ellipticgen}).

The next simplest class corresponds to manifolds for which $\mathcal{C}^{\mathrm{geom}}_0$ is hyperelliptic. In this case there are uniform expressions for a Bergman kernel in terms of the coordinates $m^2$ and $l$, and the theta values are well-known in terms of the coordinates of Weierstrass points. For curves of genus $g \geq 2$, in principle, the values of theta derivatives can be related to the theta values via the theory of Siegel modular forms and the work of \cite{ZudilinB}. The $\mathbf{5}_2$ knot and the Pretzel(-2,3,7) give rise to A-polynomial with a single component, of genus $2$ thus hyperelliptic. We leave to a future work explicit computations for A-spectral curves of genus $2$ and comparison to the perturbative invariants obtained by other methods.

We observe many times that some components of the A-polynomial of different knots are  either the same, or birationally equivalent. For instance, the A-polynomial of the $\mathbf{5}_2$ and the $\mathrm{Pretzel}(-2,3,7)$ are birationally equivalent, and one of the two factors of the A-polynomial of the $\mathbf{7}_4$ coincide with the A-polynomial of the $\mathbf{4}_1$. This remark has some interest because values of theta derivatives, which provide the corrective terms to be added to the topological recursion for comparison with the asymptotics of the colored Jones polynomial, only depend on the isomorphism class of $\mathcal{C}_0$ as a Riemann surface, i.e. only depend on $A(m,l)$ up to birational equivalence.

Since $\mathcal{C}$ is a singular curve, we do not expect a naive inequality between the degree of $A$ or of the invariant trace field (which contains the cusp field), and the genus $g$. We observe that:
\begin{itemize}
\item $g$ looks experimentally much lower than the genus of a generic smooth curve of the same degree as the $A$ polynomial.
\item the genus of the quotient $\mathcal{C}/\iota$ also drops compared to the genus of $\mathcal{C}$.
\end{itemize}
It would be interesting to have a interpretation of $g$ as well as those two observations from the point of view of representation theory. In the same vein, we can ask other open questions, here focused on geometric components:

\begin{problem}
Describe the set of elliptic curves over $\mathbb{Q}$ which are obtained as geometric components of hyperbolic $3$-manifolds. Do all elliptic curves arise in that way ?
\end{problem}

\begin{problem}
Characterize the hyperbolic $3$-manifolds so that the quotient $\mathcal{C}^{\mathrm{geom}}/\iota$ has genus $0$.
\end{problem}

\begin{problem}
For a given genus $g$, do an infinite number of non-isomorphic curves of genus $g$ arise as geometric components of a hyperbolic $3$-manifold ?
\end{problem}

These problems are already interesting in one replaces "geometric component" by the class of "$A$-spectral curves" defined in Eqn~\ref{algcurves}.

\subsection{Thetanullwerten for elliptic curves}
\label{ellipticgen}

In this section, we give a self-contained presentation to compute the theta functions and their derivatives appearing in Section~\ref{S3} and the computation of $\mathcal{J}_{\hbar}^{\mathrm{n.p.TR}}(p)$ for a genus $1$ spectral curve. For more details about elliptic modular forms, the reader may consult the recent textbook \cite[Chapter 1]{ZagierMod}.

\subsubsection{Modular forms and their derivatives}

Elliptic curves are characterized by the orbit of their period $\tau$ in the upper-half plane $\mathbb{H}$ under the modular group $\mathrm{SL}_{2}(\mathbb{Z})$. A modular form of weight $k$ for a subgroup $\Gamma \subseteq \mathrm{SL}_2(\mathbb{Z})$ is a by definition a holomorphic function $f\,:\,\mathbb{H} \rightarrow \mathbb{C}$ such that $f(\tau) = O(1)$ when $q = e^{2\mathrm{i}\pi\tau} \rightarrow 0$, and satisfying:
\beq
\label{modE}\forall \Big(\begin{array}{cc} a & b \\ c & d  \end{array}\Big) \in \Gamma\qquad f\Big(\frac{a\tau + b}{c\tau + d}\Big) = (c\tau + d)^{k} f(\tau)
\eeq
When the subgroup is not precised, it is understood that Eqn.~\ref{modE} holds for the full modular group. Obviously, modular forms are $1$-periodic functions, so have a Fourier expansion:
\beq
f(\tau) = \sum_{n \geq 0} a_n\,q^n\qquad\qquad q = e^{2\mathrm{i}\pi\tau},
\eeq
where only nonnegative indices appear owing to the growth condition when $q \rightarrow 0$. The Eisenstein series
\beq
\label{Eisendef}\forall l \geq 1\qquad E_{2l}(\tau) = \frac{1}{2\zeta(2l)} \sum_{(n,m) \in \mathbb{Z}^2\setminus\{(0,0)\}} \frac{1}{(n + m\tau)^{2l}}
\eeq
provide important examples of modular forms of weight $2l$ when $l \geq 2$. The zeta value in the denominator enforces the normalization $E_{2l}(\tau) = 1 + O(q)$ when $q \rightarrow 0$. We find convenient to absorb a factor of $\pi$ per unit weight, and introduce non standard notations $\tilde{E}_{2l} = \pi^{2l} E_{2l}$. It is well-known that the ring of modular forms is generated by $\tilde{E}_4$ and $\tilde{E}_6$. Thus, identities between modular forms of a given weight can be proved by checking that only a finite number of their Fourier coefficients match. $\tilde{E}_2$ fails to be modular, indeed one can show:
\beq
\label{E2mod}\forall \Big(\begin{array}{cc} a & b \\ c & d  \end{array}\Big) \in \mathrm{SL}_2(\mathbb{Z})\qquad \tilde{E}_2\Big(\frac{a\tau + b}{c\tau + d}\Big) = (c\tau + d)^2 \tilde{E}_2(\tau) - 6\mathrm{i}\pi\,c(c\tau + d).
\eeq
Let us define a differentation operator with an accurate normalization for our purposes:
\beq
D = 4\mathrm{i}\pi \partial_{\tau} =  2(2\mathrm{i}\pi)^2 q\partial_q.
\eeq
Obviously, derivatives of modular forms are not modular forms. If $f$ is a modular form of weight $k$ for some subgroup $\Gamma$, we rather have:
\beq
\label{modtran}\forall \Big(\begin{array}{cc} a & b \\ c & d  \end{array}\Big) \in \Gamma\qquad (Df)\Big(\frac{a\tau + b}{c\tau + d}\Big) = (c\tau + d)^{k + 2}(Df)(\tau) + 4\mathrm{i}\pi\,k\,c(c\tau + d)^{k + 1} f(\tau).
\eeq
This behavior is captured by the notion of "quasi-modular forms" and its relation with "non-holomorphic modular forms" \cite[Chapter 1]{ZagierMod}. We adopt however a more pedestrian way. It is easy to check that the combination:
\beq
\ds_k f = Df + \frac{2}{3}\,k\,\tilde{E}_2
\eeq
is modular of weight $k + 2$. $\ds_k f$ is called the \emph{Serre derivative} of $f$. Consequently, the differential closure of the ring of modular forms is generated by $E_2$, $E_4$ and $E_6$. The basic relations in the new ring are:
\begin{align}
\label{relDE1} D\tilde{E}_2 & = \frac{2}{3}(\tilde{E}_4 - \tilde{E}_2^2),\\
D\tilde{E}_4 & = \frac{8}{3}(\tilde{E}_6 -\tilde{E}_2\tilde{E}_4), \\
\label{relDE3} D\tilde{E}_6 & = 4(\tilde{E}_4^2 - \tilde{E}_2\tilde{E}_6).
\end{align}
Since the vector spaces of modular forms of weight $4$, $6$ and $8$ are $1$-dimensional, these relations can be proved by checking from Eqn.~\ref{E2mod} that $\ds_1 \tilde{E}_2$ is modular of weight $4$, hence of the form $c_4\tilde{E}_4$, and similarly $\ds_4\tilde{E}_4 = c_4\tilde{E}_6$ and $\ds_6\tilde{E}_6 = c_6\tilde{E}_6$. Then, one finds $c_{2l}$ by matching the constant Fourier coefficients of the two sides.

\subsubsection{Theta functions and their derivatives}

In genus $1$ there are $3$ even characteristics $\frac{1}{2}$, $0$ and $\frac{\tau}{2}$. The corresponding theta values are:
\beq
\vartheta_2(\tau) = \sum_{n \in \mathbb{Z}} (-1)^{n}\,q^{n^2/2}\qquad \vartheta_3(\tau) = \sum_{n \in \mathbb{Z}} q^{n^2/2}\qquad \vartheta_4(\tau) = \sum_{n \in \mathbb{Z} + 1/2} q^{n^2/2},
\eeq
and they satisfy the relation: $\vartheta_2^4 + \vartheta_4^4 = \vartheta_3^4$. The $\vartheta_i$ are modular forms of weight $1/2$, but only for a congruence subgroup $\Gamma(2)$ of $\mathrm{SL}_2(\mathbb{Z})$ (this is related to the shift of argument and the eight root of unity in Eqn.~\ref{thetatrans}). Their fourth powers build a vector modular form of weight $2$:
\beq
\label{modt}\vec{\vartheta^4}(\tau + 1) = \left(\begin{array}{ccc} -1 & 0 & 0 \\ 0 & 0 & 1 \\ 0 & 1 & 0 \end{array}\right)\cdot\vec{\vartheta^4}(\tau)\qquad \vec{\vartheta^4}(-1/\tau) = \tau^2\,\left(\begin{array}{ccc} 0 & 0 & -1 \\ 0 & 1 & 0 \\ -1 & 0 & 0 \end{array}\right)\cdot\vec{\vartheta^4}(\tau).
\eeq
It is possible to build out of $\vartheta_i^4$ expressions which are modular forms, resulting in relations to Eisenstein series upon checking a few Fourier coefficients. As before, we prefer to work with $\tilde{\vartheta}_i = \pi^{1/2}\vartheta_i$, and we obtain:
\begin{align}
\label{relEt1}\tilde{E}_4 & = \tilde{\vartheta}_2^8 + \tilde{\vartheta}_4^8 + \tilde{\vartheta}_2^4\tilde{\vartheta}_4^4 \\
 \label{relEt3}\tilde{E}_6 & = -\tilde{\vartheta}_2^{12} - \frac{3}{2}\tilde{\vartheta}_2^8\tilde{\vartheta}_2^4 + \frac{3}{2}\tilde{\vartheta}_2^4\tilde{\vartheta}_4^8 + \tilde{\vartheta}_4^{12}.
\end{align}
Combining Eqns.~\ref{relEt1}-\ref{relEt3} to Eqns.~\ref{relDE1}-\ref{relDE3}, we obtain after some algebra the basic relations in the differential ring generated by the $\tilde{\vartheta}_i$:
\begin{align}
\label{Diffexp1}\frac{D\tilde{\vartheta}_2}{\tilde{\vartheta}_2} & = \frac{1}{3}(-\tilde{\vartheta}_2^4 - 2\tilde{\vartheta}_4^4 - \tilde{E}_2), \\
\label{Diffexp2}\frac{D\tilde{\vartheta}_3}{\tilde{\vartheta}_3} & = \frac{1}{3}(-\tilde{\vartheta}_2^4 + \tilde{\vartheta}_4^4 - \tilde{E}_2), \\
\label{Diffexp3}\frac{D\tilde{\vartheta}_4}{\tilde{\vartheta}_4} & = \frac{1}{3}(2\tilde{\vartheta}_2^4 + \tilde{\vartheta}_4^4 - \tilde{E}_2).
\end{align}
From there follows the computation of the $d$-th derivative of $\tilde{\vartheta}_i$ to all orders, and we observe especially that :
\beq
D^{d}\tilde{\vartheta}_i/\tilde{\vartheta}_i = 3^{-d}P_d(\tilde{\vartheta}_2^4,\tilde{\vartheta}_4^4,\tilde{E}_2),
\eeq
where $P_d$ is a polynomial with integer coefficients. By Nesterenko's theorem, $\tilde{E}_2$ cannot be an algebraic number. So it might seem hopeless to obtain any explicit number to compute the kernels, e.g. Eqn.~\ref{psidef}. But we explained in \S~\ref{modre} how the kernels could be computed in terms of combinations of derivatives which were modular. We immediately see that the appropriate combination must be equal to $3^{-d}\,P_d(\tilde{\vartheta}_2^4,\tilde{\vartheta}_4^4,0)$. This coincides the definition of the $d^{\mathrm{th}}$ order Serre derivative\footnote{More precisely, $\ds_k$ is $-8\pi^2$ times the Serre derivative in the notations of \cite{ZagierMod}.}:
\beq
\label{Serrederk}T_{2d;i} =  \frac{\ds_{2d + 1/2}\circ\cdots\circ\ds_{5/2}\circ\ds_{1/2}\tilde{\vartheta}_i}{\tilde{\vartheta}_i} =  3^{-d}\,P_d(\tilde{\vartheta}_2^4,\tilde{\vartheta}_4^4,0).
\eeq
In the following, we focus on the computation of the $T_{2d;i}$.

\subsubsection{Application to elliptic curves}

In this paragraph we consider a curve $\mathcal{C}$ defined by an equation $\mathrm{Pol}(m^2,l) = 0$ with integer coefficients, whose smooth model $\mathcal{C}_0$ is a Riemann surface of genus $1$. Alternatively, there exists $x,y \in \mathbb{Q}(m^2,l)$ such that the defining equation of $\mathcal{C}_0$ can be brought in Weierstra{\ss} form:
\beq
\label{Weif} y^2 = 4x^3 - g_2x - g_3\qquad g_2,g_3 \in \mathbb{Q},
\eeq
and $g_2,g_3$ are called \emph{elliptic invariants}. Up to a multiplicative constant, the unique holomorphic $1$-form on $\mathcal{C}_0$ is $\dd z = \frac{\dd x}{y}$. We assume we have chosen $\mathcal{A}$ and $\mathcal{B}$ cycles on the curve, it is not necessary to be precise about this choice as we will see in a moment. If we denote $2\varpi_A = \oint_{\mathcal{A}}\dd z$ and $2\varpi_B = \oint_{\mathcal{B}} \dd z$, the holomorphic $1$-form normalized on the $\mathcal{A}$-cycle is $\dd\mathrm{a} = \frac{\dd z}{2\varpi_A}$ and the period is $\tau = \frac{\varpi_B}{\varpi_A}$. The curve $\mathcal{C}_0$ is isomorphic to $\mathbb{C}/(\mathbb{Z} \oplus\tau\mathbb{Z})$, and we can uniformize Eqn.~\ref{Weif} by
\beq
x = \wp\big(\frac{z}{2\varpi_A}\big|\tau\big)\qquad y = \wp'\big(\frac{z}{2\varpi_A}\big|\tau\big),
\eeq
where $\wp$ is the Weierstra{\ss} function:
\beq
\wp(w|\tau) = \frac{1}{w^2} + \sum_{(n,m) \in \mathbb{Z}^2\setminus\{(0,0)\}} \frac{1}{(w + n + m\tau)^2} - \frac{1}{(n + m\tau)^2}.
\eeq
Let us recall the main properties of $\wp(w|\tau)$. It is an even periodic function with periods $1$ and $\tau$, which has a double pole with coefficient $1$ and first subleading order $O(w^2)$. Its full asymptotic expansion when $w \rightarrow 0$ is:
\beq
\wp(w|\tau) = \frac{1}{w^2} + \sum_{j \geq 1} \frac{2^{2j + 2}}{(2j)!}\,\frac{(-1)^jB_{2j + 2}}{2j + 2}\,\tilde{E}_{2j + 2}(\tau)\,w^{2j},
\eeq
where $B_{2j}$ are the Bernoulli numbers. The values of the Eisenstein series for $\mathcal{C}_0$ can be expressed in terms of $g_2$ and $g_3$, by a comparison of the expansion of the left and right hand side of Eqn.~\ref{Weif} when $z \rightarrow 0$:
\beq
\label{Eiseng}\tilde{E}_4(\tau) = (2\varpi_A)^4\,\frac{3g_2}{4}\qquad\qquad\tilde{E}_6(\tau) = (2\varpi_A)^6\,\frac{27g_3}{8}.
\eeq
The equations Eqns.~\ref{relEt1}-\ref{relEt3} allow in principle the determination of:
\beq
(t_2,t_3 = t_2 + t_4,t_4) = (\vartheta_2^4,\vartheta_3^4,\vartheta_4^4).
\eeq
Yet, the precise choice of the solution depends on the choice of the basis of cycles, i.e. of a representative in the $\mathrm{SL}_2(\mathbb{Z})$-orbit of $\tau$. The modular group acts on the equation Eqns.~\ref{relEt1}-\ref{relEt3} and their solution set according to Eqn.~\ref{modt}. Reminding that $\tilde{E}_{4}$ and $\tilde{E}_6$ are modular forms, their values are preserved by the subgroup $\Gamma_+$ of upper triangular matrices with $1$'s on the diagonal. Actually, $\Gamma_+$ is the subgroup preserving the $\mathcal{A}$-cycle (see \S~\ref{algcurve}), so $\varpi_A$ is also invariant. $\Gamma_+$ acts transitively on the set of solutions of Eqns.~\ref{relEt1}-\ref{relEt3}: if $(t_2,t_3,t_4)$ is a solution, the others are
\beq
\label{lists}(-t_4,-t_3,-t_2),\quad(-t_2,t_4,t_3),\quad(-t_3,-t_4,t_2),\quad(t_4,-t_2,-t_3),\quad(t_3,t_2,-t_4).
\eeq
To summarize, any solution of Eqns.~\ref{relEt1}-\ref{relEt3} will give us the fourth powers of the theta values, maybe in desorder and with the wrong sign. But the sign does not matter to compute $D^d\tilde{\vartheta}_i/\tilde{\vartheta}_i$ from Eqns.~\ref{Diffexp1}-\ref{Diffexp3}, so the choice of another solution just results in a permutation of $i = 2,3,4$, i.e. of the label of the even characteristics.

\subsubsection{Arithmetic aspects}

The modular discriminant $\Delta(\tau) = e^{2\mathrm{i}\pi\tau}\prod_{n = 1}^{\infty}(1 - e^{2\mathrm{i}\pi n\tau})^{24}$ is another important modular form, of weight $12$. In terms of Eisenstein series:
\beq
\Delta(\tau) = \frac{E_4^3(\tau) - E_6^2(\tau)}{1728}.
\eeq
Equivalently, we find its value from Eqns.~\ref{relEt1}-\ref{relEt3} or \ref{Eiseng}:
\beq
\label{Delav}\Delta(\tau) = \frac{(2\varpi_A)^{12}}{(2\pi)^{12}}(g_2^3 - 27g_3^2) = \frac{1}{\pi^{12}}\,\frac{\tilde{\vartheta}_2^8\tilde{\vartheta}_3^8\tilde{\vartheta}_4^8}{256}.
\eeq
If we assume $g_2$ and $g_3$ rational (and this is so when $\mathcal{C}_0$ comes from an A-polynomial), we learn from Eqns.~\ref{relEt1}-\ref{relEt3} that $\tilde{\vartheta}_i^4/(2\varpi_A)^2$ for $i = 2,3,4$ are algebraic number. Even more, the reality of $g_2$ and $g_3$ imply that the complex conjugates $(t_2^*,t_3^*,t_4^*)$ must be in the list of solutions \ref{lists}, and looking case by case we infer that one of the numbers $\tilde{\vartheta}_i^4/(2\varpi_A)^4$ is real (if $\Delta > 0$) or pure imaginary (if $\Delta < 0$), while the two others are complex conjugates up to a sign. When $\Delta < 0$, we also have a privileged choice of even characteristics, namely the one for which $\tilde{\vartheta}_{i_0}^4/2\varpi_A^2$ is purely imaginary. This would remain true if $\tau$ was slightly changed by addition of an imaginary part (the $\tilde{E}_{2j}(\tau)$ would remain real). We deduce that the $d^{\mathrm{th}}$ order Serre derivatives $T_{2d;i_0}/(2\varpi_A)^{2d}$ are real, algebraic numbers. This last statement is also true for all $T_{2d;i}/(2\varpi_A)^{2d}$ when $\Delta > 0$, because the three numbers $\frac{\tilde{\vartheta}_i^4}{(2\varpi_A)^2}$ are real and we can apply directly the formulas \ref{Diffexp1}-\ref{Diffexp3}.

\subsubsection{Examples of A-spectral curves}
\label{esxP}
Among the knots we investigated, we have found five non isomorphic elliptic curves arising as components of $A$-polynomials. They turn out to have a minimal model with coefficients $0,\pm 1$. We label them by their name in Cremona classification, the equation of their minimal model ($l$ and $m^2$ are obtained by a birational transformation from $x$ and $y$), and the knot for which they arise as geometric component.

\subsubsection*{15A8: $y^2 + xy + y = x^3 + y^2$ (figure-eight knot)}

A set of elliptic invariants is $g_2 = -\frac{1}{12}$ and $g_3 = \frac{161}{216}$, and the discriminant is $\Delta = -15$. We find that $T_{2d;i_0}$ are rational numbers.
 
\begin{center}
\begin{tabular}{|l|ccc|}
\hline $\tilde{\vartheta}_i^4/(2\varpi_A)^2$ {\rule{0pt}{3.2ex}}{\rule[-1.8ex]{0pt}{0pt}} & $\frac{7 + i\sqrt{15}}{8}$ & $\frac{7 - i\sqrt{15}}{8}$ & $-\frac{i\sqrt{15}}{4}$ \\ 
\hline $T_{2;i}/(2\varpi_A)^2${\rule{0pt}{3.2ex}}{\rule[-1.8ex]{0pt}{0pt}} & $\frac{- 7 + 3i\sqrt{15}}{24}$ &  $\frac{- 7 - 3i\sqrt{15}}{24}$ & $\frac{7}{12}$ \\
$T_{4;i}/(2\varpi_A)^4$  {\rule{0pt}{3.2ex}}{\rule[-1.8ex]{0pt}{0pt}}& $\frac{47 + 21i\sqrt{15}}{96}$ & $\frac{47 - 21i\sqrt{15}}{96}$ & $-\frac{47}{48}$ \\
$T_{6;i}/(2\varpi_A)^6$ {\rule{0pt}{3.2ex}}{\rule[-1.8ex]{0pt}{0pt}}& $\frac{-665 + 9i\sqrt{15}}{1152}$ & $\frac{-665 - 9i\sqrt{15}}{1152}$ & $-\frac{301}{576}$ \\
$T_{8;i}/(2\varpi_A)^8$ {\rule{0pt}{3.2ex}}{\rule[-1.8ex]{0pt}{0pt}} & $\frac{28375 - 12999i\sqrt{15}}{13824}$ & $\frac{28375 + 12999i\sqrt{15}}{13824}$ & $-\frac{28285}{6912}$ \\
\hline
\end{tabular}
\end{center}
This curve also arises as non-geometric component in $\mathbf{7}_{4}^{(1)}$, $\mathbf{8}_{18}^{(1)}$, $\mathbf{8}_{18}^{(2)}$, $\mathbf{9}_{24}^{(1)}$, $\mathbf{9}_{37}^{(1)}$, $\mathbf{9}_{49}^{(1)}$, $\mathbf{10}_{142}^{(1)}$, $\mathbf{10}_{145}^{(1)}$, $\mathbf{10}_{146}^{(1)}$, $\mathbf{10}_{147}^{(1)}$, $\mathbf{10}_{155}^{(1)}$.

\subsubsection*{14A4: $y^2 + xy + y = x^3 - x$ ($L^2R$)}

A set of elliptic invariants is $g_2 = -\frac{25}{12}$ and $g_3 = \frac{253}{216}$, and the discriminant is $\Delta = -28$. We find again that $T_{2d;i_0}$ are rational numbers.

\begin{center}
\begin{tabular}{|l|ccc|}
\hline $\tilde{\vartheta}_i^4/(2\varpi_A)^2$ {\rule{0pt}{3.2ex}}{\rule[-1.8ex]{0pt}{0pt}}& $\frac{11 - i\sqrt{7}}{8}$ & $\frac{11 + i\sqrt{7}}{8}$ & $\frac{i\sqrt{7}}{4}$ \\ 
\hline $T_{2;i}/(2\varpi_A)^2$ {\rule{0pt}{3.2ex}}{\rule[-1.8ex]{0pt}{0pt}}& $\frac{-11 - 3i\sqrt{7}}{24}$ &  $\frac{- 11 + 3i\sqrt{7}}{24}$ & $\frac{11}{12}$ \\
$T_{4;i}/(2\varpi_A)^4$ & $\frac{71 - 33i\sqrt{7}}{96}$ {\rule{0pt}{3.2ex}}{\rule[-1.8ex]{0pt}{0pt}}& $\frac{71 + 33i\sqrt{7}}{96}$ & $-\frac{71}{48}$ \\
$T_{6;i}/(2\varpi_A)^6$ & $\frac{-1837 - 225i\sqrt{7}}{1152}$ {\rule{0pt}{3.2ex}}{\rule[-1.8ex]{0pt}{0pt}}& $\frac{-1837 + 225i\sqrt{7}}{96}$ & $\frac{319}{576}$ \\
$T_{8;i}/(2\varpi_A)^8$ & $\frac{72583 + 37509i\sqrt{7}}{13824}$ {\rule{0pt}{3.2ex}}{\rule[-1.8ex]{0pt}{0pt}} & $\frac{72583 - 37509i\sqrt{7}}{13824}$ & $-\frac{16333}{6912}$ \\
\hline
\end{tabular}
\end{center}

\subsubsection*{19A3: $y^2 + y = x^3 + x^2 + x$ ($\mathbf{9}_{35}$)}

A set of elliptic invariant is $g_2 = \frac{8}{3}$ and $g_3 = -\frac{1}{27}$, and the discriminant is $\Delta = -19$. We find that $T_{2d;i_0} \in \mathbb{Q}(\alpha)$ with:
\beq
\tilde{\vartheta}_{i_0}^4/(2\varpi_A)^2 = \frac{2i}{\sqrt{6}}(\alpha^{1/2} + \alpha^{-1/2}),\qquad\qquad\alpha = \Big(\frac{257 + 3\sqrt{57}}{4}\Big)^{1/3}.
\eeq

\subsubsection*{11A3: $y^2 + y = x^3 - x^2$ ($\mathbf{9}_{48}$)}

A set of elliptic invariants is $g_2 = -\frac{4}{3}$ and $g_3 = \frac{19}{27}$, and the discriminant is $\Delta = -11$. We find that $T_{2d;i_0} \in \mathbb{Q}(\alpha)$ with: 
\beq
\tilde{\vartheta}_{i_0}^4/(2\varpi_A)^2 = \frac{i}{\sqrt{3}}(\alpha^{1/2} - \alpha^{-1/2}),\qquad\qquad\alpha = \Big(\frac{329 + 57\sqrt{33}}{2}\Big)^{1/3}.
\eeq

\subsubsection*{43A1: $y^2 + y = x^3 + x^2$ ($\mathbf{10}_{139}$)} A set of elliptic invariants is $g_2 = -\frac{4}{3}$ and $g_3 = \frac{35}{27}$, and the discriminant is $\Delta = -43$. We find that $T_{2d;i_0} \in \mathbb{Q}(\alpha)$, with:
\beq
\tilde{\vartheta}_{i_0}^4/(2\varpi_A)^2 = \frac{i}{\sqrt{3}}(\alpha^{1/2} - \alpha^{-1/2}),\qquad\qquad\alpha = \Big(\frac{1193 + 105\sqrt{129}}{2}\Big)^{1/3}.
\eeq

In the two first examples (figure-eight knot and $L^2R$, the fact that $T_{2d;i_0}$ are rational numbers imply that the coefficients $\widetilde{\jmath}_{\chi}(p)$ (for the choice of the characteristics associated to $i_0$) for $\chi \geq 1$ sit in the same function field as the amplitudes of the topological recursion. On the contrary, for the three last examples, they will sit a priori in an extension by the element $\alpha$ of the function field where the amplitudes of the topological recursion live.

\begin{remark} The two first elliptic curves do not have complex multiplication, whereas the three last do. We thank Farshid Hajir for pointing us this property.
\end{remark}

\subsection{Bergman kernel for elliptic curves}
\label{Bergman}
In genus $1$, there is only one odd characteristics, and the corresponding theta function is:
\beq
\vartheta_1(w|\tau) = i\sum_{n \in \mathbb{Z}} e^{\mathrm{i}\pi\tau(n + 1/2)^2 + 2\mathrm{i}\pi w n}.
\eeq
It satisfies:
\beq
\vartheta_1(w + 1|\tau) = -\vartheta_1(w|\tau)\qquad\qquad\vartheta_1(w + \tau) = -e^{-\mathrm{i}\pi(2w + \tau)}\,\vartheta_1(w|\tau).
\eeq
The Bergman kernel normalized on the $\mathcal{A}$-cycle is:
\beq
B_0(z_1,z_2) = -(\ln\vartheta_1)''\big(\frac{z}{2\varpi_A}\big|\tau\big)\,\frac{\dd z_1}{2\varpi_A}\otimes\frac{\dd z_2}{2\varpi_A}.
\eeq
On the other hand, the Weierstra{\ss} function provides another natural Bergman kernel:
\beq
\label{Bkapp0}B_{\underline{\kappa}(\tau)}(z_1,z_2) = \wp\big(\frac{z_1 - z_2}{2\varpi_A}\big|\tau\big)\,\frac{\dd z_1}{2\varpi_A}\otimes\frac{\dd z_2}{2\varpi_A}.
\eeq
By "natural", we mean that it can be written in terms of the coordinates $x$ and $y$ (see Eqn.~\ref{Weif}) thanks to the addition relation for $\wp$. The result is:
\beq
\label{Bergwei}B_{\underline{\kappa}(\tau)}(z_1,z_2) = \Big[\frac{1}{4}\Big(\frac{y(z_1) + y(z_2)}{x(z_1) - x(z_2)}\Big)^2 - x(z_1) - x(z_2)\Big]\,\frac{\dd z_1}{2\varpi_A}\otimes\frac{\dd z_2}{2\varpi_A}.
\eeq
There is a well-known relation between $\wp$ and $\vartheta_1$:
\beq
\wp(w|\tau) =  - (\ln\vartheta_1)''(w|\tau) - \frac{\tilde{E}_2(\tau)}{3}.
\eeq
In other words, Eqn.~\ref{Bergwei} is the expression for a Bergman kernel normalized on the $\mathcal{A}_{\underline{\kappa}(\tau)}$-cycle, with the value:
\beq
\underline{\kappa}(\tau) = -\frac{1}{2\mathrm{i}\pi}\,\frac{\tilde{E}_2(\tau)}{3}.
\eeq

\subsection{Application to some degree $2$, elliptic A-spectral curves}
\label{sqex}
We assume in this paragraph that the spectral curve $(\mathcal{C},\ln l,\ln m)$ has a defining equation:
\beq
\label{form4}l = \frac{P_1(m^2) + P_2(m^2)\sqrt{S(m^2)}}{R(m^2)}.
\eeq
where $P_1,P_2,R$ are polynomials and $S$ is a polynomial of degree $4$ with simple roots and leading coefficient $1$. Also, it admits a smooth model $\mathcal{C}_0$ of equation:
\beq
\label{sju}\tilde{l} = \sqrt{S(m^2)}.
\eeq
We also assume that $\{m,l\}$ is $2$-torsion (i.e. $\varsigma = 1$) and that $\iota_* = -\mathrm{id}$. The spectral curve for the figure-eight knot and $L^2R$ takes this form. Then, many simplifications occur.

\subsubsection{Writing the $n$-forms}

First, the ramification points $a_i$ of the spectral curve coincide with the Weierstra{\ss} points of $\mathcal{C}_0$, i.e. with the roots of $S$. Besides, the local involution correspond to changing the sign of the squareroot $\sqrt{S(X)} \mapsto -\sqrt{S(X)}$, and is in fact defined globally on $\mathcal{C}_0$. We define a variable $z$ by integration of the holomorphic $1$-form:
\beq
\dd z = \frac{\dd X}{\sqrt{S(X)}}.
\eeq
Afterwards, it is straightforward to check the formula:
\beq
\wp\big(\frac{z_1 - z_2}{2\varpi_A}\big) = \frac{1}{4}\,\frac{(\sqrt{S(X_1)} - \sqrt{S(X_2)})^2}{(X_1 - X_2)^2} - \frac{S'(X_1) - S'(X_2)}{12(X_1 - X_2)} + \frac{(X_1 - X_2)^2}{12}.
\eeq
In particular, we find:
\beq
\label{QC}\wp\big(\frac{z - \overline{z}}{2\varpi_A}\big) = \frac{\big(S'(X)\big)^2}{16S(X)} - \frac{S''(X)}{12}.
\eeq
which intervenes in the computation of $\omega_1^{1}$. We also compute:
\begin{align}
\label{QB}\wp\big(\frac{z - a_i}{2\varpi_A}\big) & = \frac{S'(a_i)}{4(X - a_i)} + \frac{S''(a_i)}{24} \\
\label{QD}\int^{X} \wp\big(\frac{z' - a_i}{2\varpi_A}\big)\dd z' & = -\frac{\sqrt{S(X)}}{2(X - a_i)} + \frac{1}{24}\int^X \frac{\dd X\,S''(X)}{\sqrt{S(X)}}.
\end{align}
To avoid cumbersome notations, we use the same letter $a_i$ to denote the image of the ramification point $a_i$ in the Jacobian of $\mathcal{C}_0$ (on the left hand side) and the value of the $X$-coordinate at the same ramification point.
The second term in Eqn.~\ref{QD} is odd when we change the sign of the squareroot, so disappear when we integrate from $z$ to $\overline{z}$:
\beq
\label{QA}\int^{z}_{\overline{z}} \wp\big(\frac{z' - a_i}{2\varpi_A}\big)\dd z' = \frac{\sqrt{S(X)}(1/a_i - a_i)}{2(X - a_i)(X - 1/a_i)}.
\eeq
Besides, one can check by differentiating the two sides of the equality:
\beq
\label{Cieq}\int_{o,\iota(o)}^{z,\iota(z)} \wp\big(\frac{z - a_i}{2\varpi_A}\big)\,\dd z = \frac{\sqrt{S(X)}(1/a_i - a_i)}{2(X - a_i)(X - 1/a_i)} + C_i,
\eeq
where $C_i$ only depends on the basepoint $o$. This can be checked by differentiation and the fact that both sides of the equality are invariant under $\iota$. If one denotes $\{a_i,\iota(a_i),a_{\overline{i}},\iota(a_{\overline{i}})\}$ the set of ramification points, one finds that $C_i = 0$ when the basepoint $o$ is chosen as $a_{\overline{i}}$ or $\iota(a_{\overline{i}})$.
These formulas allow to complete the computation of the $\omega_{n|\underline{\kappa}(\tau)}^h$, i.e. the topological recursion with the Bergman kernel $B_{\underline{\kappa}(\tau)}$ defined in Eqn.~\ref{Bergwei}. We just need to compute expansion of the quantities in Eqns.~\ref{QA} at the branchpoints $X \rightarrow a_j$, and then take residues. The coordinate $z$ is convenient for these computations, because Taylor expanding then amounts to differentiating the Weierstra{\ss} function with respect its argument. This method yields $\omega_{n|\underline{\kappa}(\tau)}^{h}$ as a linear combination with rational coefficients of elementary $n$-forms of the type:
\beq
\label{temr}\bigotimes_{j = 1}^n \wp^{(p_j)}\big(\frac{z_j - a_{i_j}}{2\varpi_A}\big)\,\frac{\dd z_j}{(2\varpi_A)^{p_j + 1}},
\eeq
where $p_j$ are even integers. It is easy to integrate $\omega_{n|\underline{\kappa}(\tau)}^h$ over cycles with such a representation (we have to use $\ref{QA}$ for terms with some $p_j = 0$). In particular, integrating $z_j$ over $\frac{1}{2\mathrm{i}\pi}\oint_{\mathcal{B}_{\kt}}$ in Eqn.~\ref{temr} gives $0$ if $p_j > 0$, and replaces the $j^{\mathrm{th}}$-factor by $(2\varpi_A)^{-1}$ if $p_j = 0$.

$\iota_*$ has a single eigenvalued, which we assumed to be $-1$. The discussion of \S~\ref{Evenorder} applies and can be explicitly checked: we find that the amplitudes:
\beq
\label{level} G_{n|\underline{\kappa}(\tau)}^{h,(d)} = \frac{1}{n!}\,\frac{1}{(2\mathrm{i}\pi)^d\,d!}\,\underbrace{\int_{o,\iota(o)}^{p,\iota(p)}\cdots\int_{o,\iota(o)}^{p,\iota(p)}}_{n\,\,\mathrm{times}}\underbrace{\oint_{\mathcal{B}_{\kt}}\cdots\oint_{\mathcal{B}_{\kt}}}_{d\,\,\mathrm{times}} \omega_{n + d|\underline{\kappa}(\tau)}^{h}
\eeq
vanish if $d$ is odd. Then, the computation of $T_{2d;i}$ for $d$ integer as detailed in \S~\ref{ellipticgen} is all we need to evaluate $\mathcal{J}_{\hbar}^{\mathrm{n.p.TR}}(p)$.

\subsubsection{Arithmetic aspects}

Since $S$ is palindromic, we can write the results in a more compact form with a variable $w = \frac{m^2 + m^{-2}}{2}$. We also denote:
\beq
\frac{S(m^2)}{m^4} = \sigma(w).
\eeq
Notice that $w$ is a uniformization variable for the quotient $\mathcal{C}_0/\iota$ which has genus $0$ by assumption. The amplitudes $G_{n|\underline{\kappa}(\tau)}^{h,(d)}$ are rational functions in $w$ and $\sigma(w)$, with coefficients in $\mathbb{Q}[C_i]$. For positive levels, let $\mathcal{P}_{n}^h$ the set of $n$-uples $(p_1,\ldots,p_n)$ such that the monomial $\bigotimes_{j = 1}^n \wp\big(\frac{z_j - a_{i_j}}{2\varpi_A}\big)\frac{\dd z_j}{2\varpi_A}$ appears in $\omega_n^h(z_1,\ldots,z_n)$. At level $1$, we have:
\beq
\mathcal{P}_3^0 = \{(2,2,2)\}\qquad\qquad\mathcal{P}_1^1 = \{(4)\}.
\eeq
It is easy to construct recursively a set $\mathcal{Q}_n^h \supseteq  \mathcal{P}_n^h$ from the residue formula Eqn.~\ref{defwng}. At level one, we define $\mathcal{Q}_1^1 = \mathcal{P}_1^1$ and $\mathcal{Q}_3^0 = \mathcal{P}_3^0$, and if we know  all $\mathcal{Q}_n^h$ at level $\chi$, we use the following rules to define $\mathcal{Q}_n^h$ at level $\chi + 1$:
\begin{itemize}
\item if $(p,p',p_2,\ldots,p_n) \in \mathcal{Q}_{n + 1}^{h - 1}$, then $(p + p' + 2,p_2,\ldots,p_n) \in \mathcal{Q}_{n}^h$.
\item for all $(n',h') \neq (1,0),(n + 1,h)$ such that $0 \leq h' \leq h$ and $0 \leq n' \leq n$, if $(p,p_2,\ldots,p_{n' + 1}) \in \mathcal{Q}_{n' + 1}^{h'}$ and $(p',p_{n' + 2},\ldots,p_{n}) \in \mathcal{Q}_{n - n'}^{h - h'}$, then $(p + p' + 2,p_2,\ldots,p_n) \in \mathcal{Q}_{n}^h$.
\item if $(p_1,\ldots,p_n) \in \mathcal{Q}_{n}^h$ and $\sigma$ is a permutation of $n$, then $(p_{\sigma(1)},\ldots,p_{\sigma(n)}) \in \mathcal{Q}_n^h$.
\end{itemize}
We have an inclusion $\mathcal{P}_n^h \subseteq \mathcal{Q}_n^{h}$, and not an equality, because a monomial with indices $(p_1,\ldots,p_n)$ may come from several terms in the residue formula, and there might exist coincidences leading to cancellations. As a consequence:
\beq
G_{n|\underline{\kappa}(\tau)}^{h,(d)} \in \big(\sigma(w)\big)^{-3r_{n}^{h,(d)}/2}\cdot\mathbb{Q}[w,C_i],
\eeq
where:
\beq
r_{n}^{h,(d)} = \mathrm{max}\Big\{\sum_{i = 1}^{n - d} (p_{j_i} + 1)\qquad (p_1,\ldots,p_n) \in \mathcal{Q}_n^h,\,\,j_1,\ldots,j_{n - d} \in \{1,\ldots,n\}\Big\}.
\eeq
We already know $r_3^{0,(0)} = 3$, $r_3^{0,(2)} = 1$, $r_1^{1,(0)} = 3$, and by recursion one can show:
\beq
r_{n}^{h,(d)} = 3(2h - 2 + n) + d.
\eeq
Thus, our construction naturally entails:
\beq
\label{for1}\forall \chi \geq 1,\qquad \widetilde{\jmath}_{\chi}(w) \in \big(\sigma(w))^{-3\chi/2}\cdot\mathbb{Q}[w].
\eeq
At the points such that $m^2 = 1$ (in particular the reference point where the hyperbolic metric is complete), we have $w = 1$. From the definition, we see that the cusp field is $\mathbb{F} = \mathbb{Q}[\sqrt{\sigma(1)}]$, and our construction naturally entails:
\beq
\label{for2}\forall \chi \geq 1,\qquad \widetilde{\jmath}_{\chi}(w = 1) \in \big(\sigma(1)\big)^{-3\chi/2}\cdot\mathbb{Q}[C_i] \,\subseteq \mathbb{F}[C_i].
\eeq
In the examples below, we need $C_i \equiv 0$, so one can forget about $C_i$ in Eqn.~\ref{for1} and Eqn.~\ref{for2}. In this case, our Conjecture~\ref{qiq} predicts that the coefficients of the asymptotic expansion of the Kashaev invariant belong to the cusp field $\mathbb{F}$.

\subsection{$\mathbf{4}_1$ (figure-eight knot)}
\label{figure8}
Apart from the abelian factor $(l - 1)$, the A-polynomial has a unique factor (necessarily the geometric one):\beq
\mathfrak{A}(m,l) = l^2m^4 + l(-m^8 + m^6 + 2m^4 + m^2 - 1) + m^4
\eeq
It defines a curve $\mathcal{C}_0$ of genus $1$. The symbol $\{m,l\}$ is $2$-torsion, and $\iota_* = -\mathrm{id}$. The spectral curve can be put in the form of Eqn.~\ref{form4}:
\beq
\begin{split}
l & = \frac{m^8 - m^6- 2m^4 - m^2 + 1 + (m^4 - 1)\sqrt{S(m^2)}}{2m^4} \\
S(X) & = X^4 -  2X^3 - X^2 - 2X + 1,
\end{split}
\eeq
so the results of \S~\ref{sqex} can be applied, and we introduce:
\beq
w = \frac{m^2 + m^{-2}}{2} = \mathrm{ch}(2u),\qquad\frac{S(m^2)}{m^4} = \sigma(w) = 4w^2 - 4w - 3.
\eeq
The curve has $4$ ramification points, of coordinates
\beq
(m^2,l) = \big(\frac{3 \pm \sqrt{5}}{2},1\big)\qquad\mathrm{and}\qquad\big(\frac{-1 \pm i\sqrt{3}}{2},-1\big).
\eeq
The local involution $z \mapsto \overline{z}$ is defined globally on $\mathcal{C}_0$, and it corresponds to $(m,l) \rightarrow (m,1/l)$. Incidentally, it coincides with the amphichiral symmetry. The cusp field is $\mathbb{Q}[\sqrt{-3}]$. It is known \cite{Vesnin} that the hyperbolic volume of $\mathfrak{M}_u = \mathbb{S}_3\setminus\mathbf{4}_1$ with cusp angle $2\,\mathrm{Im}\,u$ and $\mathrm{Re}\,u = 0$, is:
\beq
\mathsf{Vol}(\mathfrak{M}_u) = 2\big[\Lambda(-{\rm i}u + \beta(u)/2) - \Lambda(-{\rm i}u - \beta(u)/2)\big],
\eeq
where $\beta(u) = \mathrm{arccos}\big(\mathrm{ch}(2u) - 1/2\big)$ and $\Lambda$ is the Lobachevsky function:
\beq
\Lambda(x) = - \int_{0}^x |\ln(2\sin x')|\dd x'.
\eeq
In particular, it vanishes when $u = \pm 2\mathrm{i}\pi/3$, and this value coincide with the $u$-projection of two of the four branchpoints. Hence, it we denote by $a_0$ any of these points, we find that $\mathrm{Im}\,\int_{a_0}^{p} \ln l\,\dd \ln m = 0$ is half of the hyperbolic volume with the correct additive constant.

\begin{figure}[h]
\begin{center}
\includegraphics[width=0.5\textwidth]{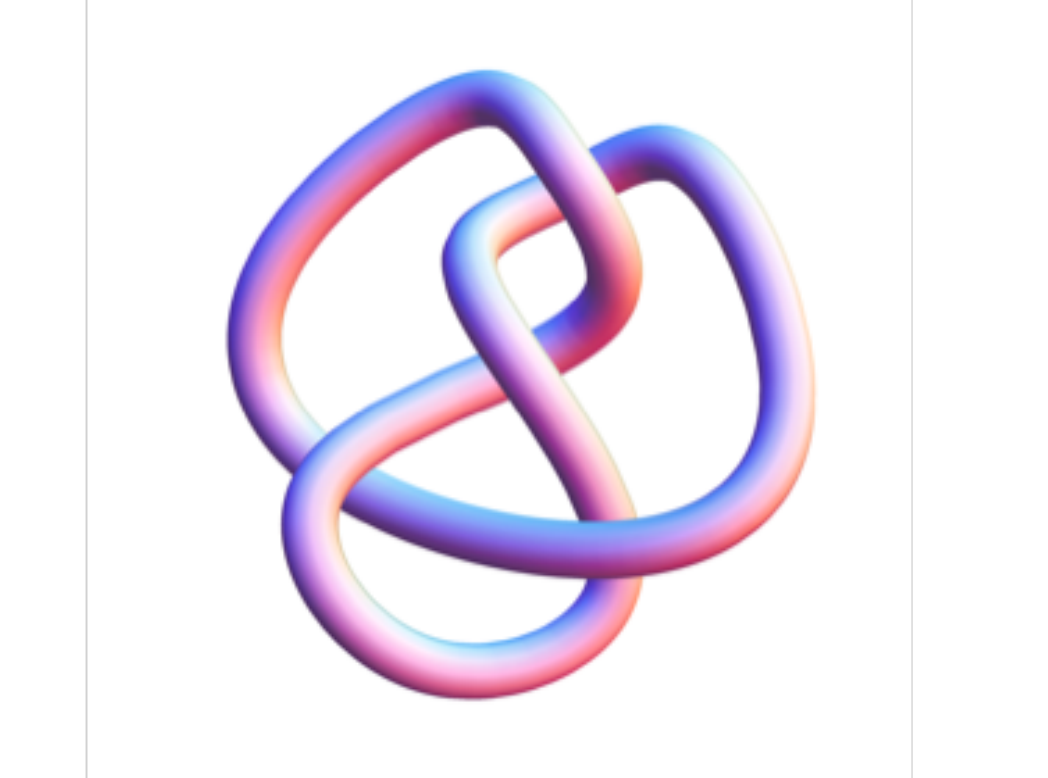}
\caption{\label{F111} The figure-eight knot}
\end{center}
\end{figure}

\subsubsection*{Amplitudes}

We now derive the three first terms of $\mathcal{J}_{\hbar}^{\mathrm{n.p.TR}}$. We choose to compute the primitive with $C_i \equiv 0$. We computed the $\omega_n^h$ up to level $3$ (i.e. $2h - 2 + n \leq 3$). We just present here the expression for the non-vanishing amplitudes $G_{n|\underline{\kappa}(\tau)}^{h,(d)}$ up to level $3$. The values of $G_{n|\underline{\kappa}(\tau)}^{h,(d = 0)}$ for level $1$ and $2$ coincide\footnote{More precisely, since \cite{DiFuji2} uses the spectral curve $2v\,\dd u$ (instead of $v\,\dd u$ here), we retrieve their amplitudes $\mathcal{F}^{(h,n)}$ by multiplying our results by $(1/2)^{\chi}$.} with the amplitudes computed in \cite[\S~3.3]{DiFuji2}. 

\beq\begin{split}
G_{3|\underline{\kappa}(\tau)}^{0,(0)} & = \frac{1}{6 \sigma^{3/2}(w)}\big(-12w^2 + 12w - 7\big), \\
G_{1|\underline{\kappa}(\tau)}^{0,(2)} & = \frac{2 (2\varpi_A)^{-2}}{15\sigma^{1/2}(w)}\big(- 4w + 3\big),  \\
G_{1|\underline{\kappa}(\tau)}^{1,(0)} & = \frac{1}{90\sigma^{3/2}(w)}\big(- 8w^3 + 44w^2 - 30w - 87 \big).
\end{split}\eeq
\hrule
\beq\begin{split}
G_{4|\underline{\kappa}(\tau)}^{0,(0)} & = \frac{1}{3\sigma^3(w)}\big(16w^5 - 32w^4  + 24w^3 + 44w^2  - 67w + 25\big), \\
G_{2|\underline{\kappa}(\tau)}^{0,(2)} & =  \frac{2(2\varpi_A)^{-2}}{135\sigma^2(w)}\big(64w^4  - 232w^3 + 156w^2 +  378w - 243\big), \\
G_{0|\underline{\kappa}(\tau)}^{0,(4)} & =  -\frac{256(2\varpi_A)^{-4}}{10125}, \\
G_{2|\underline{\kappa}(\tau)}^{1,(0)} & =  \frac{1}{1620 \sigma^3(w)}\left(\begin{array}{l}
 1280w^6  - 9088w^5 + 13136w^4 + 22176w^3 \\ - 17928w^2 - 26352w + 23193\end{array}\right), \\
G_{0|\underline{\kappa}(\tau)}^{1,(2)} & =  \frac{-368(2\varpi_A)^{-2}}{10125}.
\end{split}\eeq
\beq\begin{split}
G_{5|\underline{\kappa}(\tau)}^{0,(0)} & = \frac{1}{60\sigma^{9/2}(w)}\left(\begin{array}{l}   \\
    - 640w^8 + 1600w^7 - 1440w^6 - 9520w^5 + 18184w^4 \\ - 3988w^3 - 18542w^2 + 19071w -5644\end{array}\right),  \\
G_{3|\underline{\kappa}(\tau)}^{0,(2)} & =  \frac{2(2\varpi_A)^{-2}}{30375\sigma^{7/2}(w)}\left(\begin{array}{l} - 37888w^7 + 283424w^6 - 471088w^5 - 636000w^4 \\
 + 1368360w^3+ 174906w^2 - 1767663w+ 883791\end{array}\right),  \\
G_{1|\underline{\kappa}(\tau)}^{0,(4)} & =  \frac{4(2\varpi_A)^{-4}}{455625\sigma^{5/2}(w)}\left(\begin{array}{l} 94208w^5 - 287512w^4 + 76748w^3 \\ 370230w^2 - 112527w - 219564 \end{array}\right), \\
G_{3|\underline{\kappa}(\tau)}^{1,(0)} & =  \frac{1}{109350\sigma^{9/2}(w)}\left(\begin{array}{l}
   - 249856w^9 + 2901504w^8  - 6701952w^7 - 8240960w^6 \\
   + 8573472w^5 + 30776112w^4 - 55663848w^3 - 12104316w^2 \\
   + 71667990w - 34229709\end{array}\right),  \\
G_{1|\underline{\kappa}(\tau)}^{1,(2)} & =  \frac{8(2\varpi_A)^{-2}}{455625\sigma^{7/2}(w)}\left(\begin{array}{l}  353792w^7 - 1479360w^6 + 1280256w^5 + 1398544w^4 \\ 
- 1258392w^3 - 1990008w^2 - 10098w + 1832382\end{array}\right),  \\
G_{1|\underline{\kappa}(\tau)}^{2,(0)} & =   \frac{1}{2733750\sigma^{9/2}(w)}\left(\begin{array}{l} 13238272w^9 - 70087552w^8 + 94437312w^7 \\
+ 49067168w^6 - 177750608w^5 + 952056w^4 \\
- 78657516w^3 + 179205966w^2 + 191047329w \\
- 219603474\end{array}\right).
\end{split}\eeq

\subsubsection*{First orders of $\mathcal{J}_{\hbar}^{\mathrm{n.p.TR}}$/comparison to colored Jones}

We choose the even-half characteristics $[\mu,\nu]$ leading to real-valued theta derivatives (the last column of the table for the curve labeled $15A8$ in \S~\ref{esxP} is selected). The reader may recognize in those values the \textit{ad hoc} renormalizations of the constants $G^k$ found by the authors of \cite{DiFuji2}. We use the general expressions given in \S~\ref{firstf} to compute:
\beq\begin{split}
\widetilde{\jmath}_1 & = \frac{-1}{12\sigma^{3/2}(w)}\big(8w^3 - 4w^2 - 10w + 17\big)   \\
& = \frac{-1}{12S^{3/2}(m^2)}\big(m^{12} - m^{10} - 2m^8 + 15m^6 - 2m^4 - m^2 + 1\big),  \\
\widetilde{\jmath}_2 & = \frac{2}{\sigma^3(w)}\big(8w^3 - 4w^2 - 10w + 7\big)  \\
& = \frac{2m^6}{S^3(m^2)}\big(m^{12} - m^{10} - 2m^8 + 5m^6 - 2m^4 - m^2 + 1\big),  \\
\widetilde{\jmath}_3 & = \frac{1}{90\sigma^{9/2}(w)}\left(\begin{array}{l}  256w^8  - 512w^7 - 8704w^6 + 2048w^5 + 29792w^4 \\ - 46928w^3  + 1272w^2 + 49164w -27469\end{array}\right)  \\
& = \frac{m^2}{90S^{9/2}(m^2)}\left(\begin{array}{l} m^{32} - 4m^{30} - 128m^{28} + 36m^{26} + 1074m^{24} - 5630m^{22} \\
+ 5782m^{20} + 7484m^{18}  - 18311m^{16} + 7484m^{14} + 5782m^{12} \\
- 5630m^{10} + 1074m^8  + 36m^{6} - 128m^4 - 4m^2 + 1\end{array}\right).
\end{split}
\eeq

These coefficients are exactly those found in \cite[Section 4.2.1]{GZLD} for a power series solution of the $A$-hat recursion relation of the figure-eight knot, so we have checked our Conjecture~\ref{ahconj} up to $o(\hbar^3)$, with $C_{\hbar} = 1 + o(\hbar^3)$. These authors as well as \cite{DiFuji2} also find the same coefficients in the asymptotic expansion of a Hikami-type integral $\mathcal{J}_{\hbar}^{\mathrm{H}}(u)$ associated to the figure-eight knot. This is believed to be the correct asymptotic expansion for the colored Jones polynomial in the GVC near ${\rm i}\pi$.

\subsubsection*{Specialization to $u = {\rm i}\pi$/comparison to Kashaev invariant}

We recall that the complete hyperbolic point correspond to $w = 1$. The coefficients $\jmath_{\chi}(w = 1)$ belong to $\mathbb{Q}[\sqrt{-3}]$. We find for the first orders:
\beq\begin{split}
\widetilde{\jmath}_1(w = 1) & = -\frac{11}{12}\,(-3)^{-3/2} \\
\widetilde{\jmath}_2(w = 1) & =  2\,(-3)^{-3} \\
\widetilde{\jmath}_3(w = \mathrm{i}\pi) & = -\frac{1081}{90}\,(-3)^{-9/2}.
\end{split}\eeq
This is in agreement with the asymptotic expansion of the Kashaev invariant, proved with help of numerics \cite{GZLD} to be:
\beq
J_N(\mathfrak{K},q = e^{2{\rm i}\pi/N}) \mathop{=}_{N \rightarrow \infty} 3^{-1/4}\,N^{3/2}\,\exp\Big(\frac{N\mathrm{Vol}(\mathbf{4}_1)}{2\pi} + \frac{11\epsilon_N}{2} + 2\epsilon_N^2 - \frac{1081\epsilon_N^3}{90} + O(\epsilon_N^4)\Big)
\eeq
where $\epsilon_N = \frac{{\rm i}\pi}{(-3)^{3/2}N}$.

\subsection{Once-punctured torus bundle $L^2R$}
\label{L2R}
If $\mathfrak{t} = (\mathbb{R}^2/\mathbb{Z}^2)\setminus\{0\}$ denotes the once-punctured torus, the $3$-manifold $L^2R$ is defined as $\mathfrak{t}\times[0,1]/\sim$ where the equivalence relation is generated by:
\beq
(x,0) \sim (L\circ L \circ R(x),1)\qquad L(x) = x + 1,\quad R(x) = \frac{x}{x + 1}.
\eeq
This manifold is hyperbolic \cite{Otal}, \cite{Gueri}. Apart from the abelian factor $(l - 1)$, the A-polynomial has a unique factor (necessarily the geometric one):
\beq
\mathfrak{A}(m,l) = l^2m^4 + l(-m^6 + 2m^4 + 2m^2 - 1) + m^2.
\eeq
It defines a curve of genus $1$. The symbol $\{m,l\}$ is $2$-torsion, and $\iota_* = -\mathrm{id}$. The spectral curve can be put in the form Eqn.~\ref{form4}:
\beq
\begin{split}
l & = \frac{m^6 - 2m^4 - 2m^2 + 1 + (m^2 - 1)\sqrt{S(m^2)}}{2m^4} \\
 S(X) & = X^4 - 2X^3 - 5X^2 - 2X + 1.
\end{split}\eeq
so the results of \S~\ref{sqex} can be applied, and we introduce:
\beq
w = \frac{m^2 + m^{-2}}{2} = \mathrm{ch}(2u),\qquad \frac{S(m^2)}{m^4} = \sigma(w) = 4w^2 - 4w - 7
\eeq
The curve has $4$ ramification points, of coordinates:
\beq\begin{split}
(m^2,l) = & \Big(\frac{1 + 2\sqrt{2} \pm \sqrt{5 + 4\sqrt{2}}}{2},\frac{1 + \sqrt{2} \pm (\sqrt{2} - 1)\sqrt{5 + 4\sqrt{2}}}{2}\Big) \\ \mathrm{and}\,\,\, & \Big(\frac{1  - 2\sqrt{2} \pm i\sqrt{4\sqrt{2} - 5}}{2},\frac{1 - \sqrt{2} \pm i(\sqrt{2} + 1)\sqrt{4\sqrt{2} - 5}}{2}\Big).
\end{split}
\eeq
The local involution $z \mapsto \overline{z}$ is defined globally of $\mathcal{C}_0$. The cusp field is $\mathbb{Q}[\sqrt{-7}]$.

\subsubsection*{Amplitudes}

We again choose to compute all primitives with $C_i \equiv 0$.

\beq\begin{split}
G_{3|\underline{\kappa}(\tau)}^{0,(0)} & = -\frac{1}{24 \sigma^{3/2}(w)}\big(8w^3 + 36w^2 + 6w + 19\big) \\
G_{1|\underline{\kappa}(\tau)}^{0,(2)} & = \frac{(2\varpi_A)^{-2}}{14\sigma^{1/2}(w)}\big(-6w + 7\big),  \\
G_{1|\underline{\kappa}(\tau)}^{1,(0)} & = \frac{1}{168\sigma^{3/2}(w)}\big(40w^3 - 12w^2 - 210w - 217\big).
\end{split}\eeq

\hrule

\beq\begin{split}
G_{4|\underline{\kappa}(\tau)}^{0,(0)} & = \frac{1}{192\sigma^3(w)}\left(\begin{array}{l} 64w^6 + 832w^5 - 144w^4 + 3168w^3 \\ + 1532w^2 - 2060w + 1257\end{array}\right), \\
G_{2|\underline{\kappa}(\tau)}^{0,(2)} & = \frac{(2\varpi_A)^{-2}}{294\sigma^2(w)}\big(144w^4 - 816w^3 + 952w^2 + 1988w - 931\big), \\
G_{0|\underline{\kappa}(\tau)}^{0,(4)} & = -\frac{27(2\varpi_A)^{-4}}{1372},  \\
G_{2|\underline{\kappa}(\tau)}^{1,(0)} & = \frac{1}{28224\sigma^3(w)}\left(\begin{array}{l} 7872w^6 - 116544w^5 + 341968w^4 + 841120w^3 \\ - 443884w^2 - 350644w + 556003\end{array}\right), \\
G_{0|\underline{\kappa}(\tau)}^{1,(2)} & = -\frac{57(2\varpi_A)^{-2}}{2744}.
\end{split}\eeq

\subsubsection*{First orders of $\mathcal{J}_{\hbar}^{\mathrm{n.p.TR}}$/comparison to colored Jones}

We choose the even-half characteristics $[\mu,\nu]$ leading to real-valued theta derivatives (the last column of the table for the curve labeled $14A4$ in \S~\ref{esxP} is selected).

\beq\begin{split}
\widetilde{\jmath}_1(w) & = \frac{1}{24\sigma^{3/2}(w)}\big(-40w^3 + 44w^2 - 14w - 127\big),  \\
\widetilde{\jmath}_2(w) & = \frac{1}{128\sigma^3(w)}\left(\begin{array}{l} -64w^6 +  192w^5 + 1168w^4 + 3488w^3 \\ - 2300w^2 - 2996w + 2071\end{array}\right).
\end{split}\eeq

This is in agreement with the results of \cite{DiFuji2}, and we recognize again their \textit{ad hoc} renormalizations in the values of theta derivatives. These authors have computed the asymptotic expansion of a Hikami type integral $\mathcal{J}_{\hbar}^{\mathrm{H}}(u)$ associated to $L^2R$:
\beq
\mathcal{J}_{\hbar}^{\mathrm{H},\pm}(u) = \hbar^{\delta/2}\exp\Big(\sum_{\chi \geq -1} \jmath_{\chi}^{\mathrm{H},(\pm)}(u)\Big),
\eeq
where $\pm$ indicates the dependence of the integration contour. We have:
\beq
\widetilde{\jmath_{1}}(w^{\pm}_u) = \jmath_1^{\mathrm{H},\pm}(u),\qquad\widetilde{\jmath}_2(w^{\pm}_u) + \frac{1}{128} = \jmath_2^{\mathrm{H},\pm}(u)
\eeq
Hence, a version of Conjecture~\ref{qiq} holds up to $o(\hbar^2)$, if the colored Jones on the LHS is replaced by $\mathcal{J}_{\hbar}^{\mathrm{H},\pm}(u)$ holds. The appropriate normalization constant is  $C_{\hbar} = 1 + \frac{\hbar^2}{128} + o(\hbar^2)$.

\section{Heuristics imported from torus knots}

\label{S7}

Let $(P,Q)$ be coprime integers. The A-polynomial of the torus knot $\mathfrak K_{P,Q}$ contains a non-abelian component of the form $A(m,l) = lm^{PQ} + 1$. Since the corresponding spectral curve does not have branchpoints, its partition function and  kernels are ill--defined, so our conjecture for the Jones polynomial cannot be correct as such for torus knots for the non-abelian branch. Nevertheless, we shall see heuristically how the shape of our conjecture for any Wilson line arises in the case of torus knots. It is only at the end of this derivation, when we specialize to the Jones polynomial, that one discovers that the A-polynomial should be replaced by a blow-up of one of its deformation for the conjecture to be meaningful.

\subsection{Matrix model for torus knots}

\subsubsection{General case}

Thanks to toric symmetry, the Wilson loops of $\mathfrak{K}_{P,Q}$ can be computed by localization \cite{Witten89}, \cite{Law}, \cite{Beas}, \cite{Kallen}, and the sum over flat connections on $\mathfrak \mathbb{S}_3\setminus\mathfrak{K}_{P,Q}$ can be written as a matrix-like integral:
\beq
\label{partpq1}\mathcal{W}_{G,R}(\mathfrak K_{P,Q},\hbar)
= \frac{1}{Z_{P,Q}}\,\int \dd\mathbf{X}\, e^{-\frac{1}{\hbar}\,\frac{\mathbf{X}^2}{4PQ}}\, \prod_{\alpha>0} 4\,\mathrm{sh}\Big({\frac{\alpha\cdot\mathbf X}{2P}}\Big)\mathrm{sh}\Big(\frac{\alpha\cdot\mathbf X}{2Q}\Big)\,\, \chi_R(e^{\mathbf X}),
\eeq
where $\alpha>0$ are the positive roots of $G$, $\chi_R$ is the character of the representation $R$, and the normalization constant $Z_{P,Q}$ is:
\beq
\label{partpq}Z_{P,Q}=\,\int_{{\mathbb R}^{\mathrm{n}}} \dd\mathbf{X}\,e^{-\frac{1}{\hbar}\,\frac{\mathbf{X}^2}{4 PQ}}\, \prod_{\alpha > 0} 4\,\mathrm{sh}\Big(\frac{\alpha\cdot\mathbf{X}}{2P}\Big)\,\mathrm{sh}\Big(\frac{\alpha\cdot\mathbf{X}}{2Q}\Big).
\eeq
This can be written even more explicitly, using Weyl's formula for the characters:
\beq
\chi_R(e^{\mathbf{X}})= \frac{\sum_{\mathbf w\in {\rm Weyl}(G)}\, \epsilon(\mathbf w)\,\, e^{\mathbf w(\rho+\Lambda_R)\cdot\mathbf X}}{\prod_{\alpha>0} 2\,\mathrm{sh}\big(\frac{\alpha\cdot\mathbf X}{2}\big)},
\eeq
where $\rho=\frac{1}{2}\sum_{\alpha>0} \alpha$ is the vector of Weyl's constants, $\Lambda_R$ is the highest weight associated to $R$.

\subsubsection{$\mathrm{SU}(\mathrm{n})$ case}
\label{aqi}
For $SU(\mathrm{n})$, the positive roots are $\alpha_{i,j} = \mathbf{e}_i-\mathbf{e}_j$ with $i<j$ and where $\mathbf{e}_i=(0,\dots,0,1,0,\dots,0)$ with $1$ in the $i^{\rm th}$ position, and $\rho=\sum_{i = 1}^{\mathrm{n}} (\frac{\mathrm{n}+1}{2}-i)\mathbf{e}_i $. The Weyl group is the symmetric group $\mathfrak S_{\mathrm{n}}$. Irreducible representations $R$ are in correspondence with Young diagrams with n rows, or partitions $\lambda = (\lambda_1\geq \lambda_2 \geq \dots \geq \lambda_{\mathrm{n}}\geq 0)$, and we have $\Lambda_R=(\lambda_1,\dots,\lambda_{\mathrm{n}})$. The character associated to the representation indexed by $\lambda$ is the Schur polynomial:
\beq\begin{split}
s_{\lambda}(e^{\mathbf{X}}) & = \frac{\sum_{\sigma \in \mathfrak{S}_{\mathrm{n}}} \epsilon(\sigma)\,\prod_{i = 1}^{\mathrm{n}} e^{X_{\sigma(i)}(\lambda_i - i + \frac{\mathrm{n} + 1}{2})}}{\prod_{1 \leq i < j \leq \mathrm{n}} 2\,\mathrm{sh}\big(\frac{X_i - X_j}{2}\big)} \\
& = \frac{\mathrm{det}(e^{H_iX_j})}{\prod_{1 \leq i < j \leq \mathrm{n}} 2\,\mathrm{sh}\big(\frac{X_i - X_j}{2}\big)}\,\prod_{i = 1}^n e^{((\mathrm{n} + 1)/2 - c)X_i},
\end{split}\eeq
where $H_i = \lambda_i - i + c$, $c$ is an arbitrary constant. From Harish-Chandra formula, we also have:
\beq
s_{\lambda}(e^{\mathbf{X}}) = \frac{\Delta(\mathbf{H})\,\Delta(\mathbf{X})\,\prod_{i = 1}^{\mathrm{n}} e^{((\mathrm{n} + 1)/2 - c)X_i}}{\prod_{1 \leq i < j \leq \mathrm{n}} 2\,\mathrm{sh}\big(\frac{X_i - X_j}{2}\big)}\,\int_{\mathrm{U}(\mathrm{n})} \dd U\,e^{\mathrm{Tr}\,\mathbf{H}U\mathbf{X}U^{\dagger}},
\eeq
where $\dd U$ is the Haar measure on $\mathrm{U}(\mathrm{n})$ with total mass $1$,
\beq
\mathbf{H} = \mathrm{diag}(H_1,\ldots,H_{\mathrm{n}})\qquad\qquad \mathbf{X} = \mathrm{diag}(X_1,\ldots,X_{\mathrm{n}}),
\eeq
and $\Delta(\mathbf{X}) = \prod_{1 \leq i < j \leq \mathrm{n}} (X_i - X_j)$ is the Vandermonde determinant. Thus, Eqn.~\ref{partpq1} gives:
\beq\begin{split}
\label{pros}\mathcal{W}_{\mathrm{SU}(\mathrm{n}),R}(\mathfrak K_{P,Q},\hbar)
&= \frac{\Delta(\mathbf{H})}{(PQ)^{\mathrm{n}(\mathrm{n}-1)/2}\,Z_{P,Q}}\,\int_{\mathbb{R}^{\mathrm{n}}\times\mathrm{U}(\mathrm{n})} \dd \mathbf{X}\,\dd U\,\big(\Delta(\mathbf{X})\big)^2 \\
& e^{-\frac{1}{\hbar}\,\mathrm{Tr}\,\frac{\mathbf{X}^2}{4 PQ} + (\frac{\mathrm{n} + 1}{2} - c)\mathrm{Tr}\,X + \mathrm{Tr}\,\mathbf{H}U\mathbf{X}U^{\dagger}}\,e^{\mathcal{V}(\mathbf{X}/P) + \mathcal{V}(\mathbf{X}/Q) - \mathcal{V}(\mathbf{X})},
\end{split}\eeq
where we have defined the potential $\mathcal{V}(\mathbf{X})$ as:
\beq
e^{\mathcal{V}(\mathbf{X})} = \frac{\prod_{1 \leq i < j \leq \mathrm{n}} 2\,\mathrm{sh}\big(\frac{X_i - X_j}{2}\big)}{\Delta(\mathbf{X})}.
\eeq
Notice that $\mathcal{V}(\mathbf{X})$ is invariant under translation of $\mathbf{X}$ by a matrix proportional to the identity matrix $\mathbf{1}_{\mathrm{n}}$. We may include the factors $i = j$ since they are equal to $1$, and rewrite:
\beq\begin{split}
\mathcal{V}(\mathbf{X}) & = \mathrm{Tr}\,\ln\Big[\frac{2\,\mathrm{sh}\big(\mathbf{1}_{\mathrm{n}}\otimes \mathbf{X} - \mathbf{X}\otimes\mathbf{1}_{\mathrm{n}}\big)}{\mathbf{1}_{\mathrm{n}}\otimes \mathbf{X} - \mathbf{X}\otimes\mathbf{1}_{\mathrm{n}}}\Big] \\
& = \frac{1}{2}\sum_{l,m \geq 1} (-1)^m\,\frac{B_{l + m}}{l + m}\,\frac{\mathrm{Tr}\,\mathbf{X}^l}{l!}\,\frac{\mathrm{Tr}\,\mathbf{X}^m}{m!},
\end{split}\eeq
where $B_l$ the $l^{\mathrm{th}}$ Bernoulli number. To get rid of the linear term in the exponential in Eqn.~\ref{pros}, we shift:
\beq
\mathbf{X} \rightarrow \mathbf{X} + \hbar\,PQ\,\big(\frac{\mathrm{n} + 1}{2} - c\big)\,\mathbf{1}_n.
\eeq
The contour of integration $\mathbb{R}^n$ is shifted to $\mathcal{C}^n_{R,\hc}$ accordingly:
\beq
\mathcal{C}_{R,\hc} = \mathbb{R} + \hbar\,PQ\,\big(\frac{\mathrm{n} + 1}{2} - c\big).
\eeq
We define the normal matrix $M = U\mathbf{X}U^{\dagger}$, and the invariant measure on the space $\mathcal{H}_{\mathrm{n}}(\mathcal{C}_{\hbar,R})$ of normal matrices with eigenvalues on the contour $\mathcal{C}_{R,\hbar}$ is:
\beq
\dd M = \frac{\mathrm{Vol}[\mathrm{U}(\mathrm{n})]}{(2\pi)^{\mathrm{n}}\,\mathrm{n}!}\,\big(\Delta(\mathbf{X})\big)^2\,\dd \mathbf{X}\,\dd U.
\eeq
Therefore:
\beq\begin{split}
\mathcal{W}_{\mathrm{SU}(\mathrm{n}),R}(\mathfrak{K}_{P,Q},\hbar) & = \frac{\mathrm{Vol}[\mathrm{U}(\mathrm{n})]}{(2\pi)^{\mathrm{n}}\,\mathrm{n}!}\,\frac{e^{\hbar\,PQ\,\mathrm{n}((\mathrm{n} + 1)/2 - c)^2/2}}{(PQ)^{\mathrm{n}(\mathrm{n} - 1)}\,Z_{P,Q}}\,\Delta(\mathbf{H})\,e^{2PQ\hbar((\mathrm{n} + 1)/2 - c)\,\mathrm{Tr}\,\mathbf{H}},  \\
\label{matrii}& \times \int_{\mathcal{H}_{\mathrm{n}}(\mathcal{C}_{R,\hbar})} \dd M\,e^{-\frac{1}{\hbar}\,\mathrm{Tr}\,\frac{M^2}{4PQ}}\,e^{\mathcal{V}(M/P) + \mathcal{V}(M/Q) - \mathcal{V}(M)}.
\end{split}\eeq
The Vandermonde of the $H$'s is related to the dimension of the representation:
\beq
\mathrm{dim}\,R = \frac{\Delta(\mathbf{H})}{\prod_{i = 1}^{\mathrm{n}} (H_i + \mathrm{n} - c)!},
\eeq
and the trace of $\mathbf{H}$ is related to the number of boxes in $\lambda$:
\beq
\mathrm{Tr}\,\mathbf{H} = |\lambda| + \mathrm{n}\Big(c - \frac{\mathrm{n} + 1}{2}\Big).
\eeq
So far, the constant $c$ was arbitrary, in particular it can depend on $\lambda$. The choice:
\beq
c = -\frac{|\lambda|}{\mathrm{n}} + \frac{\mathrm{n} + 1}{2}
\eeq
allows to have $\mathrm{Tr}\,\mathbf{H} = 0$, and we now stick to it. The normalization constant $Z_{P,Q}$ is computed with the same steps for the trivial representation $R_{\emptyset}$ of $\mathrm{SU}(\mathrm{n})$.
\beq
\forall i \in \{1,\ldots,\mathrm{n}\}\qquad \lambda_i^{\emptyset} = 0\,\qquad\, H^{\emptyset}_i = - i + \frac{\mathrm{n} + 1}{2}.
\eeq
Thus:
\beq
\label{auq}\mathcal{W}_{\mathrm{SU}(\mathrm{n}),R}(\mathfrak{K}_{P,Q},\hbar) = D_R\,\,\frac{\int_{\mathcal{H}_{\mathrm{n}}(\mathcal{C}_{R,\hbar})} \dd M\,e^{-\frac{1}{\hbar}\,\mathrm{Tr}\,\frac{M^2}{4PQ}}\,e^{\mathcal{V}(M/P) + \mathcal{V}(M/Q) - \mathcal{V}(M)}\,e^{\mathrm{Tr}\,\mathbf{H}M}}{\int_{\mathcal{H}_n(\mathbb{R})} \dd M\,e^{-\frac{1}{\hbar}\,\mathrm{Tr}\,\frac{M^2}{4 PQ}}\,e^{\mathcal{V}(M/P) + \mathcal{V}(M/Q) - \mathcal{V}(M)}\,e^{\mathrm{Tr}\,\mathbf{H}^{\emptyset} M}},
\eeq
where the multiplicative constant is given by:
\beq
D_R = e^{\hbar\,PQ\,|\lambda|^2/\mathrm{n}}\,\frac{\prod_{j = 1}^{\mathrm{n}} \big(H_j + \frac{\mathrm{n} - 1}{2} + \frac{|\lambda|}{\mathrm{n}}\big)!}{\prod_{j = 1}^{\mathrm{n} - 1} j!}\,\mathrm{dim}\,R.
\eeq
The integral in the numerator is similar to that in the denominator, except for the \emph{external field} $\mathbf{H}$ (resp. $\mathbf{H}^{\emptyset}$) encoding the highest weight associated to $R$ (resp. to the trivial representation). We also shifted the contour from $\mathbb{R}$ to $\mathcal{C}_{R,\hbar} = \mathbb{R} + \frac{2\hbar PQ}{\mathrm{n}}\,|\lambda|$. Since the discussion to come remains at a formal level, we shall move it back to $\mathbb{R}$, and assume the range of integration to be the space of hermitian matrices $\mathcal{H}_{\mathrm{n}}(\mathbb{R})$ in Eqn.~\ref{auq}.

\subsection{Computation from the topological recursion}

\subsubsection{Principle}

For matrix integrals (with or without external potential) of the form Eqn.~\ref{matrii}, we have \cite{EOFg}
\beq
\label{sai}Z = \Tau_{\hbar}(\spcurve_{\mathrm{n},\hbar}),
\eeq
where $\spcurve_{\mathrm{n},\hbar}=(\curve,x,y)$ is the spectral curve of the matrix integral, which in general depend on $\hbar$ and the size of the matrix n, and $\Tau_{\hbar}$ is the non-perturbative partition function defined in Section \ref{S3}. We have define somewhat arbitrarily\footnote{This line of reasoning does not tell us the scale of $\hbar$, because changing $\hc$ to $\alpha \hc$ amounts to rescaling $y\,\dd x$ to $\alpha\,y\dd x$, and we are not precise enough to identify $y$ and $x$ to (some multiple of) the meridian and longitude eigenvalues $m$ and $l$ of knot theory. Outside Section~\ref{S7}, the quantization condition satisfied by A-spectral curves provided a good argument in favor of this choice.}
Eqn.~\ref{sai} means that the asymptotic expansion of the left hand side is given by the right hand side (which we defined as a formal asymptotic series). Adding an external field in the form $e^{\Tr\,\mathbf{H}\,M}$ amounts (see for instance \cite{NicoO}) to modify the spectral curve by addition of simple poles $p_1,\ldots,p_{\mathrm{n}} \in \curve$ to $x$ with residue $\hc$ with respect to $\dd y$, and such that $y(p_j) = \hbar\,H_j$, and some other simple poles $o_1,\ldots,o_{\mathrm{n}} \in \curve$ with residue $-\hc$.
\beq
(\curve,x,y) \mapsto  \Big(\curve, x + \hc \sum_{j = 1}^{\mathrm{n}} \frac{\dd S_{o_j,p_j}}{\dd y},y\Big).
\eeq
Similarly, we denote $p_j^{\emptyset}$ the poles associated to the external field $\mathbf{H}^{\emptyset}$, i.e. we have $y(p_j^{\emptyset}) = \hbar\big(-i + \frac{\mathrm{n} + 1}{2}\big)$. We thus find:
\beq\label{trouseq1}
\mathcal{W}_{\mathrm{SU}(\mathrm{n}),R}(\mathfrak{K}_{P,Q},\hbar) = D_R\,\,\frac{\Tau_{\hc}\Big(\curve, x + \hbar\sum_{j = 1}^{\mathrm{n}} \dd S_{o_j,p_j}/\dd y,y\Big)}{\Tau_{\hc}\Big(\curve, x +\hbar \sum_{j = 1}^{\mathrm{n}} \dd S_{o_j,p_j^{\emptyset}}/\dd y,y\Big)}.
\eeq
Then, the symplectic invariance (see \S~\ref{sympeq}) allows to exchange\footnote{This exchange is possible when the set of zeroes of $\dd y$ is not empty, i.e. when the spectral curve $(\mathcal{C},y,x)$ has at least one ramification point.}  the role of $x$ and $y$:
\beq\label{trouseq2}
\mathcal{W}_{\mathrm{SU}(\mathrm{n}),R}(\mathfrak{K}_{P,Q},\hbar) = D_R\,\,\frac{\Tau_{\hc}\Big(\curve,y,x + \hbar \sum_{j = 1}^{\mathrm{n}} \dd S_{o_j,p_j}/\dd y\Big)}{\Tau_{\hc}\Big(\curve,y,x +\hbar \sum_{j = 1}^{\mathrm{n}} \dd S_{o_j,p_j^{\emptyset}}/\dd y\Big)}.
\eeq
Now, the simple poles are added to the second function in the spectral curve, and we recognize the $\mathrm{n}|\mathrm{n}$-kernel described in \S~\ref{spinker}, for the spectral curve $\hat{\mathcal{S}}_{\mathrm{n},\hbar} = (\mathcal{C},y,x)$ for some basepoint $o$:
\beq
\label{saiqq}\mathcal{W}_{\mathrm{SU}(\mathrm{n}),R}(\mathfrak{K}_{P,Q},\hbar) = D_R\,\,\frac{\psi_{\hbar}^{[\mathrm{n}|\mathrm{n}]}\big(p_1,o_1\,;\ldots;\,p_{\mathrm{n}},o_{\mathrm{n}})}{\psi_{\hbar}^{[\mathrm{n}|\mathrm{n}]}\big(p_1^{\emptyset},o_1\,;\ldots ;\,p_{\mathrm{n}}^{\emptyset},o_{\mathrm{n}})}.
\eeq

\subsubsection{$\mathrm{SU}(2)$ case: Jones polynomial}

For $\mathrm{SU}(\mathrm{n} = 2)$ and the representation $R$ associated to $(\lambda_1,\lambda_2) = (N - 1,0)$, we retrieve the colored Jones polynomial (see Eqn.~\ref{Jonesdef}). It is thus computed from the $2$-kernel with points $p_1$ and $p_2$ of projections:
\beq
\label{jsu}y(p_1) = 2\hbar H_1 = N\hbar \qquad\qquad y(p_2) = 2\hbar\,H_2 = - N\hbar,
\eeq
and for the trivial representation:
\beq
y(p_1^{\emptyset}) = \hbar,\qquad y(p_2^{\emptyset}) = -\hbar.
\eeq
This leads to:
\beq
\label{sjoa}J_N(\mathfrak{K}_{P,Q},q = e^{2\hbar}) = N\,e^{\hbar\,PQ\,(N - 1)^2}\,\frac{\mathrm{sh}\,\hbar}{\mathrm{sh}\,N\hbar}\,\frac{\psi_{\hbar}^{[2|2]}(p_1,p_2)}{\psi_{\hbar}^{[2|2]}(p_1^{\emptyset},p_2^{\emptyset})}.
\eeq
and we insist that the kernels are computed for the spectral curve of the matrix integral after exchange of $x$ and $y$.

\subsubsection{Spectral curve for the torus knots}

Let $P',Q'$ be integers such that $P'Q - Q'P =1$. The spectral curve $\mathcal{S}_{\mathrm{n},\hbar} = (\curve,x,y)$ of the matrix integral in Eqn.~\ref{partpq} was derived in \cite{BEMknots}, in the regime when $\mathrm{n}\hbar$ is of order $1$:
\beq
\spcurve:\qquad
\label{sapap}e^{-(P+Q)y} -  e^{-\mathrm{n}\hc}\,e^{-(P'x + Qy)} - e^{-(Q'x + Py)} + e^{\mathrm{n}\hc}\,e^{-(P' + Q')x} = 0.
\eeq
This curve can be uniformized with a variable $z \in \mathbb{C} \equiv \mathcal{C}$:
\beq
e^{x} = e^{(P - Q)\mathrm{n}\hc}\,e^{-Pz}\,\Big(\frac{e^{2\mathrm{n}\hbar} - e^z}{1 - e^z}\Big)^{Q}\qquad e^y = e^{(P' - Q' - 1/Q)\mathrm{n}\hbar}\,e^{-P'z}\,\Big(\frac{e^{2\mathrm{n}\hbar} - e^z}{1 - e^z}\Big)^{Q'}.
\eeq
hence its genus is $0$, and there is no theta function in the definition of its partition function and kernels. It was argued in \cite{BEMknots} that the topological recursion for this curve reproduces the torus knots invariants. 

Here, we are interested in the regime where $\hbar$ is small and $\mathrm{n}$ is fixed (and in particular $\mathrm{n} = 2$). If we keep $x$ and $y$ of order $1$, the curve is trivial:
\beq
\label{saiq}e^{Py} = e^{P'x}.
\eeq
in the sense that it does not have ramification points. So, the partition function $\mathcal{T}_{\hbar}$ of this spectral curve is ill-defined. Actually, when the limit spectral curve is ill-defined, the information about the unstable terms (i.e. the terms decaying with $\hbar \rightarrow 0$) are actually obtained contained in the blow-up of $\mathcal{S}_{n,\hbar}$ at its basepoint $o$. It is realized by setting $x = \sqrt{2\mathrm{n}\hbar}\,\tilde{x}$ and $y = \sqrt{2\mathrm{n}\hbar}\,\tilde{y}$ with $\tilde{x}$ and $\tilde{y}$ of order $1$, and retaining the first non trivial order in Eqn.~\ref{sapap} when $\mathrm{n}\hbar \rightarrow 0$, we find:
\beq
\label{spqo}PQ\,\tilde{y}^2 - (P'Q + PQ')\tilde{x}\tilde{y} + P'Q'\,\tilde{x}^2 + 1 = 0.
\eeq
The formulae \ref{sai} and \ref{saiqq} are expected to be correct if applied to the spectral curve of Eqn.~\ref{spqo} at least for the terms of order $o(1)$ when $\hbar \rightarrow 0$. The non decaying terms are rather given by the limit spectral curve itself, and thus are trivial. This is in agreement with the fact \cite{Kashtor} that there is no exponential growth of the Jones polynomial of torus knots (they are not hyperbolic), in other words $\widetilde{\jmath}_{-1} \equiv 0$.

\subsection{General mechanism}

We expect that the mechanism described for torus knots complements is general (see for instance the conjecture in \cite[Section 6]{Law}), and our proposal should essentially compute Wilson lines for more general $3$-manifold $\mathfrak{M}$, provided the spectral curve is well-chosen. A scenario would be that:
\beq
\label{forsm}\mathcal{W}_{G,R}(\partial\mathfrak{M},\hbar) = \frac{\int \dd \mathbf{V}\,\chi_{R}(\mathbf{V})\,e^{S(\mathbf{V})}}{\int \dd \mathbf{V}\,e^{S(\mathbf{V})}},
\eeq
where $\mathbf{V}$ consists of eigenvalues of the holonomy operator $\mathrm{P}\exp\oint_{\partial\mathfrak{M}} \mathcal{A}$, and $S(\mathbf{V})$ the effective action for $\mathbf{V}$ after integrating out the other degrees of freedom against the Chern-Simons action. If a formula like \ref{forsm} holds, the steps of \S~\ref{aqi} can be repeated to find:
\beq\label{eqWchext1}
\mathcal{W}_{\mathrm{SU}(\mathrm{n})}(\partial\mathfrak{M},\hbar/2) \propto \frac{\int \dd M\,e^{S(M)}\,e^{\mathrm{Tr}\,\mathbf{H}\,M}}{\int \dd M\,e^{S(M)}},
\eeq
where $M = U\,\mathbf{X}\,U^{\dagger}$, $U \in \mathrm{SU}(\mathrm{n})$ and $H_i = \lambda_i - i + c$ for some constant $c$. Then, if the integral $\int \dd M\,e^{S(M)}$ is a Tau function with spectral curve $\mathcal{S}_{\mathrm{n},\hc} = (\mathcal{C},x,y)$, we would find again:
\beq\label{eqWsources2}
\mathcal{W}_{\mathrm{SU}(\mathrm{n})}(\partial\mathfrak{M},\hbar) \propto \frac{\mathcal{T}_{\hbar}\big(\mathcal{C},y,x + \hbar\sum_{i = 1}^{\mathrm{n}} \dd S_{o,p_i}/\dd y\big)}{\mathcal{T}_{\hbar}\big(\mathcal{C},y,x + \hbar\sum_{i = 1}^{\mathrm{n}} \dd S_{o,p_i^{\emptyset}}/\dd y\big)}  \propto \frac{\psi_{\hbar}^{[\mathrm{n}|\mathrm{n}]}(p_1,\ldots,p_{\mathrm{n}})}{\psi_{\hbar}^{[\mathrm{n}|\mathrm{n}]}(p_1^{\emptyset},\ldots ,p_{\mathrm{n}}^{\emptyset})},
\eeq
where the $\mathrm{n}|\mathrm{n}$-kernel is computed for the spectral curve $\hat{\mathcal{S}}_{\mathrm{n},\hbar} = (\mathcal{C},y,x)$ after the exchange $(x\leftrightarrow y)$. For hyperbolic manifolds, one expects to find as spectral curve a deformation of the A-polynomial (or at least of subcomponents of it), which reduces to the A-polynomial in the $\mathrm{SU}(\mathrm{n} = 2)$ case, i.e. to $m \propto e^{2y}$, $l \propto e^{x}$ and $A(m,l) = 0$. This is plausible because it is known that the A-polynomial can be obtained as the saddle point equation obtained by elimination from the Neumann-Zagier potential \cite{Hikami}. In our argument, we see then from Eqn.~\ref{jsu} that the points $p_1$ and $p_2$ needed to compute the Jones polynomial have $m$-projection:
\beq
\label{saiii}\ln m(p_1) = N\hbar + \mathrm{cte}\qquad\qquad \ln m(p_2) = -N\hbar + \mathrm{cte},
\eeq
and one recognizes\footnote{Again, the argument for the overall scaling of $\hbar$ does not come from the discussion of this Section.} (up to constant shift) the identification between the hyperbolic structure parameter $m = e^{u}$ and the quantum group parameter $q = e^{2\hbar}$ and $N$ appearing in Eqn.~\ref{Jonesdef}.

\section{Perspectives}

We have constructed a formal asymptotic series $\mathcal{J}_{\hbar}^{\mathrm{n.p.TR}}(p)$ depending on a point on the $\mathrm{SL}_2(\mathbb{C})$ character variety of a hyperbolic $3$-manifolds with $1$-cusp, which has interesting properties \textit{per se}. It depends on a choice of characteristics $\mu,\nu \in \mathbb{C}^g$ (which might be restricted to even-half characteristics) and basepoint $o$ for the computation of iterated primitives. Provided a accurate choice is made for those data, we have conjectured that it computes the asymptotic expansion of the colored Jones polynomial, discarding roots of unity. We have made a non-trivial check to first orders for the figure-eight knot. A weaker conjecture is that our series is a formal solution of the $A$-hat recursion relation. We made a closely related check to first orders for the $L^2R$. We think that working on the A-hat recursion relation satisfied by the colored Jones polynomial is a good approach in an attempt to prove our conjecture (or a slight modification of it).

The intuition behind our construction comes from the theory of integrable systems and its relations to loop equations. $\mathcal{J}_{\hbar}^{\mathrm{n.p.TR}}(p)$ was defined formally by introducing an infinite number of infinitesimal deformations of the A-polynomial curve, and one may wonder if this can be interpreted as an integrable perturbation of the Wess-Zumino-Witten CFT. 

The main interest of our proposal is rather structural than computational. Although we do have an algorithm, it requires the use of a basis of meromorphic forms which behaves well under integration, and the computation of theta functions and derivatives, so is less efficient than other methods. Yet, the expression in terms of the topological recursion suggest possible connections between knot theory and respectively integrable systems \cite{EORev}, \cite{BEInt}, intersection theory on the moduli space \cite{Einter} and enumerative geometry, which deserve further investigation.

This conjecture gives also a framework for the study of the arithmetic properties of perturbative knot invariants. It would particularly interesting to compare our predictions for the expansion of the Kashaev invariant (i.e. at the complete hyperbolic point) to those of \cite{GarouDimo} obtained by a gluing procedure.

When the quotient of (a component of) the $\mathrm{SL}_2(\mathbb{C})$ character variety by the involution $\iota(m,l) = (1/m,1/l)$ is a genus $g_{\iota} = 0$ curve, $\mathcal{J}_{\hbar}^{\mathrm{n.p.}TR}$ is a formal power series, although it takes into account non-perturbative effects. When this property does not hold, it is rather an asymptotic series which contains fast oscillations to all orders when $\hbar \rightarrow 0$. However, the K-theoretical properties of the A-polynomial imply that, if we specialize $\hbar$ to sequences ${\rm i}\pi/k$ with $k$ an integer of fixed congruence, we retrieve a formal power series. It would be interesting to know if this wild behavior can be seen in the asymptotics of the colored Jones polynomial for knots such that $g_{\iota} \neq 0$. The simplest examples of this kind we know are $\mathbf{8}_{21}$ and the $\mathrm{Pretzel}(-2,3,9)$, and are currently under investigation. Actually, earlier experiments on the asymptotics of the colored Jones have been performed to our knowledge only for knots with $g_{\iota} = 0$. If Conjecture~\ref{qiq} contains some part of truth, new phenomena may be discovered. Else, one would need to understand how it should be modify to preserve the matching for $\mathbf{4}_1$ and $L^2R$.

Our conjectures could be generalized in several directions.
\begin{itemize}
\item One may wonder if the $\mathrm{n}|\mathrm{n}$ kernels can be identified to asymptotics of other relevant knot invariants. Notice also that Hirota equations imply determinantal formulas \cite{BEInt}:
\beq
\psi^{[\mathrm{n}|\mathrm{n}]}_{\hbar}(p_1,q_1;\ldots;p_{\mathrm{n}},q_{\mathrm{n}}) \propto \mathrm{det}_{i,j} \psi^{[1|1]}_{\hbar}(p_i,q_j),
\eeq
where $\propto$ means equality up to a factor involving prime forms. A naive guess, inspired by Section~\ref{S7}, would be to compare $\psi_{\hbar}^{[\mathrm{n}|\mathrm{n}]}$ to Wilson lines in a Chern-Simons theory with $\mathrm{SU}(\mathrm{n}|\mathrm{n})$ gauge group in the limit of large representations. The rescaled size of the representation would be in correspondence (see \S~\ref{CStheo} in the case of $\mathrm{SU}(\mathrm{n})$ with points $p_1,\ldots,p_{2\mathrm{n}}$ on the character variety.
\item One may wish to study the asymptotics of the colored Jones when $q \rightarrow \zeta_d$ (a root of unity, instead of $q \rightarrow 1$ here). It is natural to propose a conjecture similar to \ref{qiq}, with the curve of equation $\lim_{q \rightarrow \zeta_d} \hat{\mathfrak{A}}(e^{u},e^{v},q)$ replacing the $A$-polynomial.
\item One may wish to study the $a$- (or $Q$-) deformation of the knot invariants, considered recently in \cite{Mina}. In the regime when $q \rightarrow 1$ but keeping $a$ finite, this amounts to study asymptotics for gauge groups of large rank. We guess that the non-topological recursion for the $a$-deformed $A$-polynomial will come into play.
\item And, at the top of the hierarchy, one may consider the categorified knot invariants, which results from another deformation with a variable $t$ \cite{GukD}. These knot invariants can be seen as generating series of BPS invariants. They are conjectured to be annhilated by an operator $\mathfrak{A}(e^{u},e^{\hbar\partial_u},a,q,t)$, called the super-A-hat polynomial, which is explicitly known in a few examples \cite{GHS}, \cite{GHSa}.
In this case, although a $t$-deformed spectral curve can be defined, we think that a "deformed topological recursion" should be used in order to compute something meaningful about their asymptotics. This intuition is based on the fact that Schur polynomial have to be replaced by Macdonald polynomials under this deformation, and ongoing work suggests that the analysis of the matrix model of Eqn.~\ref{partpq1} requires a deformation of the topological recursion.
\end{itemize}
Although the identification for the colored Jones polynomial ($N\hbar = \ln m(p_u)$) was rather simple, appropriate and non-trivial "mirror maps" (like in topological strings \cite{TopoVO}, \cite{BKMP}) could be necessary to make any of those generalizations effective.

In yet another direction, the (generalized) volume conjecture can also be formulated for links with $L$ components. The $\mathrm{SL}_{\mathrm{n}}(\mathbb{C})$ character variety has local complex dimension $(\mathrm{n} - 1)L$ at a generic point \cite{Dimvar}. It is a challenging problem to reduce - if only possible -  the asymptotics of $3$-manifolds with $L$ cusps to algebraic geometry on this variety.

\subsubsection*{Acknowledgments}

We thank warmly David Cimasoni, Tudor Dimofte, Stavros Garoufalidis, Rinat Kashaev, Marcos Mari\~{n}o and Don Zagier for fruitful discussions, as well as the organizers of the BIRS workshop "New recursion formulae and integrability for Calabi-Yau spaces" in Banff, where this work was initiated. G.B. thanks the participants of the Oberwolfach workshop "Low-dimensional topology and number theory", whose comments led to improve a later version of the article, and the MPI Bonn for hospitality. B.E. thanks the university of Geneva and Stanislav Smirnov for their hospitality, and the Centre de Recherche Math\'ematiques de Montr\'eal. G.B. benefited from the support of Fonds Europ\'een S16905 (UE7 - CONFRA). B.E. is supported by the ANR project GranMa "Grandes Matrices Al\'{e}atoires" ANR-08-BLAN-0311-01 and by the European Science Foundation through the Misgam program.

\appendix

\section{Diagrammatic representation for the non-perturbative topological recursion}
\label{appdiag}

\subsection{Non-perturbative partition function}

The non-perturbative Tau function $\mathcal{T}_{\hc}$ was defined in Eqn.~\ref{taudef}. We had:
\beq
\mathcal{T}_{\hc} = e^{\hc^{-2} F_0 + F_1}\,\vartheta\big[{}^{\mu}_{\nu}\bigr](\zeta|\tau)\,\hat{\mathcal{T}}_{\hc},
\eeq
with:
\beq\begin{split}\label{taudefapp}
\hat{\mathcal{T}}_{\hc} & = \exp\Big(\sum_{h \geq 2} \hc^{2h - 2}\,F_h\Big)\,\,\, \frac{1}{\vartheta\bigl[{}^{\mu}_{\nu}\bigr](\zeta|\tau)} \\
& \times \Big\{1 + \sum_{r \geq 1} \frac{1}{r!} \sum_{\substack{h_j \geq 0,\,\,d_j \geq 1 \\ 2h_j - 2 + d_j > 0}} \hspace{-10pt} \hc^{\sum_j 2h_j - 2+ d_j}\,\bigotimes_{j = 1}^r \frac{F^{(d_j)}_{h_j}\cdot\nabla^{\otimes d_j}}{(2\mathrm{i}\pi)^{d_j}\,d_j!}\Big\}\vartheta\bigl[{}^{\mu}_{\nu}\bigr](\zeta|\tau). 
\end{split}\eeq
We also recall that, owing to special geometry, the $k^{\rm th}$ derivative of $\om_n^h$ with respect to filling fractions is:
\beq
\om_n^{h,(k)} = \overbrace{\oint_\bcycle\dots\oint_\bcycle}^{k\,\,{\rm integrals}} \om_{n+k}^h.
\eeq
For $2-2h-n<0$, we represent $\om_n^h$ by a surface with $h$ handles and $n$ legs, and we represent $\nabla^{\otimes k} \vartheta/\vartheta$ by a black vertex with $k$ legs.

Then, with those diagrammatic notations, Eqn.~\ref{taudefapp} is represented as a sum of graphs.
Each graph has exactly one black vertex, whose legs are attached to the legs of a product of $\om_n^h$'s, such that all legs are paired.
\beq
\hat{\mathcal{T}}_{\hc}  = 1+
\sum \frac{\hc^{-\chi_{\rm Euler}}}{\#\,{\rm Aut}}\times\mathrm{weight} \left(\begin{array}{c}
\includegraphics[width=0.4\textwidth]{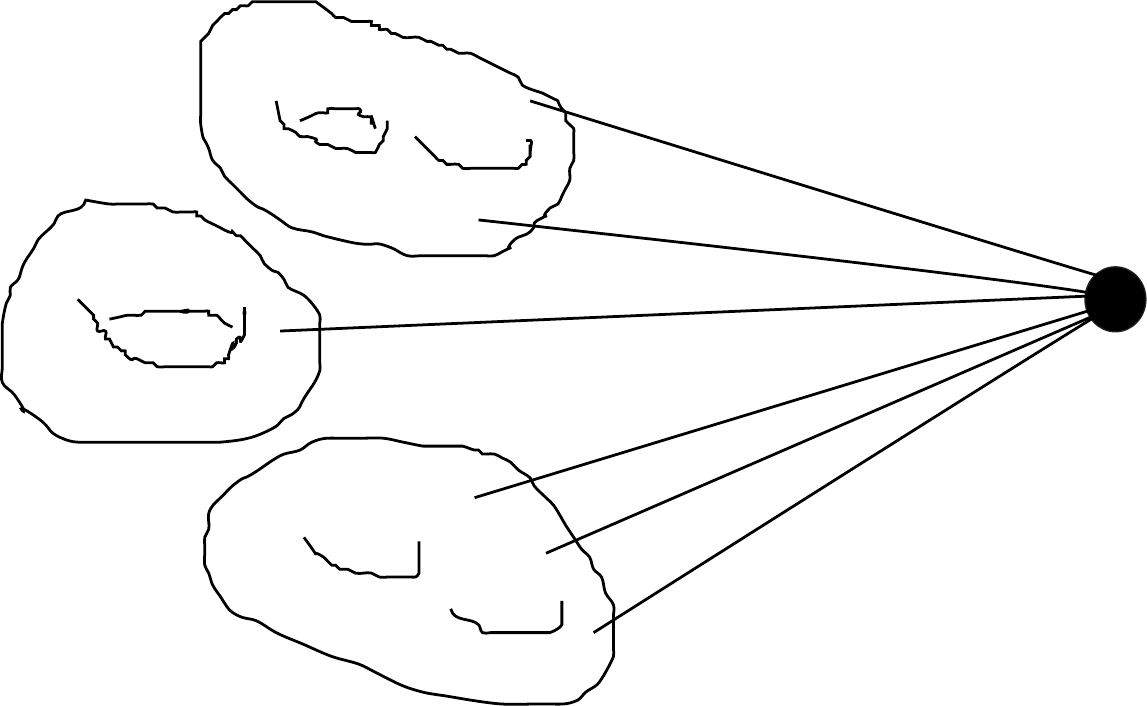}\end{array}\right),
\eeq
where $\chi_{\rm Euler}$ is sum of the Euler characteristics of all punctured surfaces of the graph (each of them having a negative Euler characteristics), and $\#\,{\rm Aut}\in \mathbb N^*$ is the symmetry factor of the graph.

\subsection{Logarithm of the non-perturbative partition function}

Notice that the generating function for the derivatives of $\ln\vartheta$, is related to the generating function for the derivatives of $\vartheta$, by keeping the cumulants. If we represent $\nabla^{\otimes k} \ln\vartheta$ by a white vertex with $k$ legs, we have that the black vertex is the sum of all possible products of white vertices having the same legs.
\beq
\frac{\nabla^{\otimes k}\vartheta\bigl[{}^{\mu}_{\nu}\bigr]}{\vartheta\bigl[{}^{\mu}_{\nu}\bigr]} =  \begin{array}{c}\includegraphics[width=0.2\textwidth]{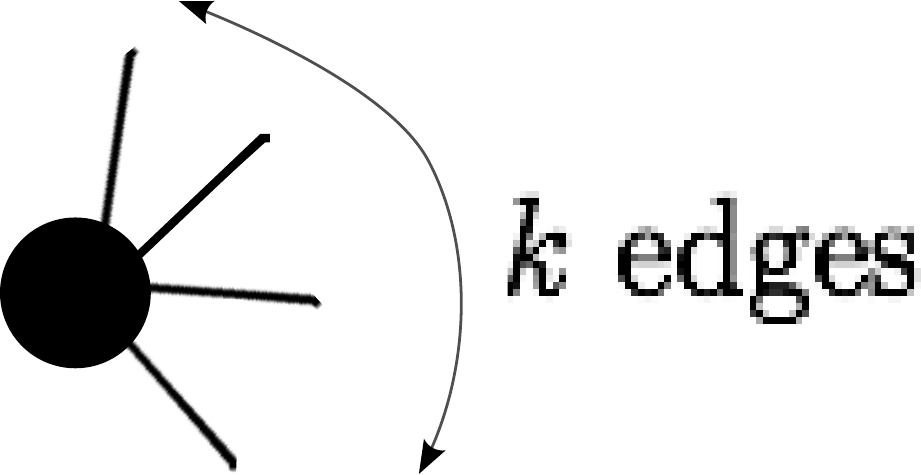}\end{array},\qquad \nabla^{\otimes k}\ln\vartheta\bigl[{}^{\mu}_{\nu}\bigr] = \begin{array}{c} \includegraphics[width=0.2\textwidth]{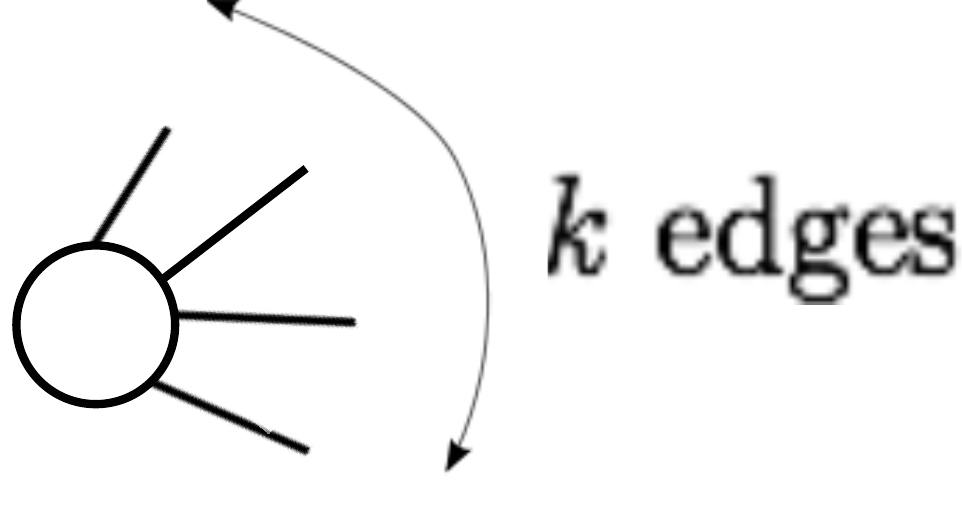}\end{array},
\eeq
and:
\beq
\includegraphics[width=0.4\textwidth]{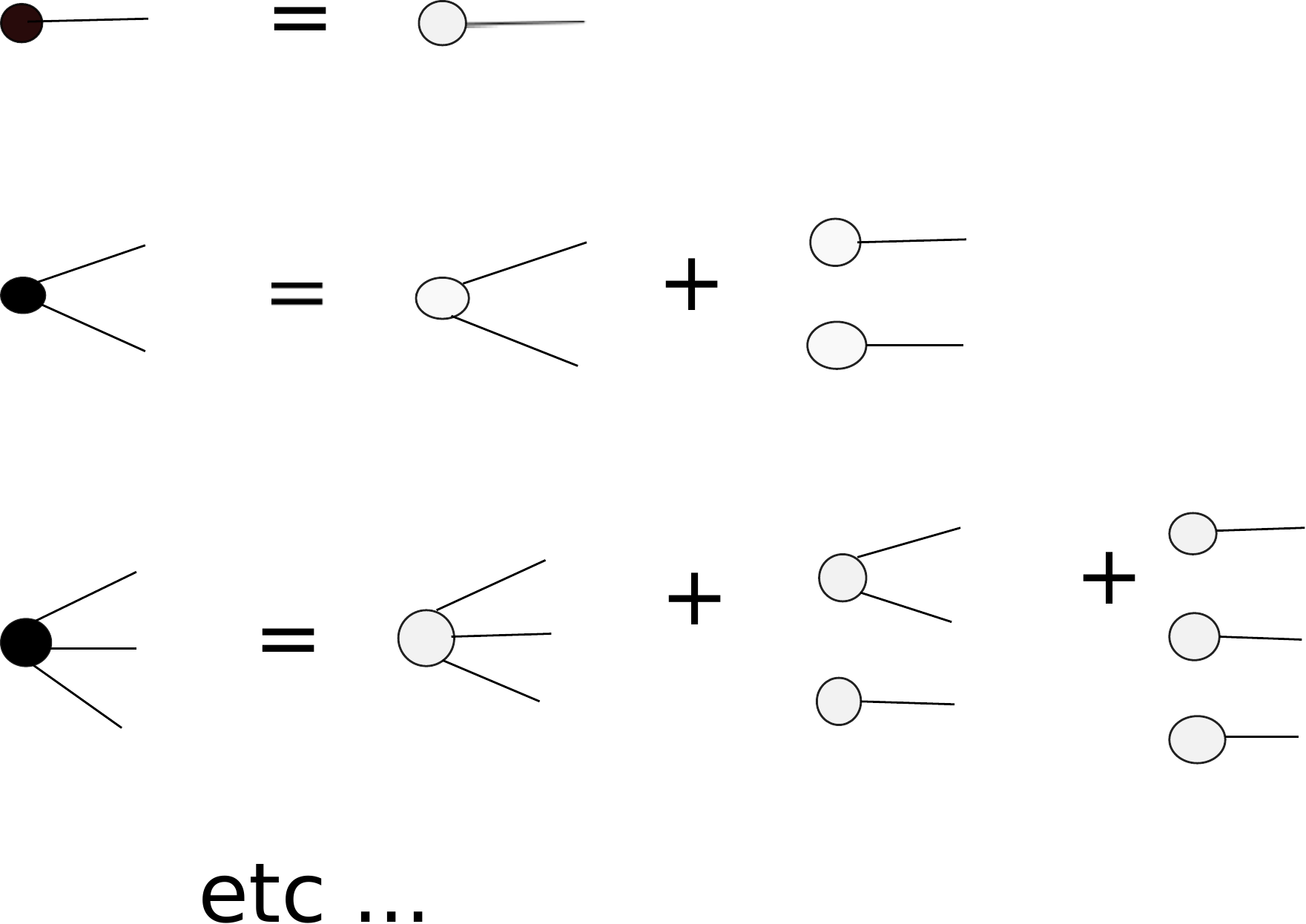}
\eeq
This means that the diagrammatic representation of $\hat{\mathcal{T}}_{\hc}$, in terms of white vertices is a sum of all graphs, not necessarily connected, whose vertices are either surfaces with handles and punctures, or white vertices, and whose edges connect the punctures to white vertices. Then, taking the logarithm of $\hat{\mathcal{T}}_{\hc}$ has the same diagrammatic representation, but keeping only connected graphs:
\beq
\ln\hat{\mathcal{T}}_{\hc}  = 
\sum \frac{\hc^{-\chi_{\rm Euler}}}{\#\,{\rm Aut}}\times\mathrm{weight}\left(\begin{array}{c}
\includegraphics[width=0.4\textwidth]{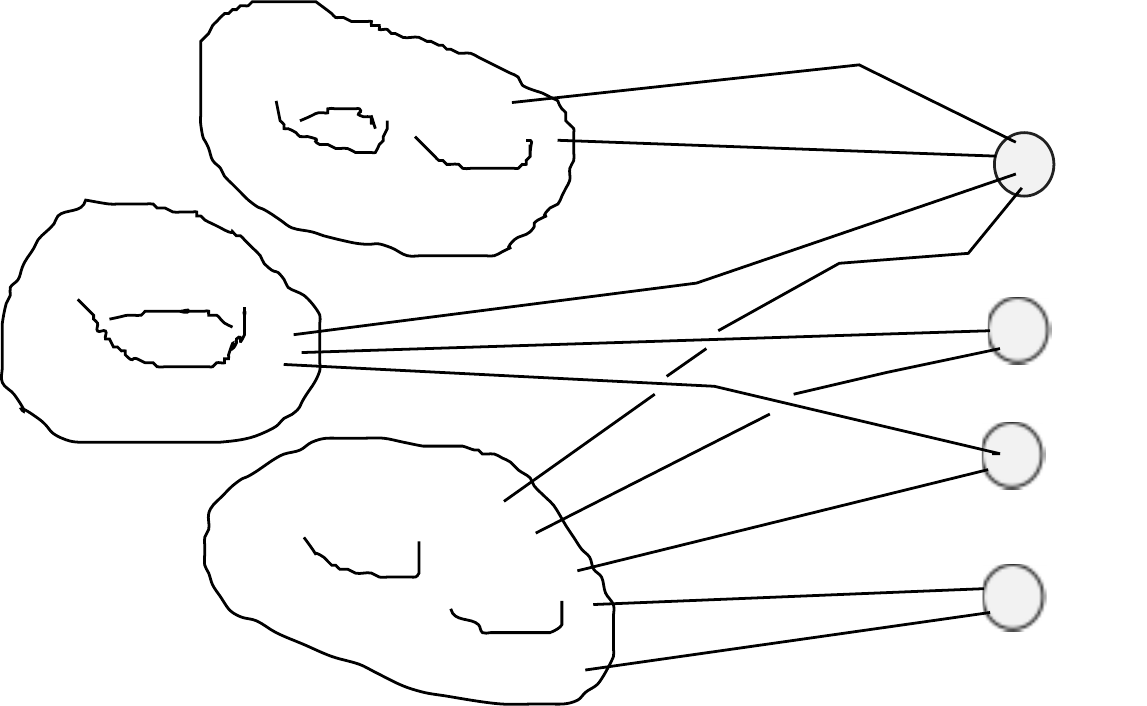}\end{array}\right),
\eeq
where $\chi_{\rm Euler}$ is the sum of the Euler characteristics of the punctured surfaces appearing in the graph.

\subsection{$\mathrm{n}|\mathrm{n}$ kernels}

We take the example of the $2|2$ kernel since it is the one which can be compared to the Jones polynomial. It was defined in Eqn.~\ref{dkea} by:
\beq
\psi_{\hc}^{[2|2]}(p_1,o_1\,;\,p_2,o_2) = \frac{{\mathcal{T}}_{\hc}\big[v\,\dd u \rightarrow v\,\dd u + \hc\,\dd S_{o_1,p_1} + \hc\,\dd S_{o_2,p_2}\big]}{\mathcal{T}_{\hc}\big[v\,\dd u\big]}.
\eeq
Notice that, when we add in the numerator simples poles $p_1,o_1$, $p_2,o_2$ to the spectral curve, we obtain:
\beq
\label{qaq}\om_n^h \to \sum_{r=0}^\infty \frac{\hc^{r}}{r!}\,\overbrace{\int_{o_1,o_2}^{p_1,p_2}\dots\int_{o_1,o_2}^{p_1,p_2}}^{r\,\,{\rm times}} \om_{n+r}^h,
\eeq
and thus
\beq
\label{qaq2}\om_n^{h,(d)} \to  \sum_{r=0}^\infty \frac{\hc^{r}}{r!}\,\overbrace{\oint_\bcycle \dots \oint_\bcycle}^{d\,\,{\rm times}} \overbrace{\int_{o_1,o_2}^{p_1,p_2}\dots\int_{o_1,o_2}^{p_1,p_2}}^{r\,\,{\rm times}} \om_{n+r+d}^h.
\eeq
Notice that the formula for ${\mathcal{T}}_{\hc}$ involves only $F_h^{(d)} = \om_0^{h,(d)}$ with $2-2h-d<0$. In particular, Eqns.~\ref{qaq}-\ref{qaq2} do not yield the terms with $h=0$ and $d = 1,2$, but they produce all the other terms. Besides, the argument of the theta function and its matrix of periods is also shifted by the addition of simple poles in the spectral curve. We obtain:

\vspace{-0.5cm}

\begin{small}
\beq\begin{split}
\vartheta\bigl[{}^{\mu}_{\nu}\bigr](\mathbf{w}|\tau) &\to 
\vartheta\bigl[{}^{\mu}_{\nu}\bigr]\Big(\mathbf w+ \ab(p_1) - \ab(o_1) + \ab(p_2) - \ab(o_2) + \sum_{r\geq 2} \frac{\hc^{r-1}}{r!}\, \oint_\bcycle \overbrace{\int_{o_1,o_2}^{p_1,p_2}\dots\int_{o_1,o_2}^{p_1,p_2}}^{r\,\,{\rm times}} \om_{r+1}^0\, \Big| \\
& \label{shif}\qquad \tau+\sum_{r\geq 1} \frac{\hc^r}{r!}\, \oint_\bcycle\oint_\bcycle \overbrace{\int_{o_1,o_2}^{p_1,p_2}\dots\int_{o_1,o_2}^{p_1,p_2}}^{r\,\,{\rm times}} \om_{r+2}^0 \Big).
\end{split}\eeq
\end{small}

\vspace{-0.3cm}

\noindent and its Taylor expansion in $\hc$ generates the terms with $h=0$, $d=1,2$ which were missing in the expansion of $\om_0^{h,(d)}$. Eventually, taking into account the finite shift appearing in Eqn.~\ref{shif}, the vertices are now associated with derivatives of theta functions evaluated at:
\beq
\mathbf{w}_{\bullet} = \ab(p_1) - \ab(o_1) + \ab(p_2) - \ab(o_2) + \zeta,
\eeq
namely:
\beq
\frac{\nabla^{\otimes k}\vartheta_{\bullet}\bigl[{}^{\mu}_{\nu}\bigr]}{\vartheta_{\bullet}\bigl[{}^{\mu}_{\nu}\bigr]} =  \begin{array}{c}\includegraphics[width=0.2\textwidth]{blackv}\end{array},\qquad \nabla^{\otimes k}\ln\vartheta_{\bullet}\bigl[{}^{\mu}_{\nu}\bigr] = \begin{array}{c} \includegraphics[width=0.2\textwidth]{whitev}\end{array}.
\eeq
Thus, the result can be represented diagrammatically as follows:
$$
\psi_{\hc}^{[2|2]}(p_1,o_1;p_2,o_2)  = \frac{\ee{\frac{1}{\hc} \left(\int_{o_1}^{p_1} v\,\dd u + \int_{o_2}^{p_2} v\,\dd u \right)}\,E(p_1,p_2)\,E(o_1,o_2)}{E(p_1,o_1)\,E(p_2,o_2)\,E(p_1,o_2)\,E(o_1,p_2)}\,\frac{\vartheta_\bullet\bigl[{}^{\mu}_{\nu}\bigr]}{\vartheta\bigl[{}^{\mu}_{\nu}\bigr]}\,\,
\frac{\hat\psi_{\hc}^{[2|2]}(p_1,o_1\,;\,p_2,o_2)}{\hat{\mathcal{T}}_{\hc}},
$$
with:
\beq
\label{diaia} \ln\hat\psi_{\hc}^{[2|2]}(p_1,o_1\,;\,p_2,o_2)  
= \sum \frac{\hc^{-\chi_{\rm Euler}}}{\#\,{\rm Aut}}\times\mathrm{weight}\left(\begin{array}{c}
\includegraphics[width=0.4\textwidth]{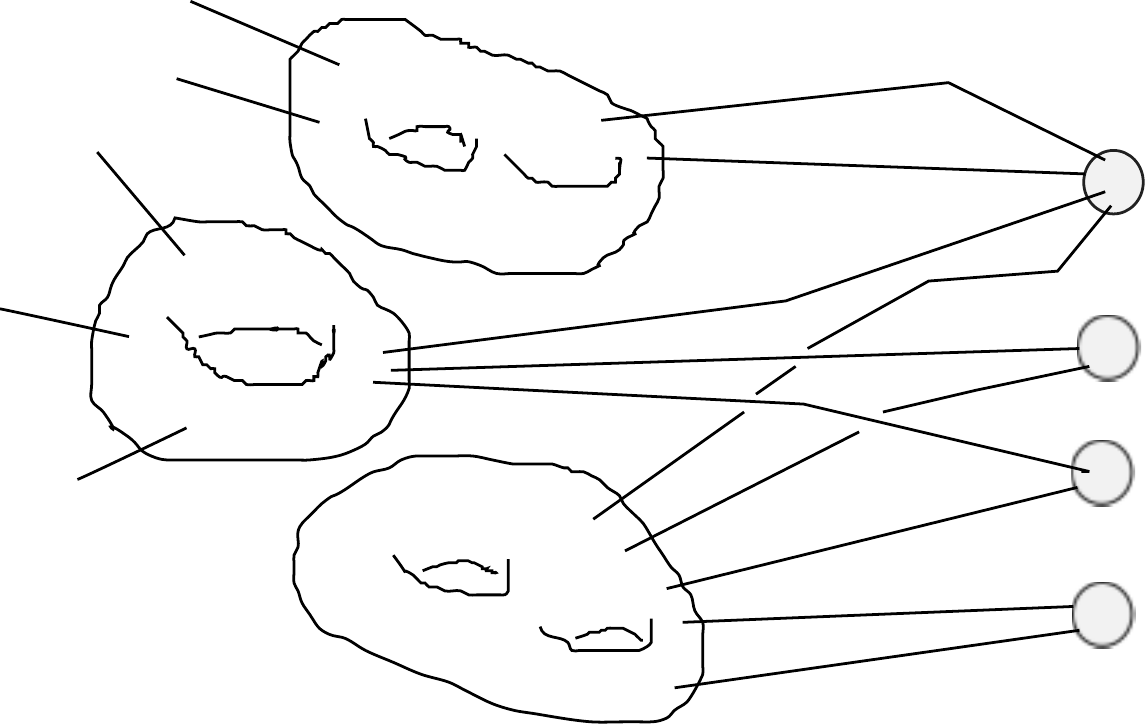}\end{array}\right),
\eeq
where now some legs attached to surfaces are not contracted with legs of white vertices. Those open legs are associated to $\int_{o_1,o_2}^{p_1,p_2} = \int_{o_1}^{p_1} + \int_{o_2}^{p_2}$ (or more generally $\int_{o_1}^{p_1} + \cdots \int_{o_{\mathrm{n}}}^{p_{\mathrm{n}}}$ if we wanted to compute the $\mathrm{n}|\mathrm{n}$ kernels). At each order $\chi$, the sum consists of a finite number of connected graphs whose sum of Euler characteristics of surfaces is $\chi_{\mathrm{Euler}}$. Notice that a surface with $n$ punctures and $h$ handles is associated to an $\omega_n^h$, which can itself be expressed as a sum over pants decomposition (or equivalently skeleton graphs) of that surface (see Fig.~\ref{toprec} and for a more detailed description \cite[Section 3]{EORev}).

\subsection{Perturbative knot invariants to first orders}

The central object in our conjecture concerning the asymptotics of the colored Jones was:
\beq
\ln \hat{\psi}_{\hc}^{[2|2]}(p,o\,;\,\iota(p),\iota(o)) = 2 \sum_{\chi \geq 0} \hc^{-\chi}\,\widetilde{\jmath}_{\chi}(p).
\eeq
and from Eqn.~\ref{diaia} they acquire a diagrammatic representation. When $\iota_* = -\mathrm{id}$, $\omega_n^{h,(d)}$ vanish whenever $d$ is odd, so the only graphs with non-zero weight are those where each surface is contracted with an even number of legs incident to a white vertex. We give below the two first orders in diagrams in this case.
\beq
\begin{split}
2\,\widetilde{\jmath}_1(p) & = \begin{array}{c}\includegraphics[width=0.4\textwidth]{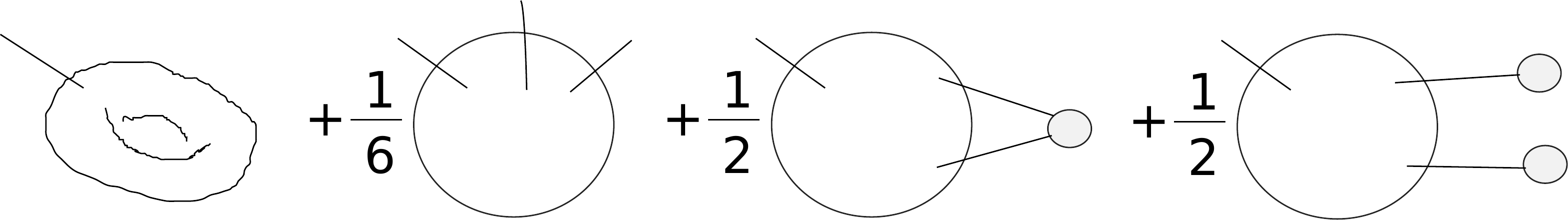}\end{array}, \\
2\,\widetilde{\jmath}_2(p) & = \begin{array}{c}\includegraphics[width=0.7\textwidth]{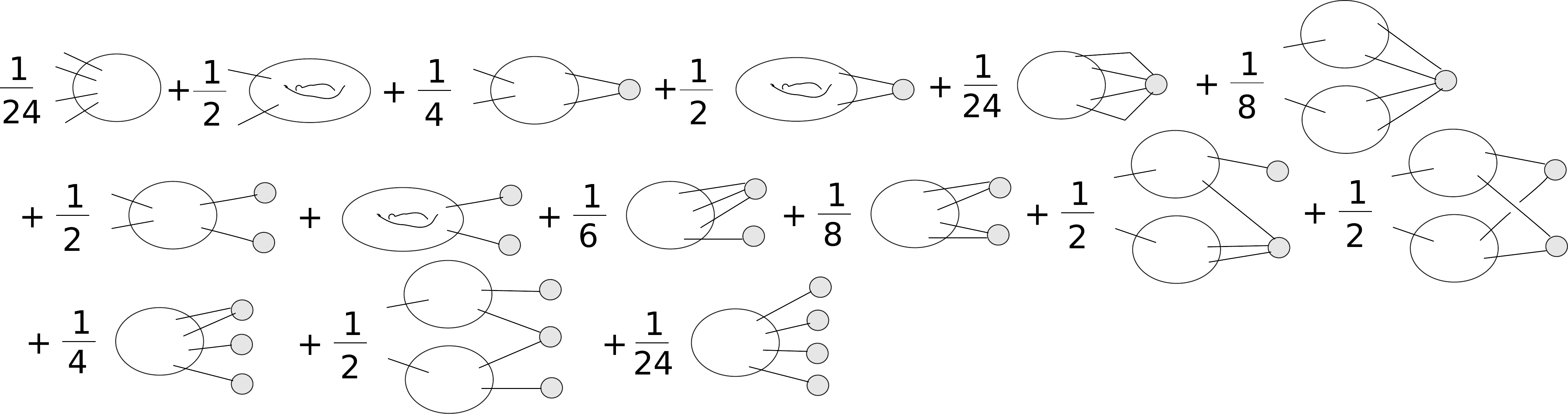}\end{array}.
\end{split}\eeq
and we can compare this diagrammatic representation to the (black terms in) the expressions given in \S~\ref{conmi}.

\newpage

\section{Some A-spectral curves}

\subsection{In Rolfsen classification}
\label{Fig1}

\begin{center}
\vspace{10pt}
\begin{tabular}{|c|l|}
\hline genus & knot complements \\ \hline
0 & $\mathbf{10}_{152}^{(1)}$ \\
1 & $\mathbf{4}_{1}\quad\mathbf{7}_{4}^{(1)}\quad\mathbf{8}_{18}^{(1)}\quad\mathbf{8}_{18}^{(2)}\quad\mathbf{8}_{18}^{(3)}\quad\mathbf{9}_{24}^{(1)}\quad\mathbf{9}_{35}^{(1)}\quad\mathbf{9}_{37}^{(1)}\quad\mathbf{9}_{48}^{(1)}\quad\mathbf{9}_{49}^{(1)}$ \\
& $\mathbf{10}_{139}\quad\mathbf{10}_{142}^{(1)}\quad\mathbf{10}_{145}^{(1)}\quad \mathbf{10}_{146}^{(1)}\quad\mathbf{10}_{147}^{(1)}\quad\mathbf{10}_{155}^{(1)}\quad L^2R\quad m129(0,3)$ \\
2 & $\mathbf{5}_2\quad \mathrm{Pretzel}(-2,3,7)\quad\mathbf{7}_4^{(2)}\quad\mathbf{7}_7^{(1)}\quad\mathbf{8}_5^{(1)}\quad\mathbf{9}_{37}^{(2)}\quad\mathbf{10}_{136}^{(1)}\quad\mathbf{10}_{154}^{(1)}$ \\
& $\mathbf{10}_{160}^{(1)}\quad\mathbf{10}_{163}^{(1)}$ \\ 
3 & $\mathbf{6}_{1}\quad\mathbf{7}_7^{(2)}\quad\mathbf{8}_5^{(2)}\quad\mathbf{9}_{35}^{(2)}\quad\mathbf{9}_{47}^{(1)}\quad\mathbf{9}_{48}^{(2)}\quad LR^3$ \\
4 & $\mathbf{7}_{2}\quad\mathbf{8}_{21}\quad\mathbf{9}_{10}^{(1)}\quad\mathbf{9}_{23}^{(1)}\quad\mathbf{9}_{46}\quad\mathbf{9}_{49}^{(2)}\quad\mathbf{10}_{61}^{(1)}\quad\mathbf{10}_{145}^{(2)}\quad\mathbf{10}_{146}^{(2)}$ \\
5 & $\mathbf{8}_{1}\quad\mathbf{8}_{20}\quad\mathbf{9}_{10}^{(2)}\quad\mathbf{9}_{17}^{(1)}\quad\mathbf{9}_{41}^{(1)}\quad\mathbf{9}_{47}^{(2)}\quad\mathbf{10}_{142}^{(2)}\quad\mathbf{10}_{144}^{(1)}$ \\
6 & $\mathbf{8}_{16}^{(1)}\quad\mathbf{9}_{17}^{(2)}\quad\mathbf{9}_{31}^{(1)}\quad\mathbf{10}_{136}^{(2)}\quad\mathbf{10}_{152}^{(2)}$ \\
7 & $\mathbf{6}_3\quad\mathbf{8}_2\quad\mathbf{9}_2\quad\mathbf{9}_{16}^{(1)}\quad\mathbf{9}_{41}^{(2)}$ \\
8 & $\mathbf{9}_{37}^{(3)}\quad\mathbf{9}_{42}\quad\mathbf{10}_1\quad\mathbf{10}_{61}^{(2)}\quad\mathbf{10}_{138}^{(1)}\quad\mathbf{10}_{138}^{(2)}\quad\mathbf{10}_{140}\quad\mathbf{10}_{141}^{(2)}$ \\
9 & $\mathbf{7}_5\quad\mathbf{8}_3\quad\mathbf{8}_4\quad\mathbf{8}_{16}^{(2)}\quad\mathbf{9}_3\quad\mathbf{9}_{10}^{(1)}\quad\mathbf{9}_{16}^{(2)}\quad\mathbf{9}_{28}^{(1)}\quad\mathbf{10}_{62}^{(1)}\quad\mathbf{10}_{154}^{(2)}\quad\mathbf{10}_{155}^{(2)}$ \\
10 & $\mathbf{10}_{2}\quad\mathbf{10}_{125}\quad\mathbf{10}_{132}$ \\
11 & $\mathbf{9}_{4}\quad\mathbf{9}_{23}^{(11)}$ \\
12 & $\mathbf{7}_{6}\quad\mathbf{8}_{15}\quad\mathbf{9}_{38}^{(1)}$ \\
13 & $\mathbf{7}_{3}\quad\mathbf{8}_{11}\quad\mathbf{9}_{43}\quad\mathbf{10}_{144}^{(2)}$ \\
14 & $\mathbf{8}_{7}\quad\mathbf{9}_5\quad$ \\
15 & $\mathbf{8}_{9}\quad\mathbf{8}_{10}\quad\mathbf{9}_6\quad\mathbf{10}_{3}\quad\mathbf{10}_{62}^{(2)}\quad\mathbf{10}_{128}\quad\mathbf{10}_{146}^{(3)}\quad\mathbf{10}_{161}\quad\mathbf{10}_{162}$ \\
16 & $\mathbf{8}_{6}\quad\mathbf{9}_{28}^{(16)}\quad\mathbf{9}_{29}^{(1)}\quad\mathbf{10}_{4}$ \\
17 & $\mathbf{9}_{31}^{(2)}\quad\mathbf{10}_{8}$ \\
18 & $\mathbf{10}_{126}\quad\mathbf{10}_{143}$ \\
19 & $\mathbf{8}_{8}\quad\mathbf{9}_{44}$ \\
20 & $\mathbf{9}_{9}\quad\mathbf{10}_{5}\quad\mathbf{10}_{160}^{(2)}$ \\
21 & $\mathbf{8}_{12}\quad\mathbf{9}_{7}\quad\mathbf{9}_{29}^{(2)}$ \\
22 & $\mathbf{10}_{153}$ \\
23 & $\mathbf{7}_4\quad\mathbf{8}_{13}\quad\mathbf{10}_{147}^{(2)}$ \\
24 & $\mathbf{9}_8$ \\
25 & $\mathbf{9}_{11}\quad\mathbf{10}_{133}$ \\
26 & $\mathbf{8}_{14}\quad\mathbf{9}_{38}^{(2)}\quad\mathbf{9}_{46}$ \\
28 & $\mathbf{9}_{12}$ \\
\hline
\end{tabular}
\vspace{10pt}
\end{center}

\noindent We present list of hyperbolic knot complement sorted by genus of their components, which is exhaustive up to 8 crossings (knots which do not appear have components of higher genus). Some knots with $9$ and $10$ crossings knots and once-punctured torus bundle (knot complement in lens spaces) were included. We also added $m129(3,0)$, which is studied in \cite{Villegas}, and is the orbifold obtained by $(0,3)$ Dehn surgery on the first cusp of $m129$. When the A-polynomial have several components which are not of the form $(lm^{a} + b)$, we indicate their label $(\alpha)$ in exponent.

\subsection{In tetrahedron census}

We present an exhaustive list of hyperbolic $3$-manifolds triangulated with atmost $6$ tetrahedra. Those which are complement of a knot with atmost 10 crossings were rather included in Fig.~\ref{Fig1}. For all the knots in this table, we observe that the A-polynomial has a single component.\label{Fig2}.

\begin{center}
\begin{tabular}{|c|l|}
\hline  genus & knot complements \\ \hline
 2 & $\mathbf{k3}_{1}$ \\ 
5 & $\mathbf{k4}_3\quad\mathbf{k4}_{4}\quad\quad\mathbf{k5}_{11}$ \\
6 & $\mathbf{k5}_{1}$ \\
7 &  $\mathbf{k6}_{32}$ \\
8 & $\mathbf{k5}_{5}\quad\mathbf{k5}_{10}\quad\mathbf{k5}_{14}\quad\mathbf{k5}_{15}\quad\mathbf{k5}_{16}\quad\mathbf{k6}_2\quad\mathbf{k6}_{33}$\\
9 &  $\mathbf{k5}_4\quad\mathbf{k5}_6\quad\mathbf{k6}_{42}$\\
10 &  $\mathbf{k5}_7\quad\mathbf{k6}_{19}$ \\
11 &  $\mathbf{k5}_{18}\quad\mathbf{k6}_5\quad\mathbf{k6}_{39}$ \\
12 & $\mathbf{k5}_{13}\quad\mathbf{k5}_{17}$ \\
13 & $\mathbf{k6}_3\quad\mathbf{k6}_{22}\quad\mathbf{k6}_{31}\quad\mathbf{k6}_{38}$ \\
14 & $\mathbf{k6}_{10}\quad\mathbf{k6}_{11}\quad\mathbf{k6}_{14}\quad\mathbf{k6}_{29}$ \\
15 & $\mathbf{k6}_4\quad\mathbf{k6}_6$ \\
16 & $\mathbf{k6}_{16}$ \\
17 & $\mathbf{k6}_{15}\quad\mathbf{k6}_{17}\quad\mathbf{k6}_{25}\quad\mathbf{k6}_{27}$ \\
18 & $\mathbf{k6}_7\quad\mathbf{k6}_{18}$ \\
20 & $\mathbf{k6}_{13}\quad\mathbf{k6}_{37}$ \\
21 & $\mathbf{k6}_8\quad\mathbf{k6}_{12}$ \\
22 & $\mathbf{k6}_9\quad\mathbf{k6}_{30}$ \\
24 & $\mathbf{k6}_{21}\quad\mathbf{k6}_{35}$ \\
\hline
\end{tabular}
\end{center}

\subsection{Properties}

\label{Fig34}

We present properties of spectral curves for various knots. Each block collect equivalent curves modulo birational transformations. Notice that $l \rightarrow C\, l^{\pm}m^a$ implies $\omega_n^h \rightarrow (\pm)^{n}\omega_n^h$. The column $g$ gives the genus of the curve, and the column $g_{\iota}$ gives the genus of the quotient curve $\mathcal{C}/\iota$, i.e. the number of $+1$ eigenvalues of $\iota_*$. In the column H, we indicate if $\iota$ coincide or not with the hyperelliptic involution. If this is the case, we necessarily have $\iota_* = -\mathrm{id}$. $|\{a\}|$ indicates the number of ramification points. They are all simple, except when we indicate with a superscript $+1$ the presence of one extra ramification point of order $3$ at $(m,l) = (-1,1)$. We then indicate the minimal positive integer $\varsigma$ such that $2\varsigma\cdot\{m,l\} = 0 \in$ in $\mathrm{K}_2(\mathcal{C})$. When the knot is amphichiral and when the component is stable under $\alpha(m,l) = (1/m,l)$, we indicate the number of $+1$ eigenvalues of the induced map $\alpha_*$ in homology. We put question marks when we could not obtain the answer in a reasonable time with \textsc{maple}.

\begin{center}
\vspace{1cm}
\begin{tabular}{|l|cccc|c|cc|}
\hline knot {\rule{0pt}{2.5ex}}{\rule[-1.2ex]{0pt}{0pt}}& $g$ & $g_{\iota}$ & H & $|\{a\}|$ & $\varsigma$ & amphicheiral & $\alpha_{*}$ \\\hline\hline
$\mathbf{4}_1$ {\rule{0pt}{2.5ex}} & 1 & 0 & yes & $4$ & 1 & yes & $0$ \\
$\mathbf{7}_4^{(1)}$ & \multicolumn{4}{|l|}{deduced from $\mathbf{4}_1$ by $l \rightarrow lm^4$} & 1 & no & \\
$\mathbf{8}_{18}^{(1)}$ & 1 & 0 & no & $4$ & 2 & yes &  \\
$\mathbf{8}_{18}^{(2)}$ & \multicolumn{4}{|l|}{deduced from $\mathbf{8}_{18}^{(1)}$ by $l \rightarrow l^{-1}$} & 2 & yes &    \\
$\mathbf{9}_{24}^{(1)}$ & \multicolumn{4}{|l|}{idem $\mathbf{4}_1$}  & 1 & no & \\
$\mathbf{9}_{37}^{(1)}$ & \multicolumn{4}{|l|}{idem $\mathbf{4}_1$} & 1 & no & \\
$\mathbf{9}_{49}^{(1)}$ & \multicolumn{4}{|l|}{idem $\mathbf{4}_1$ with $l \rightarrow lm^8$} & 1 & no &   \\
$\mathbf{10}_{142}^{(1)}$ & \multicolumn{4}{|l|}{idem $\mathbf{4}_1$ with $l \rightarrow lm^{12}$} & 1 & no & \\
$\mathbf{10}_{145}^{(1)}$ & \multicolumn{4}{|l|}{idem $\mathbf{4}_1$ with $l \rightarrow -lm^{-2}$} & 2 & no & \\
$\mathbf{10}_{146}^{(1)}$ & \multicolumn{4}{|l|}{idem $\mathbf{4}_1$ with $l \rightarrow -lm^{-6}$} & 2 & no &  \\
$\mathbf{10}_{147}^{(1)}$ & \multicolumn{4}{|l|}{idem $\mathbf{4}_1$ with $l \rightarrow -l^{-1}m^2$} & 2 & no & \\
$\mathbf{10}_{155}^{(1)}$ {\rule[-1.2ex]{0pt}{0pt}}& \multicolumn{4}{|l|}{idem $\mathbf{8}_{18}^{(1)}$ with $l \rightarrow l^{-1}m^{-4}$} & 2 & no &  \\
\hline
$\mathbf{8}_{18}^{(3)}$  {\rule{0pt}{2.5ex}}{\rule[-1.2ex]{0pt}{0pt}} & 1 & 0 & ? & $6$ & 1 & yes & $0$ \\ \hline 
$\mathbf{9}_{35}^{(1)}$  {\rule{0pt}{2.5ex}}{\rule[-1.2ex]{0pt}{0pt}} & 1 & 0 & yes & $4^{+ 1}$ & 2 & no & \\ \hline
$\mathbf{9}_{48}^{(1)}$ {\rule{0pt}{2.5ex}}{\rule[-1.2ex]{0pt}{0pt}} & 1 & 0 & no & $2^{+ 1}$ & 1 & no & \\ \hline
$\mathbf{10}_{139}$ {\rule{0pt}{2.5ex}}{\rule[-1.2ex]{0pt}{0pt}} & 1 & 0 & yes & $6$ & 2 & no & \\ \hline
$L^2R$ {\rule{0pt}{2.5ex}}{\rule[-1.2ex]{0pt}{0pt}} & 1 & 0 & yes & $4$ & 1 & no & \\
\hline\hline
$\mathbf{5}_2$ {\rule{0pt}{2.5ex}} & 2 & 0 & yes & $8$ & 2 & no & \\
$\mathbf{7}_{7}^{(1)}$ & \multicolumn{4}{|l|}{idem $\mathbf{5}_2$ with $l \rightarrow lm^{-4}$} & 2 & no &\\
$\mathbf{8}_5^{(1)}$ & \multicolumn{4}{|l|}{idem $\mathbf{5}_{2}$ with $l \rightarrow l^{-1}m^{-12}$} & 2 & no & \\
$\mathbf{10}_{154}^{(1)}$ & 2 & 0 & yes & 8 & 2 & no &  \\
$\mathbf{10}_{160}^{(1)}$ & \multicolumn{4}{|l|}{idem $\mathbf{5}_2$ with $l \rightarrow lm^8$} & 2 & no & \\
$\mathbf{10}_{163}^{(1)}$ & \multicolumn{4}{|l|}{idem $\mathbf{5}_2$ with $l \rightarrow lm^{-8}$} & 2 & no &  \\
$\mathrm{P}(-2,3,7)$ & \multicolumn{4}{|l|}{idem $\mathbf{10}_{154}^{(1)}$ with $l \rightarrow l^{-1}m^{-26}$} & 2 & no &\\
$\mathbf{k3}_1$ {\rule[-1.2ex]{0pt}{0pt}} & \multicolumn{4}{|l|}{idem $\mathbf{10}_{154}^{(1)}$ with $l \rightarrow -lm^{-26}$} & 2 & no & \\
\hline
$\mathbf{7}_4^{(2)}$ {\rule{0pt}{2.5ex}} & 2 & 0 & yes & $6$ & 2 & no &\\
$\mathbf{9}_{37}^{(2)}$ {\rule[-1.2ex]{0pt}{0pt}} & \multicolumn{4}{|l|}{idem $\mathbf{7}_{4}^{(2)}$ with $l \rightarrow l^{-1}m^{-4}$} & 2 & no & \\
\hline
$\mathbf{10}_{136}^{(1)}$ {\rule{0pt}{2.5ex}}{\rule[-1.2ex]{0pt}{0pt}}& 2 & 0 & yes & $6$ & 2 & no &  \\ \hline\hline
\end{tabular}
\end{center}

\label{Fig35} 
\begin{center}
\begin{tabular}{|l|ccc|c|cc|}
\hline knot {\rule{0pt}{2.5ex}}{\rule[-1.2ex]{0pt}{0pt}}& $g$ & $g_{\iota}$ & $|\{a\}|$ & $\varsigma$ & amphicheiral & $\alpha_{*}$ \\ \hline
$\mathbf{6}_1$ {\rule{0pt}{2.5ex}}& 3 & 0 & $12$ & 1 & no &  \\
$\mathbf{9}_{35}^{(2)}$   & \multicolumn{3}{|l|}{idem $\mathbf{6}_1$ with $l \rightarrow lm^4$} & 1 & no & \\
$\mathbf{9}_{47}^{(1)}$ & \multicolumn{3}{|l|}{idem $\mathbf{6}_1$ with $l \rightarrow lm^8$} & 1 & no &\\
$\mathbf{9}_{48}^{(2)}$ {\rule[-1.2ex]{0pt}{0pt}}  & \multicolumn{3}{|l|}{idem $\mathbf{6}_1$ with $l \rightarrow l^{-1}m^{-4}$} & 1 & no &  \\
\hline  $\mathbf{7}_{7}^{(2)}$  {\rule{0pt}{2.5ex}}{\rule[-1.2ex]{0pt}{0pt}} & 3 & 0 & $8$ & 2 & no & \\
\hline $\mathbf{8}_{5}^{(2)}$ {\rule{0pt}{2.5ex}}{\rule[-1.2ex]{0pt}{0pt}} & 3 & 0 & $14$ & 2 & no &  \\
\hline $LR^3$ {\rule{0pt}{2.5ex}}{\rule[-1.2ex]{0pt}{0pt}} & 3 & 0 & $10$ & 1 & no & \\
\hline\hline
$\mathbf{7}_2 $ {\rule{0pt}{2.5ex}}{\rule[-1.2ex]{0pt}{0pt}} & 4 & 0  & $16$ & 2 & no & \\
\hline
$\mathbf{8}_{21}$ {\rule{0pt}{2.5ex}}{\rule[-1.2ex]{0pt}{0pt}} & 4 & 1 & $12$ & 2 & no &\\ \hline
$\mathbf{9}_{10}$ {\rule{0pt}{2.5ex}}{\rule[-1.2ex]{0pt}{0pt}} & 4 & 0 & $12$ & 2 & no & \\ \hline
$\mathbf{9}_{23}^{(1)}$ {\rule{0pt}{2.5ex}}{\rule[-1.2ex]{0pt}{0pt}} & 4 & 0 & $12$ & 2 & no &\\ \hline
$\mathbf{9}_{46}$ {\rule{0pt}{2.5ex}}{\rule[-1.2ex]{0pt}{0pt}} & 4 & 1 & $10$ & 1 & no &\\ \hline
$\mathbf{9}_{49}^{(2)}$ {\rule{0pt}{2.5ex}}{\rule[-1.2ex]{0pt}{0pt}} & 4 & 1 & $2$ & 2 & no &\\ \hline
$\mathbf{10}_{61}^{(1)}$  {\rule{0pt}{2.5ex}}  & 4 & 0 & $16$ & 2 & no &\\
$\mathbf{10}_{146}^{(2)}$ {\rule[-1.2ex]{0pt}{0pt}} & \multicolumn{3}{|l|}{idem $\mathbf{10}_{61}^{(2)}$ with $l \rightarrow lm^{-8}$} & 2 & no &\\
\hline
$\mathbf{10}_{145}^{(2)}$ {\rule{0pt}{2.5ex}}{\rule[-1.2ex]{0pt}{0pt}}  & 4 & 1 & $14$ & 2 & no &\\
\hline\hline
$\mathbf{8}_1$ {\rule{0pt}{2.5ex}}{\rule[-1.2ex]{0pt}{0pt}}& 5 & 0 & $20$ & 1 & no &
\\ \hline
$\mathbf{8}_{20}$ {\rule{0pt}{2.5ex}}{\rule[-1.2ex]{0pt}{0pt}} & 5 & 1 & $12$ & 2 & no & \\
 \hline
 $\mathbf{9}_{17}^{(1)}$ {\rule{0pt}{2.5ex}} & 5 & 0 & $18$ & 2 & no &\\
 $\mathbf{10}_{144}^{(1)}$ {\rule[-1.2ex]{0pt}{0pt}}& \multicolumn{3}{|l|}{idem $\mathbf{9}_{17}^{(2)}$ with $l \rightarrow l^{-1}m^8$} & 1 & no &\\ \hline
 $\mathbf{10}_{142}^{(2)}$ {\rule{0pt}{2.5ex}}{\rule[-1.2ex]{0pt}{0pt}} & 5 & 1 & $18$ & 2 & no & \\
 \hline
 $\mathbf{k4}_3$ {\rule{0pt}{2.5ex}}{\rule[-1.2ex]{0pt}{0pt}} & 5 & 0 & $16$ & 2 & no &  \\
 \hline
 $\mathbf{k4}_4$  {\rule{0pt}{2.5ex}}{\rule[-1.2ex]{0pt}{0pt}} & 5 & 0 & $18$ & 2 & no &\\
 \hline
 $\mathbf{k5}_{11}$  {\rule{0pt}{2.5ex}}{\rule[-1.2ex]{0pt}{0pt}} & 5 & 0 & $14$ & 1 & no & \\
\hline\hline
$\mathbf{10}_{136}^{(2)}$ {\rule{0pt}{2.5ex}}{\rule[-1.2ex]{0pt}{0pt}} & 6 & 2 & $16$ & 2 & no & \\
\hline\hline
$\mathbf{6}_3$ {\rule{0pt}{2.5ex}} & 7 & 1 & $12$ & 2 & yes & 3 \\
$\mathbf{9}_{16}^{(1)}$ {\rule[-1.2ex]{0pt}{0pt}}  & \multicolumn{3}{|l|}{idem $\mathbf{6}_3$ with $l \rightarrow lm^{12}$} & 1 & no & \\
\hline
$\mathbf{9}_2$ {\rule{0pt}{2.5ex}}{\rule[-1.2ex]{0pt}{0pt}} & 7 & 0 & $24$ & 2 & no & \\
\hline\hline
\end{tabular}
\end{center}

\newpage 
\frenchspacing
\bibliographystyle{QT}

\end{document}